\documentclass[a4paper,11pt]{article}

\usepackage[a4paper,
  left=2.5cm, right=2.5cm,
  top= 3cm, bottom=4cm]{geometry}
\usepackage{amsmath}
\usepackage{oldgerm}
\usepackage{amssymb}
\usepackage{bbm}
\usepackage[dvips]{graphicx}
\usepackage{epsfig}
\usepackage{color}
\usepackage{cite}
\usepackage{epic}
\usepackage{hyperref}

\usepackage{empheq}

\numberwithin{equation}{section}

\usepackage[utf8]{inputenc}

\usepackage{graphicx}
\usepackage{mathrsfs}
\usepackage{amsmath}
\usepackage{amssymb}
\usepackage{braket} 
\usepackage{hyperref}
\usepackage{frontespizio}
\usepackage{comment}
\usepackage{epsfig}
\usepackage{float}
\usepackage{color}
\usepackage[title]{appendix}
\usepackage{multirow}
\usepackage{tikz}
\usetikzlibrary{decorations.pathmorphing}
\usepackage{varwidth}
\usepackage{tikz}
\usetikzlibrary{backgrounds,patterns,calc}
\usepackage{xparse}
\usepackage{subfigure}
\usepackage{caption}
\usepackage{enumitem}
\setlist[itemize]{leftmargin=*}

\usetikzlibrary{decorations.pathmorphing}

\tikzset{zigzag/.style={decorate, decoration=zigzag}}

\newcommand{\citep}{\cite}

\newcommand{\bea}{\begin{eqnarray}}
\newcommand{\eea}{\end{eqnarray}}
\newcommand{\be}{\begin{equation}}
\newcommand{\ee}{\end{equation}}
\newcommand{\ba}{\begin{align}}
\newcommand{\ea}{\end{align}}

\usepackage{subfigure}

\begin{document}

\begin{titlepage}

\vskip -0.5cm 
 
\begin{flushright}

\end{flushright}
 
\vskip 1cm
\begin{center}
 
{\huge \bf Lorentzian Vacuum Transitions:\vskip 0.2cm
    Open or Closed Universes?} 
 
 \vskip 1.2cm
 
 {\bf { Sebasti\'an C\'espedes$^{a}$, Senarath P. de Alwis$^b$, Francesco Muia$^c$, Fernando Quevedo$^c$ }}

 \vskip 1cm
 
 { {\it $^{a}$\small \it Instituto de F\'{i}sica Te\'orica UAM-CSIC
C/ Nicol\'as Cabrera 13-15, \\ Campus de Cantoblanco, 28049 Madrid, Spain}}  \\ 
  {\small {\it $^{b}$ Physics Department, University of Colorado, Boulder, CO 80309 USA}}\\
{\small {\it $^{c}$ DAMTP, Centre for Mathematical Sciences, Wilberforce Road,  Cambridge, CB3 0WA, UK}}
 
 \vskip 2.2cm
 
\abstract{\small{We consider the generalisation of quantum tunneling transitions in the WKB approximation to  the  time-independent functional Schr\"odinger and Wheeler-DeWitt equations. Following a Lorentzian approach, we compute the transition rates among different scalar field vacua and compare with  those performed by Coleman and collaborators using the Euclidean approach. For gravity, we develop a general formalism for computing transition rates in  Wheeler's superspace.   This is then applied to computing decays in flat space and then to transitions in the presence of gravity. In the latter case we point out the complexities arising from having non-positive definite kinetic terms illustrating them in the simplified context of mini-superspace. This corresponds to a generalisation of the well-known `tunneling from nothing' scenarios. While we can obtain the leading term for the transitions  obtained by Euclidean methods we also point out some differences and ambiguities. We show that there is no obstruction to keeping the spherically ($SO(4)$) symmetric closed slicing for the new vacuum after a de Sitter to de Sitter transition. We argue that this is the natural Lorentzian realisation of the  Coleman-De Luccia instanton and that a closed universe is also obtained if the mini-superspace assumption is relaxed. This is contrary to the open universe
predicted by Coleman-De Luccia which relies on an analytic continuation  performed {\it after} bubble nucleation. Our findings may have important   cosmological implications related to the origin of inflation and to the string landscape. In particular, they question the widespread belief that evidence for a closed universe would rule out the string landscape.
} }

\end{center}

\end{titlepage}

\tableofcontents

\newpage

\section{Introduction}
\label{sec:Introduction}

Vacuum decay may  be the cause of both the beginning and end of our universe. It is also  one of the very few  systems in which the quantum aspects of gravity are crucial in order to have a proper description of the physical process. Euclidean methods, developed mostly by Coleman and collaborators~\cite{Coleman:1977py, Callan:1977pt, Coleman:1980aw}, have been used to extend the well understood WKB quantum mechanics techniques  to field theory and gravity. Decay rates have been computed for simple systems, corresponding to scalar field potentials with several minima, mostly in the thin-wall approximation. Related investigations by Brown and Teitelboim involved studying the nucleation of branes interpolating between vacua of different values of the cosmological constant~\cite{Brown:1988kg}. However, while the quantum mechanical calculations are well under  control, the extensions to field theory and gravity rely on  extrapolations such as analytic continuations and approximations, such as dilute instantons, that are not fully justified especially in the presence of gravity (for a review see for instance~\cite{Weinberg:2012pjx} and for a recent comprehensive and critical discussion see~\cite{Andreassen:2016cvx}).

A Hamiltonian approach to vacuum transitions that describes them directly without the need to use Euclidean techniques was developed by Fischler, Morgan and Polchinski (FMP)~\cite{Fischler:1989se,Fischler:1990pk}. Solving the Hamiltonian constraints and the Israel matching conditions for the system of two spacetimes with different cosmological constants, together with the bubble  wall (brane) separating them,  allowed them to compute the transitions from a Schwarzschild black hole (i.e. spherically symmetric asymptotically Minkowski) space-time to a de Sitter (dS)  space-time, but the formalism applies to all vacuum transitions considered by Coleman-De Luccia (CDL). The motivation for their calculation was the series of papers by Guth and collaborators on the theme of creating an inflating ‘universe in the lab’ culminating in the FGG work~\cite{Farhi:1989yr}. The latter was based on a Euclidean instanton construction whose validity was somewhat questionable (as pointed out by the authors of the paper themselves). The reason was that this instanton  was singular and the question of whether it should be included in the Euclidean functional integral became an issue. However the  fact that the Hamiltonian calculation of the decay rate agrees with the Euclidean approach, implied (as FMP argued)  that the final results for the transition rates are robust despite the fact that  Euclidean singular instanton calculation was not well defined.

One important difference between the Euclidean and Hamiltonian approaches is the fact that in the Euclidean approach a series of analytic continuations (which go beyond what may be justified by WKB quantum mechanics) are needed in order to find the  Lorentzian geometry after the transition. Even though the starting point may be a closed universe with $SO(4)$ spherical symmetry, the resulting geometry (inside the light cone of an observer at the centre of the nucleated bubble) after the transition turns out to correspond to an open universe with hyperbolic symmetry. However in the Hamiltonian approach of FMP there is no such implication. The entire analysis is done within the context of a spherically symmetric ($SO(3)$) ansatz   but the calculation does not imply the CDL argument for an open universe\footnote{As is well known dS space allows several foliations including those that correspond to closed, open and flat slicings. In CDL the symmetries  of the scalar field determine the preferred slicing after analytic continuation. However the latter cannot be justified by standard WKB arguments and this argument is not meaningful in the Hamiltonian formalism.}. Furthermore the natural Lorentzian mini-superspace calculation corresponding to the CDL Euclidean calculation would have $SO(4)$ symmetry and hence necessarily gives rise to a closed universe as we discuss later. This is an important difference especially if we consider the possibility that our own universe could be the result of a vacuum transition. Actually, based on the CDL result, it has been claimed that this is the only generic prediction of the string landscape~\cite{Freivogel:2005vv, Kleban:2012ph}\footnote{For a different view based on the ``no boundary wave function'', see the extensive work of Hawking, Hartle and Hertog (for example \cite{Hartle:2013oda}  and references therein). In particular in \cite{Hawking:2017wrd} it has been argued based on a dS/CFT conjecture that the probability of observing negative curvature on exit from eternal inflation is exponentially suppressed.
}. 

To the best of our knowledge, the fact that the Hamiltonian formalism developed by FMP can give rise to a closed universe has not been emphasised so far. Probably because the approach of FMP was originally developed in order to address the question of Minkowski black hole to dS transition that was posed  by~\cite{Farhi:1989yr} in the Euclidean approach. However, the FMP formalism applies also to all other potential vacuum transitions (dS to dS, Minkowski to dS as well as transitions involving AdS~\cite{Freivogel:2005qh,Bachlechner:2016mtp, deAlwis:2019dkc,Fu:2019oyc,Mirbabayi:2020grb}). One limitation of this approach is that it only describes transitions among two spacetimes differing by the vacuum energy  separated by a brane, whereas in general the CDL approach is formulated in terms of scalar potentials with different minima and barriers between them. However, in actual practice most such calculations reverted to the thin-wall approximation so that there was no essential difference to having the spaces separated by a brane as in the Brown-Teitelboim~\cite{Brown:1988kg} (BT) calculation. Furthermore the landscape of string theory results from transitions due to the nucleation of branes, which in the EFT approach are of string scale thickness, and hence effectively a thin wall so, as discussed by Bousso and Polchinski~\cite{Bousso:2000xa} for instance, the BT process (and hence FGG/FMP)  is more relevant for the landscape of string theory than the scalar field process of CDL. The latter is more relevant for questions of eternal inflation and other transitions such as the transition towards decompactification, however.

A generalisation of the FMP formalism to include explicit scalar field potentials is still an open question. Here, we are only partially successful in addressing this since we do it only in a mini-superspace model in which the metric and scalar field only depend on time. Nevertheless we find several interesting results essentially extending the `tunneling from nothing' arguments of Hartle-Hawking, Vilenkin and Linde~\cite{Vilenkin:1982de, Hartle:1983ai, Vilenkin:1984wp,Linde:1983mx}, and then compare with the Euclidean approach\footnote{Of course in the context of the  ``no boundary wave function'' Hawking and collaborators have long advocated for a landscape of closed universes. See for example~\cite{Hartle:2013oda} and references therein.}. We start by revisiting the extension of the WKB approximation to field theory by solving the functional, time-independent Schr\"odinger equation in the WKB approximation. We reproduce the Euclidean results for the (exponential term in the) decay rates but within a totally different method. The pre-factor however is different and there is no problem with negative modes \footnote{For a recent discussion of the negative modes issues see for instance~\cite{Jinno:2020zzs} and references therein.}. Next we include gravity albeit in a mini-superspace model and then we discuss the extension (for the case of a brane) to an $SO(3)$ symmetric situation following FMP and our earlier work. One of our concrete results is to confirm the fact that in the Hamiltonian approach the end result for the geometry of the remaining universe may be a closed rather than an open universe. We then study the physical implications of this result comparing with previous studies of CDL transitions.

The article is organised as follows:
\begin{itemize}
\item In Sec.~\ref{sec:WKBandWDW} we develop the formalism to address quantum transitions in Wheeler's superspace adapting the semi-classical WKB approximation in a  covariant superspace approach. Given the nature of the Hamiltonian constraint leading to  the Wheeler-DeWitt equation, the corresponding Schr\"odinger wave functional is time independent. The transition probabilities  are ratios of squares of wave functionals for the different configurations. Expanding on standard WKB techniques in quantum mechanics and on previous approaches towards field theory~
\cite{Banks:1973ps, Banks:1974ij, Gervais:1977nv, Bitar:1978vx,  Tanaka:1993ez}, we find general expressions, covariant on a generalised Wheeler superspace, for the leading and next order corrections to the wave functions. In particular, we find a general closed expression for the semi-classical wave functional with the pre-factor given by the analog of the Van Vleck determinant.

\item In Sec.~\ref{sec:FlatSpace} we apply the formalism of Sec.~\ref{sec:WKBandWDW} to the flat space field theoretical case of a scalar field potential neglecting the effects of gravity. In particular, following the textbook quantum mechanics matching conditions we explicitly compute the S-matrix as well as the lifetime of the corresponding resonances and obtain the decay rate.
In contrast to the calculations of Coleman and collaborators~\cite{Coleman:1977py, Callan:1977pt, Coleman:1980aw}, in our calculation we find that there is no issue with negative modes or analytic continuation of a manifestly real amplitude to get an imaginary part to the energy that can be interpreted as a decay width. Instead in our calculation the decay width is identified in the standard way as the imaginary part of a complex pole in the S-matrix.

\item In Sec.~\ref{sec:DecayCurvedSpace} we include gravity but in order to have explicit results we concentrate  on the mini-superspace model in order to compare with the CDL Euclidean approach.  In this case superspace reduces to a two-dimensional space with coordinates the scalar field $\phi(t)$ and the metric scale factor $a(t)$. We find a general expression for the decay rate which in the thin-wall approximation  gives exactly the CDL result but with an unclear interpretation. Furthermore, the fact that the metric in superspace is not positive definite allows for classical paths to connect the two dS minima without the need to pass through the barrier modifying substantially the results for the transition rate as compared with CDL. We argue that both approaches may be addressing different questions.
We emphasise the difference between the two approaches. Even though the calculations in both cases can be said to be done in mini-superspace with an $SO(4)$ symmetry, in CDL, after analytic continuation the standard picture of vacuum transition with a wall separating the two dS spacetimes emerges turning the original $SO(4)$ symmetry into $SO(3,1)$. Whereas in the Lorentzian approach the $SO(4)$ symmetry remains. We also point out the main difference: the fact that CDL implies an open universe whereas we, as with the ‘tunneling from nothing’ scenarios and FMP, find a closed universe.

\item In Sec.~\ref{sec:OpenClosed} we quickly review the relevant points of FMP to extend the results of the previous sections beyond mini-superspace, though with the restriction that the matter sector includes only the two cosmological constants and no scalar field. We describe the trajectory of the wall after nucleation and find it similar to CDL with the curious fact that the speed of the wall reaches a maximum which is less than the speed of light. Then we review the CDL arguments to obtain an open universe and revise the other implications for early universe cosmology as addressed for instance in~\cite{Freivogel:2005vv}: impact on CMB, the number of inflation e-folds,  etc. We then concentrate on the implications of having a closed rather than an open universe after the transition. We point out the physical differences not only regarding the potential for measuring the curvature of the universe, but also on how inflation is obtained after the transition, address the constrains on the number of e-foldings and the effect on density perturbations. 

\item Finally we discuss open questions and give a general outlook in the concluding section.
\end{itemize}

\section{WKB for field theory and quantum gravity}
\label{sec:WKBandWDW}

In this section we develop a general formalism to generalise the WKB formalism for vacuum decay to the Wheeler-DeWitt wave function $\Psi$ in Wheeler's superspace. The Wheeler-DeWitt (WDW) equation is a constraint equation on the space of wave functionals that describe a gravitational system. The ratios of absolute squares of the WDW wave functionals for the different configurations ${\mathcal M}_1$ and ${\mathcal M}_2$ can then be interpreted as relative  probabilities for realising them.

\begin{equation}
{\mathcal P}({\mathcal M}_1\rightarrow {\mathcal M}_2) = \frac{|\Psi({\mathcal M}_2)|^2}{|\Psi({\mathcal M}_1)|^2}
\end{equation}

 The discussion below may be trivially specialised to the case of field theory in which case the (spatial integral of the) WDW equation is essentially the time independent Schr\"odinger equation.

\subsection{Semi-classical expansion  for the WDW equation}

We assume that space-time can be foliated into a family of non-intersecting spacelike three-slices that can be seen (at least locally) as the level surfaces of a scalar function $t$. The function $t$ can be interpreted as a global time function. Given a line element of the generic form
\begin{equation}
\label{eq:GeneralLineElement}
ds^2 = \left(-N_t^2 + N_i N^i\right) dt^2 + 2 N_i dt dx^i + \gamma_{ij} dx^i dx^j \,,
\end{equation}
where $N_t$, $N_i$ are lapse and shift, while $\gamma_{ij}$ is the spatial three-dimensional metric, the gravitational contribution to the Lagrangian of the system can be written as
\begin{equation}
\label{eq:GeneralLagrangian}
L_g = \int d^3x \, N_t \sqrt{\gamma} \left(K_{ij} K^{ij} - K^2 + {}^{(3)} R\right) \,,
\end{equation}
where $K_{ij} = \frac{1}{2 N_t} \left(\partial_i N_j + \partial_j N_i - \partial_0 \gamma_{ij}\right)$ is the extrinsic curvature, $K = \gamma^{ij} K_{ij}$ and ${}^{(3)} R$ is the intrinsic curvature of the three-dimensional slice. We denote the canonically conjugate momentum to a field $\Phi^M$  by $\pi_M$. The primary constraints of the system arising from Eq.~\eqref{eq:GeneralLagrangian} are $\pi_{N_{t}}\approx0,\,\pi_{N_{i}}\approx 0$, where $\approx$ means that they are constraints on the classical solutions. In the quantum case we correspondingly have constraints on the space of wave-functionals $\Psi(\Phi)$, where $\Phi$ collectively denotes the three-metric and matter fields (and their spatial derivatives up to second order) present in the system. Note that the classical constraints $\pi_{N_{t}}\approx0,\,\pi_{N_{i}}\approx 0$ imply that the wave function $\Psi$ is independent of $N_{t}$, $N_{r}$. The full system is described by $L_g + L_{\rm mat}$ where $L_{\rm mat}$ is the Lagrangian that described the matter present in the system.\\

The Hamiltonian constraint takes the general form
\begin{equation}
\label{eq:GeneralHamiltonian}
{\cal H=}\frac{1}{2}G^{MN}(\Phi)\pi_{M}\pi_{N}+f[\Phi]\approx0 \,.
\end{equation}
where $G_{MN}$ is the metric on the $d$-dimensional field space, including all the components of $\gamma_{ij}$ and the matter fields, and can be read from the kinetic terms of $L_g + L_{\rm mat}$. At the same time, the momentum constraint ${\cal P}_{i}\approx 0$ has to hold.  Up to operator ordering ambiguities which are fixed by demanding the derivatives with respect to the components $\Phi^{M}$ of $\Phi$ are covariant with respect to the metric $G_{MN}$ (which we emphasise is not positive definite in the presence of gravity), we have for the WDW equation (replacing $\pi_{M}\rightarrow-i\hbar\nabla_{M}$)

\begin{equation}
{\cal H} \Psi(\Phi) = \left[-\frac{\hbar^{2}}{2}G^{MN}(\Phi)\nabla_{M}\nabla_{N}+f[\Phi]\right]\Psi(\Phi) =0 \,.\label{eq:WD}
\end{equation}
As usual we can write
\begin{equation}
\Psi[\Phi]=e^{\frac{i}{\hbar}S[\Phi]} \,,\label{eq:Psi}
\end{equation}
and define the semi-classical expansion
\begin{equation}
S[\Phi]=S_{0}[\Phi]+\hbar S_{1}[\Phi]+O(\hbar^{2}).\label{eq:Sexpn}
\end{equation}
Substituting Eq.~\eqref{eq:Psi} and Eq.~\eqref{eq:Sexpn} in the WDW equation we can in principle determine recursively the semi-classical expansion coefficients. The lowest two orders give
\begin{eqnarray}
\frac{1}{2}G^{MN}\frac{\delta S_{0}}{\delta\Phi^{M}}\frac{\delta S_{0}}{\delta\Phi^{N}}+f[\Phi] & = & 0\label{eq:HJ} \,,\\
2G^{MN}\frac{\delta S_{0}}{\delta\Phi^{M}}\frac{\delta S_{1}}{\delta\Phi^{N}} & = & iG^{MN}\nabla_{M}\nabla_{N}S_{0} \,. \label{eq:prefactoreqn}
\end{eqnarray}
In Hamilton-Jacobi theory, which corresponds to the classical limit of the quantum calculation, $\pi_M=\frac{\partial S}{\partial\Phi^M}$. Observe that at a turning point $\pi_{M}=\frac{\delta S_{0}}{\delta\Phi^{M}}=0$ for all $M$, and
the semi-classical expansion breaks down since $S_{1}$ cannot be
determined. 

Let us now introduce on the selected spatial slice a set of integral curves, parametrised
by $s$, on the field manifold
\begin{equation}
C(s)\frac{d\Phi^{N}}{ds}=G^{MN}\frac{\delta S_{0}}{\delta\Phi^{M}}.\label{eq:curves}
\end{equation}
Given the constraints, the classical action (on a classical trajectory) becomes
\begin{eqnarray}
S_{0}[\Phi_{s}] & = & \int^{\Phi_{s}}\int_{X}\pi_{M}d\Phi^{M}  \label{eq:HJ2}\nonumber \\
 & = & \int^{s}ds'\int_{X}\frac{\delta S_{0}}{\delta\Phi^{M}}\frac{d\Phi^{M}}{ds'}  \nonumber \\
 & = & \int^{s}ds'C^{-1}(s')\int_{X}\frac{\delta S_{0}}{\delta\Phi^{M}}G^{MN}\frac{\delta S_{0}}{\delta\Phi^{N}}  \nonumber \\
 & = & -2\int^{s}ds'C^{-1}(s')\int_{X}f[\Phi_{s'}] \,.\label{eq:S0intf}
\end{eqnarray}
We have used Eq.~\eqref{eq:curves} in the third line and Eq.~\eqref{eq:HJ} in the fourth. Similarly from Eq.~\eqref{eq:prefactoreqn} after integrating over the spatial
slice $X$ we have
\begin{equation}
\frac{dS_{1}}{ds}=\int_{X}\frac{d\Phi^{N}}{ds}\frac{\delta S_{1}}{\delta\Phi^{N}}=\frac{i}{2}C^{-1}(s)\int_{X}\nabla^{2}S_{0}\label{eq:dS1}
\end{equation}
giving
\begin{equation}
S_{1}[\Phi_{s}]=\frac{i}{2}\int^{s}ds'C^{-1}(s')\int_{X}\nabla^{2}S_{0}[\Phi_{s'}] \,. \label{eq:S1}
\end{equation}
For a given parametrisation $C(s)$ one can in principle  solve the first order differential equation in Eq.~\eqref{eq:curves} (with $S_0$ given by Eq.~\eqref{eq:S0intf} to get $\Phi_{s}$ as a function of $s$ and substituting in Eq.~\eqref{eq:S1} the semi-classical correction $S_{1}$ is determined away from any turning points (caustics). 

One can also choose a parameter $\tau$ to be the distance function along the trajectories, defined as
\begin{equation}
d\tau^{2}\equiv\int_{X}\delta\Phi^{M}G_{MN}\delta\Phi^{N} \,,\label{eq:dtau}
\end{equation}
we get using Eq.~\eqref{eq:curves} and Eq.~\eqref{eq:HJ}
\begin{equation}
\left(\frac{d\tau}{ds}\right)^{2}=\int_{X}\frac{\delta\Phi^{M}}{ds}G_{MN}\frac{\delta\Phi^{N}}{ds}=-2C^{-2}(s)\int_{X}f[\Phi] \,.\label{eq:dtauds}
\end{equation}

Solving this for $C(s)$ we have from Eq.~\eqref{eq:S0intf} for the classical action $S_{0}$ (with an arbitrary parametrisation of the integration trajectory) 

\begin{equation}
S_{0}[\Phi_{s}]-S_{0}[\Phi_{0}]=\int_{0}^{s}ds'\left[\int_{X}\frac{d\Phi_{s'}^{M}}{ds'}G_{MN}\frac{d\Phi_{s'}^{N}}{ds'}\right]^{1/2}\left[\int_{X}(-2f[\Phi_{s'}]\right]^{1/2} \,.
\end{equation}

The classical path along which this has to be evaluated is of course
the one which extremises this action with the end points fixed. The
variational derivative may be worked out easily by observing that the variation of the first factor gives the left hand side of the geodesic equation on superspace. Introducing the metric compatible connection on superspace (Vilkovsky connection) $D/D\tau=\frac{d\Phi^{M}}{d\tau}\nabla_{M}$ we get
\begin{equation}
G_{PN} \, \Sigma[\Phi_{\tau}]\frac{D}{D\tau} \left(\Sigma[\Phi_{\tau}]\frac{d\Phi_{\tau}^{N}}{d\tau} \right)+\frac{\delta f[\Phi_{\tau}]}{\delta\Phi^{P}}=0 \,,
\end{equation}
where after doing the variation we have set $s=\tau$ and defined $\Sigma[\Phi_{\tau}]\equiv\sqrt{-2\int_{X}f[\Phi_{\tau}]}$. 
 Or defining the affine parameter $\sigma$ by $d\sigma=d\tau/\Sigma[\Phi_{\tau}]$ (note that this corresponds to $s$ if we choose $C(s)=1$ in Eq.~\eqref{eq:dtauds}) we
have the classical equations of motion (that would follow also directly from the Hamiltonian in Eq.~\eqref{eq:GeneralHamiltonian})
\begin{equation}
\label{eq:GeneralEOM}
\frac{D}{D\sigma}\frac{d\Phi_{\sigma}^{N}}{d\sigma}+G^{NP}\frac{\delta f[\Phi_{\sigma}]}{\delta\Phi^{P}}=0 \,.
\end{equation}
We remark in passing that this equation of motion does not have an obvious interpretation as Lorentzian or Euclidean since in the presence of gravity (as we stressed before) the superspace metric $G_{MN}$ is not positive definite. We will see the consequences of this explicitly when we discuss the mini-superspace example.
Now going back to Eq.~\eqref{eq:dtauds} and putting $s=\tau$,
\begin{equation}
C^{2}(\tau)=-2\int_{X}f[\Phi].\label{eq:ctau}
\end{equation}
Hence in terms of $\tau$ we have 
\begin{empheq}[box=\fbox]{align}
\quad S_{0}[\Phi_{\tau}] &=\int^{\tau}d\tau'\sqrt{-2\int_{X}f[\Phi_{\tau'}]} \,.\quad \nonumber \\ 
\quad S_{1}[\Phi_{\tau}] & =\frac{i}{2}\int^{\tau}d\tau'\frac{1}{\sqrt{-2\int_{X}f[\Phi_{\tau'}]}}\int_{X}\nabla^{2}S_{0}[\Phi_{\tau'}] \,,\quad\label{eq:S1tau}
\end{empheq}

\subsection{Wave Function and van Vleck Determinant}
\label{sec:VanVleckDeterminant}

In the corresponding multi-dimensional quantum mechanical case the expression for $S_{1}$ is given by the VanVleck determinant~\citep{VanVleck:1928zz} (see also~\cite{Brown:1971zzc}). To see the connection
we first observe (essentially generalising an argument in~\cite{Tanaka:1993ez})
the following. Changing to the coordinates defined with respect to the orthonormal basis we have for a variation around a trajectory
\begin{equation}
\delta\Phi^{M}(x)=\delta\tau t^{M}(x)+\delta\lambda^{\bar{P}}\frac{\partial\Phi^{M}}{\partial\lambda^{\bar{P}}} \,,\label{eq:deltaPhi}
\end{equation}
where $t^M=\partial \Phi^M/\partial \tau$, while 
\begin{equation}
\label{eq:OrthogonalVects}
\frac{\partial}{\partial\lambda^{\bar{P}}} = \int_{X} \frac{\partial\Phi^{M}}{\partial\lambda^{\bar{P}}}\frac{\partial}{\partial \Phi^{M}} \,.
\end{equation}
The vectors defined in Eq.~\eqref{eq:OrthogonalVects} are orthogonal to the vector $\partial/\partial\tau=\int_{X}\left(\partial\Phi^{M}/\partial\tau\right)\partial/\partial \Phi^{M}$.

Let us denote the components in the original coordinates (including
the spatial position $x$ which is to be treated as an index as $A=\{M,x\}$. In other words it is convenient to use DeWitt's condensed notation treating the set of fields (including metric components) $\{\Phi^{M}\}$ as a set of ``coordinates'' i.e. $q^{i}\rightarrow\Phi^{Mx}$ where the spatial variable $x$  is treated as an index with the understanding that sums over $x$ are integrals and Kronecker deltas are replaced by Dirac delta functions. The superspace metric in the new coordinate system $(\bar{A}=\tau,\lambda^{\bar{P}})$ is 
\begin{equation}
\bar{G}_{\bar{A}\bar{B}}=\frac{\partial\Phi^{A}}{\partial\lambda^{\bar{A}}}G_{AB}\frac{\partial\Phi^{B}}{\partial\lambda^{\bar{B}}} \,,\label{eq:metricnew}
\end{equation}
therefore
\begin{eqnarray}
G^{AB}\nabla_{A}\nabla_{B}S_{0} & = & G^{\bar{A}\bar{B}}\nabla_{\bar{A}}\nabla_{\bar{B}}S_{0}=\frac{1}{\sqrt{\bar{G}}}\partial_{\bar{A}}\left(\sqrt{\bar{G}}G^{\bar{A}\bar{B}}\partial_{\bar{B}}S_{0}\right)\nonumber \\
 & = & \frac{1}{\sqrt{\bar{G}}}\partial_{\bar{A}}\left(\frac{\partial\lambda^{\bar{A}}}{\partial\Phi^{A}}\sqrt{\bar{G}}G^{AB}\partial_{B}S_{0}\right)=\frac{1}{\sqrt{\bar{G}}}\partial_{\bar{A}}\left(\frac{\partial\lambda^{\bar{A}}}{\partial\Phi^{A}}\sqrt{\bar{G}}C(\tau)\frac{\partial\Phi^{A}}{\partial\tau}\right)\nonumber \\
 & = & \frac{1}{\sqrt{\bar{G}}}\partial_{\bar{A}}\left(\sqrt{\bar{G}}C(\tau)\frac{\partial\lambda^{\bar{A}}}{\partial\tau}\right)=\frac{1}{\sqrt{\bar{G}}}\partial_{\tau}\left(\sqrt{\bar{G}}C(\tau)\right)\,. \label{eq:Laplacian}
\end{eqnarray}
In the second line we used Eq.~\eqref{eq:curves}. From Eq.~\eqref{eq:dS1} we have
\begin{equation}
\frac{dS_{1}}{d\tau}=\frac{i}{2}\partial_{\tau}\ln\left(C(\tau)\sqrt{\bar{G}}\right)=\frac{i}{2}\partial_{\tau}\ln\left(C(\tau)\det\frac{\partial\Phi^{A}}{\partial\lambda^{\bar{A}}}\sqrt{G}\right) \,.\label{eq:dS1-2}
\end{equation}
Noting that the line element on superspace may be rewritten as 
\begin{equation}
ds^{2}=\bar{G}{}_{\bar{A}\bar{B}}d\lambda^{\bar{A}}d\lambda^{\bar{B}}=d\tau^{2}+G_{AB}\frac{d\Phi^{A}}{d\lambda^{\bar{N}}}\frac{d\Phi^{B}}{d\lambda^{\bar{M}}}d\lambda^{\bar{N}}d\lambda^{\bar{M}} \,,\label{eq:lineelement}
\end{equation}
we have (using also Eq.~\eqref{eq:ctau}) 
\begin{equation}
S_{1}[\Phi_{\tau}]=\frac{i}{2}\ln\sqrt{-2\int_{X}f[\Phi_{\tau}]}+\frac{i}{2}\ln\sqrt{\left(\det G_{AB}\frac{d\Phi^{A}}{d\lambda^{\bar{N}}}\frac{d\Phi^{B}}{d\lambda^{\bar{M}}}\right)_{\tau}}+{\rm constant}.\label{eq:S1-2}
\end{equation}
Thus the integral in the second term of Eq.~\eqref{eq:S1tau} for $S_1$ is in fact the log of the determinant of the superspace metric in the orthogonal directions to the trajectory defined by $\partial/\partial\tau$. Thus the semi-classical wave function may be written as 
\begin{equation}
\boxed{\quad \Psi[\Phi_{\tau}]=\frac{\left[-2\int_{X}f[\Phi_{0}]\right]^{1/4}}{\left[-2\int_{X}f[\Phi_{\tau}]\right]^{1/4}}\frac{\left(\det G_{AB}\frac{d\Phi^{A}}{d\lambda^{\bar{N}}}\frac{d\Phi^{B}}{d\lambda^{\bar{M}}}\right)_{0}^{1/4}}{\left(\det G_{AB}\frac{d\Phi^{A}}{d\lambda^{\bar{N}}}\frac{d\Phi^{B}}{d\lambda^{\bar{M}}}\right)_{\tau}^{1/4}}e^{\frac{i}{\hbar}\left(S_{0}\left[\Phi_{\tau}\right]-S_{0}[\Phi_{0}]\right)}\Psi[\Phi_{0}] \,,\quad}\label{eq:Psi1}
\end{equation}
 with $S_{0}$ given by Eq.~\eqref{eq:S1tau}. This formula generalises one obtained for many particle Quantum Mechanics and canonical QFT in~\cite{Banks:1973ps,Banks:1974ij, Gervais:1977nv, Tanaka:1993ez}.

Alternatively we can rewrite the expression for this in terms of the
(generalisation of) the VanVleck determinant. To see this we go back to Eq.~\eqref{eq:dS1} and choose the parameter $s$ such that $C(s)=1$. Then the calculation in Eq.~\eqref{eq:Laplacian} shows that 
\begin{equation}
S_{1}[\Phi_{s}]-S_{1}\left[\Phi_{0}\right]=\frac{i}{2}\ln\left(\det\frac{\delta\Phi^{A}}{\delta\lambda^{\bar{A}}}\sqrt{G}\right) \,.
\end{equation}
Assuming that the complete integral of the Hamilton-Jacobi equation depends on a set of parameters $\alpha_{\bar{A}}$, we can identify the conjugate variables with our parameters $\lambda^{\bar{A}}$ , i.e.
\begin{equation}
\lambda^{\bar{A}}=\frac{\delta S_{0}[\Phi_{s};\alpha]}{\delta\alpha^{\bar{A}}} \,.\label{eq:lambdaconj}
\end{equation}
The VanVleck matrix can then be written as 
\begin{equation}
\left[\frac{\delta^{2}S_{0}}{\delta\Phi^{A}\delta\alpha^{\bar{A}}}\right]=\left[\frac{\delta\lambda^{\bar{A}}}{\delta\Phi^{A}}\right]=\left[\frac{\delta\Phi^{A}}{\delta\lambda^{\bar{A}}}.\right]^{-1}
\end{equation}
Therefore

\begin{equation}
S_{1}[\Phi_{s}]=-\frac{i}{2}\ln\left(\det\left[\frac{\delta^{2}S_{0}}{\delta\Phi^{A}\delta\alpha^{\bar{A}}}\right]\sqrt{G}\right)_{s}+{\rm constant} \,.\label{eq:S1-S0}
\end{equation}
Finally, the wave function with semi-classical corrections take the form
\begin{empheq}[box=\fbox]{align}
& \nonumber \\ 
\qquad \Psi[\Phi_{s}] &= \frac{1}{P_{0}}\sqrt[4]{G_{s}}\det\left[\frac{\delta^{2}S_{0}}{\delta\Phi^{A}\delta\alpha^{\bar{A}}}\right]_{s}e^{\frac{i}{\hbar}S_{0}\left[\Phi s\right]}\Psi[\Phi_{0}] \,,\,
\qquad \nonumber \\
\qquad S_{0}[\Phi_{s}] & =  \int_{o}^{s}ds'\sqrt{-2\int_{X}f[\Phi_{s'}]ds'}+{\rm constant} \,,\qquad
\\ \nonumber
\label{eq:Psi2}
\end{empheq}
with the constant $1/P_{0}$ fixed such that the left hand side at $s=0$ agrees with the right hand side at $s=0$.

A formula similar to this in the context of canonical field theory has been given by Bitar and Chang~\citep{Bitar:1978vx}. However the pre-factor was obtained there, not by  following the WKB method for getting it, but by switching to a functional integral over the fluctuations around the classical path. This gives them a pre-factor which is the inverse of the VanVleck pre-factor above. Furthermore these authors (in contrast to those of~\citep{Gervais:1977nv, Tanaka:1993ez}) claim agreement with the pre-factor in the Euclidean instanton analyis of Coleman et al.~\cite{Coleman:1977py, Callan:1977pt, Coleman:1980aw}. However we fail to see this. For instance the latter depended on the dilute gas approximation and relied crucially on the presence of a single negative mode in the fluctuations around it. Clearly in the above formula the issue of negative modes do not play a special role. In the next two sections we will  elaborate on these differences.

\section{WKB in flat space}
\label{sec:FlatSpace}

\begin{figure}[h!] 
\begin{center} 
\includegraphics[scale=0.7]{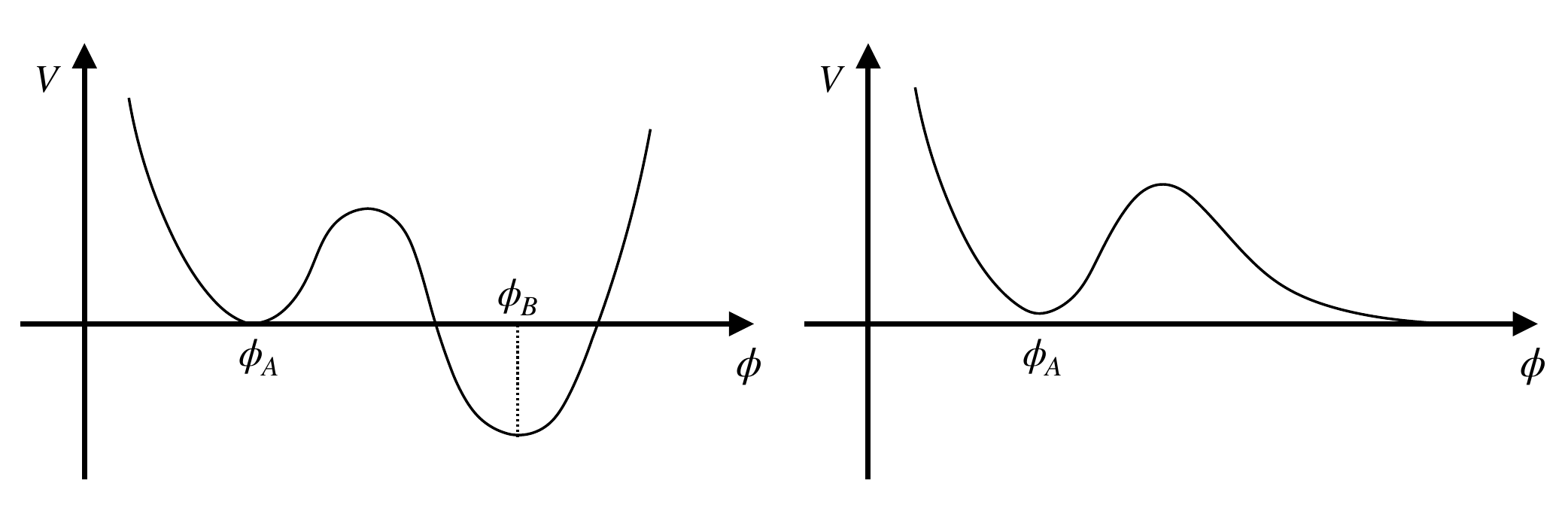}
\caption{Potential.\label{fig:PotentialFlatSpace}}
\end{center} 
\end{figure}

In this section we will recall how to use WKB for the study of vacuum decay in field theory, in flat space. We will analyse two different situations, corresponding to the two potentials in Fig.~\ref{fig:PotentialFlatSpace}. The left panel corresponds to the process of vacuum decay from a false vacuum to a true vacuum characterised by a lower energy density. In this context, we will show explicitly how the leading order final result corresponds to the CDL bounce, and how the analytic continuation is well justified by the fact that there is an under the barrier integral, which is equivalent to analytically continuing time to an imaginary variable. In this case, the WKB formalism can be used to compute the transmission coefficient $T^2 = \frac{|\psi(\phi_B)|^2}{|\psi(\phi_A)|^2}$, that gives a measure of the decay rate. In the case corresponding to the right panel of Fig.~\ref{fig:PotentialFlatSpace} we are able to be more precise; as the potential asymptotically goes to zero, we can define the S-matrix for such a system and we will show how to compute the decay rate exactly from the interpretation of a resonance as a complex pole of the S-matrix.\\

Consider the potential in Fig.~\ref{fig:PotentialFlatSpace}. We need to solve the flat space version of Eq.~\eqref{eq:WD}. Now we have a global constraint - classically it is a constraint on the Hamiltonian (rather than the density), 
\begin{equation}
H=\int_{X}\left[\frac{\pi_{\phi}^{2}}{2}+\frac{1}{2}\boldsymbol{\nabla}_{x}^{2}\phi+V(\phi)\right]=E \,,\label{eq:Hphi}
\end{equation}
 and hence the Schr\"odinger equation
\begin{equation}
\int_{X}\left[-\frac{\hbar^{2}}{2}\frac{\delta^{2}}{\delta\phi(x)^{2}}+\frac{1}{2}\boldsymbol{\nabla}_{x}^{2}\phi+V(\phi)\right]\Psi[\phi]=E\Psi[\phi].
\end{equation}
Using $G^{\phi\phi}=1$ in Eq.~\eqref{eq:HJ} and Eq.~\eqref{eq:prefactoreqn}
\begin{equation}
\int_{X}f[\phi]=\int_{X}\left[\frac{1}{2}\boldsymbol{\nabla}_{x}^{2}\phi+V(\phi)\right]-E\equiv U[\phi]-E,\label{eq:Uphi}
\end{equation}
so Eq.~\eqref{eq:S0intf} becomes 
\begin{equation}
S_{0}\left[\phi\right]=-2\int^{s}ds'C^{-1}(s)\left[U[\phi_{s'}]-E\right] \,.
\end{equation}

Note that in order to make these expressions well-defined we choose the field configuration $\phi$ such that at large $|{\bf x}|$ it goes asymptotically to $\phi_{IV}$ rapidly enough to make all the integrals above finite. Alternatively as we will do in the next section, we can work in a compact space such as a three-sphere.

At this point we can write, in general
\begin{align}
k & =\sqrt{2\left[E-U[\phi(\tau)]\right]}\,,\qquad {\rm for}\,E>U\left(\phi(\tau)\right) \,,\label{eq:ktau}\\
\kappa & =\sqrt{2\left[U[\phi(\tau)]-E\right]}\,,\qquad {\rm for}\,E<U(\phi(\tau)) \,.\label{eq:kappatau}
\end{align}
The leading order wave-functionals in the classically non-allowed and allowed regions respectively are determined by the WKB matching conditions. Consider the case in which the classically forbiden region is located at $\tau > \tau_0$, then the wave functionals in the classically allowed and in the classically forbidden regions are respectively
\begin{eqnarray}
\Psi[\phi] &=& \frac{2 A}{\sqrt{k} } \cos\left(\int_\tau^{\tau_0} k(\phi(\tau)) d\tau - \frac{\pi}{4}\right) - \frac{B}{\sqrt{k}} \sin\left(\int_\tau^{\tau_0} k(\phi(\tau)) d\tau - \frac{\pi}{4}\right) \,, \nonumber \\
\Psi[\phi] &=& \frac{A}{\sqrt{\kappa} } \exp\left(-\int_{\tau_0}^{\tau} \kappa(\phi(\tau)) d\tau\right) + \frac{B}{\sqrt{\kappa}} \exp\left(\int^\tau_{\tau_0} \kappa(\phi(\tau)) d\tau \right) \,,
\label{eq:WKB1}
\end{eqnarray}

If the classically forbidden region is located at $\tau < \tau_0$, the wave-functionals are
\begin{eqnarray}
\Psi[\phi] &=& \frac{A}{\sqrt{\kappa} } \exp\left(-\int_{\tau_0}^{\tau} \kappa(\phi(\tau)) d\tau\right) + \frac{B}{\sqrt{\kappa}} \exp\left(\int^\tau_{\tau_0} \kappa(\phi(\tau)) d\tau\right) \,, \nonumber \\
\Psi[\phi] &=& \frac{2 A}{\sqrt{k} } \cos\left(\int_\tau^{\tau_0} k(\phi(\tau)) d\tau - \frac{\pi}{4}\right) - \frac{B}{\sqrt{k}} \sin\left(\int_\tau^{\tau_0} k(\phi(\tau)) d\tau - \frac{\pi}{4}\right) \,,
\label{eq:WKB2}
\end{eqnarray}
In the above we've only kept the pre-factor corresponding to  longitudinal fluctuations of the field i.e. corresponding to the first term in Eq.~\eqref{eq:S1-2}. The second term coming from transverse fluctuations is not explicitly written since it plays no role in the further discussion. 

\begin{figure}[h!] 
\begin{center} 
\includegraphics[scale=0.7]{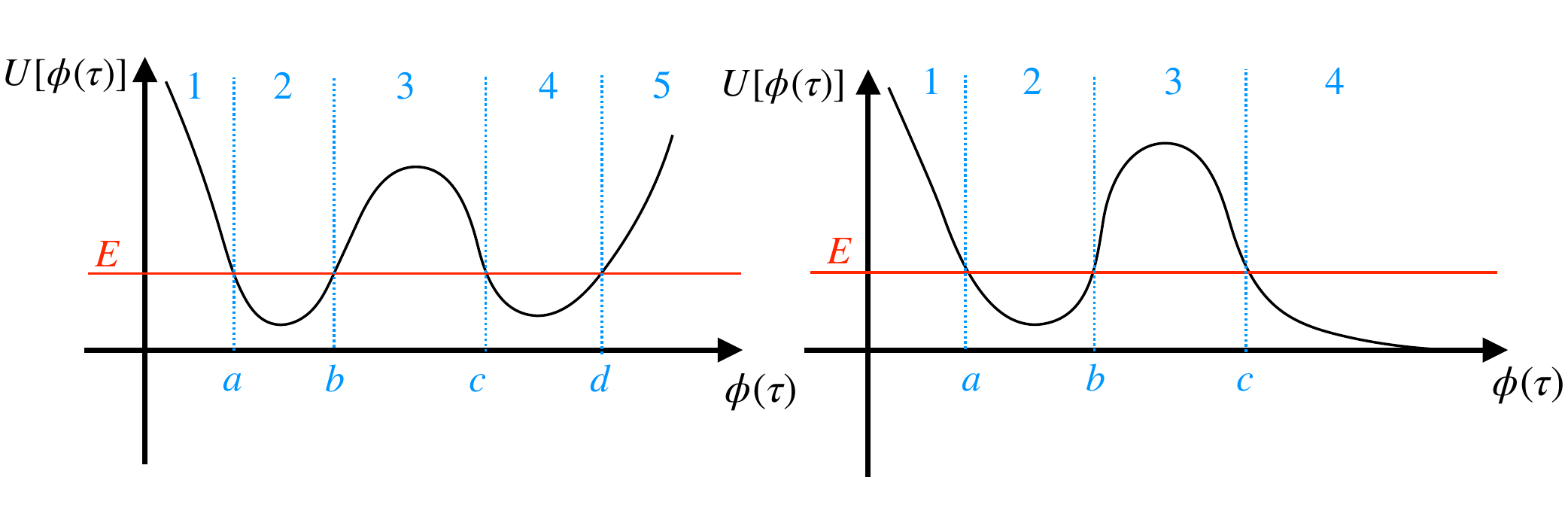}
\caption{Potential.\label{fig:PotentialUFlatSpace}}
\end{center} 
\end{figure}

Now recall that the classically allowed and forbidden regions are defined in terms of the potential $U[\phi]$, that depends on the specific path in the field space chosen to perform the integration. Hence they cannot be visualised in the potentials of Fig.~\ref{fig:PotentialFlatSpace}. However, we can expect that for a generic path in field space, it would take a form that is similar to that shown in Fig.~\ref{fig:PotentialFlatSpace}, with a finite barrier in the middle and infinite barriers on both sides for the case of the left panel of Fig.~\ref{fig:PotentialFlatSpace}, and on the left side only for the case of the right panel of Fig.~\ref{fig:PotentialFlatSpace}, see Fig.~\ref{fig:PotentialUFlatSpace}. At this point we will distinguish between the two cases.

\subsection{WKB for decay in a two-vacua potential}

In the case of the left panel of Fig.~\ref{fig:PotentialUFlatSpace}, we can identify five different regions, three of which are classically disallowed ($1$, $3$ and $5$) while two are classically allowed ($2$ and $4$). We impose that in region $1$ the decaying component of the wave-functional is absent, hence $B_1 = 0$ and
\be
\Psi_1[\phi] = \frac{A_1}{\sqrt{\kappa}} \exp\left(- \int_\tau^{a} \kappa d\tau \right) \,.
\ee
Using Eq.~\eqref{eq:WKB2} we can easily find that
\be
A_2 = \left(\cos\theta e^{-i\pi/4} + \sin\theta e^{i\pi/4}\right) A_1 \,, \qquad B_2 = \left(\cos\theta e^{i\pi/4} +\sin\theta e^{-i\pi/4}\right) A_1 \,,
\ee
where
$\theta = \int_{a}^{b} k d\tau \,, $ 
so that
\be
\Psi_2[\phi] = \frac{A_2}{\sqrt{k}} \exp\left(i \int_b^\tau k d\tau\right) + \frac{B_2}{\sqrt{k}} \exp\left(-i \int_b^\tau k d\tau\right) \,.
\ee
The connection between the regions $2$ and $4$ can be easily found by using the following connection formula, see~\cite{Merzbacher:1998}
\be
\begin{pmatrix}
A_4 \\ B_4
\end{pmatrix} = \frac{1}{2}
\begin{pmatrix}
\frac{1}{2 \lambda} + 2 \lambda & i \left(\frac{1}{2 \lambda} - 2 \lambda\right) \\ - i \left(\frac{1}{2 \lambda} - 2 \lambda\right) & \frac{1}{2 \lambda} + 2 \lambda
\end{pmatrix} 
\begin{pmatrix}
A_2 \\ B_2
\end{pmatrix} \,,
\ee
where
$\lambda = \exp\left(\int_b^c \kappa d \tau\right)$.
We find that
\be
\label{eq:A4B4}
A_4 = e^{i\pi/4} \frac{A_2}{2 \lambda} (\sin\theta - 4 i \cos\theta \lambda^2) \,, \qquad B_4 = e^{i\pi/4} \frac{A_2}{2 \lambda} (-i\sin\theta + 4 \cos\theta \lambda^2) \,,\nonumber
\ee
so that the wave-functional takes the form
\be
\Psi_4[\phi] = \frac{A_4}{\sqrt{k}} \exp\left(i \int_c^\tau k d\tau\right) + \frac{B_4}{\sqrt{k}} \exp\left(-i \int_c^\tau k d\tau\right) \,.
\ee
Using Eq.~\eqref{eq:WKB1} we find that
\be
A_5 = \frac{A_4 e^{i \omega} + B_4 e^{-i\omega}}{2} \,, \qquad B_5 = i(A_4 e^{i\omega} - B_4 e^{-i\omega}) \,,\nonumber
\ee
where
$\omega = \int_c^d k d\tau \, $, 
so that
\be
\Psi_5[\phi] = \frac{A_5}{\sqrt{\kappa}} \exp\left(-\int_d^\tau \kappa d\tau \right) + \frac{B_5}{\sqrt{\kappa}} \exp\left(\int_d^\tau \kappa d\tau\right) \,.
\ee
Of course, we need to require that the rising component of the wave-functional is absent in region $5$, namely that $B_5 = 0$. This implies
\be
- 4 \lambda^2 \left(1 + i e^{2 i \omega}\right) \cos\theta + i \left(1 - i e^{2 i \omega}\right) \sin\theta = 0 \,, \nonumber
\ee
which is satisfied if
\begin{eqnarray}
\cos\theta &=& 0 \quad \text{and} \quad \omega = -\frac{\pi}{4} + n \pi \quad (n \in \mathbb{Z}) \,, \quad \text{or} \nonumber  \\
\sin\theta &=& 0 \quad \text{and} \quad \omega = \frac{\pi}{4} + n \pi \quad (n \in \mathbb{Z}) \,. \nonumber
\end{eqnarray}
However, note that if we require that $\sin\theta = 0$, the coefficients $A_4$ and $B_4$ in Eq.~\eqref{eq:A4B4} would be enhanced with respect to $A_2$ and $B_2$ by a factor $\propto \lambda$. If the initial condition of the process is a homogeneous configuration with the field in the false vacuum, we expect that the coefficients $A_4$ and $B_4$ are suppressed with respect to $A_2$ and $B_2$, therefore we need to impose $\cos\theta = 0$. Note that the leading order of the transmission coefficient $T^2 = \frac{|\Psi_4[\phi]|^2}{|\Psi_2[\phi]|^2}$ is given by the factor $1/\lambda^2$ and gives a measure of the decay rate, despite in this case it is not possible to formally define it in terms of the S-matrix, unlike the case discussed in the next section. At leading order then
\be
\label{eq:DecayRateWKB}
\Gamma \sim T^2 \sim \frac{1}{\lambda^2} \propto \exp\left(-2 \int_b^c\kappa d\tau\right) \,,
\ee
which is equivalent to the CDL result.

It is particularly interesting to notice that the result in Eq.~\eqref{eq:DecayRateWKB} is equivalent to the Euclidean action evaluated on the bounce solution, upon subtraction of the background action. Let us make this statement more explicit: between the two turning points $b$ and $c$, the potential energy is larger than the total energy $E$ of the system. As the total energy is given by the sum of kinetic energy and potential energy, the kinetic energy has to be negative. This can be achieved by rotating the time variable to the Euclidean time, that gains an imaginary unit and makes the kinetic energy negative. Hence, in the under-the-barrier region (between $b$ and $c$) it is totally justified to rotate to Euclidean time $s = i t$, where $t$ is the usual Lorentzian time. The Euclidean action is
\be
\label{eq:EuclideanAction}
S_E[\phi(s)] = \int ds \left[ \int d^3 x \, \left( \frac{1}{2} \left(\frac{d\phi(s)}{ds}\right)^2 \right) + U[\phi(s)]\right] \,.
\ee
The Euclidean energy is conserved, which implies
\be
\label{eq:EuclideanEnergyConservation}
\int d^3 x \, \left(\frac{1}{2} \left(\frac{d\phi(s)}{ds}\right)^2\right) - U[\phi(s)] = -E \,,
\ee
where we take $E = U[\phi_A]$\footnote{Note that in the case of the left panel of Fig.~\ref{fig:PotentialFlatSpace}, $E = U[\phi_A] = 0$.} (despite the WKB argument holds for a general $E$), assuming that the initial state is a homogeneous field configuration $\phi = \phi_A$. Using Eq.~\eqref{eq:EuclideanEnergyConservation}, and noting that, in the under-the-barrier region
\be
\frac{d\tau}{ds} = \sqrt{2(U[\phi(s)] - U[\phi_A])} \,,
\ee
the Euclidean action in Eq.~\eqref{eq:EuclideanAction} becomes simply
\begin{eqnarray}
S_E[\phi(s)] &=& \int ds \left[ \int d^3 x \, \left( \frac{1}{2} \left(\frac{d\phi(s)}{ds}\right)^2 \right) + U[\phi(s)]\right] = \nonumber \\
&=& \int ds \left[2 (U[\phi(s)] - U[\phi_A])\right] + \int ds \, U[\phi_A] = \nonumber \\
&=& \int_b^c d\tau \sqrt{2(U[\phi(s)] - U(\phi_A))} + S_E^{\rm back} \,,
\label{eq:EuclideanAction2}
\end{eqnarray}
where $S_E^{\rm back}$ is the background Euclidean action evaluated on the homogeneous solution $\phi = \phi_A$. Therefore
\be
\boxed{\quad S_E[\phi(s)] - S_E^{\rm back} = \int_b^c d\tau \kappa \,,\quad }
\ee
which shows the equivalence between the Euclidean action evaluated on the bounce solution (upon subtracting the background action) and the usual WKB factor of Eq.~\eqref{eq:DecayRateWKB}.

Of course, the resort to the use of the Euclidean action is just a trick that sometimes makes the computation easier and is completely justified quantum mechanically, as it reproduces exactly the WKB result. However, there is no intrinsic reason to use the bounce solution for the post-nucleation phase: up to Eq.~\eqref{eq:DecayRateWKB} we have not introduced Euclidean time and the Euclidean action, and these are actually not needed to get to the final result. The quantum mechanics problem only knows about the symmetries that are put in the problem from the very beginning: for instance one can require from the start that the problem has a spherical $O(3)$ symmetry. There is no way for the post-nucleation solution to gain a $\sqrt{t^2 - |x|^2}$ dependence, as it is usually obtained by naively rotating back to Lorentzian signature the bounce solution.

\subsection{WKB for decay in run-away potential}

In the case of the right panel of Fig.~\ref{fig:PotentialFlatSpace} we can give an explicit definition of the decay rate, as we can discuss  the question in terms  of an S-matrix. We have in mind a state which comes in from the right, tunnels through the barrier, is reflected off the wall on the left and tunnels through the barrier back to give an outgoing state to the right. The resonances that can be formed in this process constitute the decaying state that we are interested in. The S-matrix is a phase ($S=A_{4}/B_{4}$) with complex poles corresponding to the bound states in region 2. To identify the decay widths of the corresponding resonances we first identify the bound states - these correspond to 
\begin{equation}
\cos\theta = 0 \qquad \Rightarrow \qquad \theta = \left(n+\frac{1}{2}\right)\pi  \quad (n \in \mathbb{Z})\,.
\end{equation}
Note that the last expression determines the possible discrete values of the energy $E_n$: in fact the energy appears both in the limits of integration (as it determines the turning points) and in the integrand (see Eq.~\eqref{eq:ktau} and Eq.~\eqref{eq:kappatau}). Considering the lowest energy state $E_0$ and expanding around this point we can write (see~\cite{Merzbacher:1998})
\be
\cos \theta\simeq\mp(E-E_{0})\left(\frac{\partial \theta}{\partial E}\right)\bigg|_{E = E_0} \,, \qquad \sin \theta\bigg|_{E = E_0} \simeq 1 \,, \nonumber
\ee
and get
\be
S\equiv\frac{A_4}{B_4}=\frac{E-E_{0}-i\left[ 1/\left(4\theta{}^{2}\left(\frac{\partial \theta}{\partial E}\right)|_{E = E_0}\right)\right] }{E-E_{0}+i\left[ 1/(\left(4\theta{}^{2}\left(\frac{\partial \theta}{\partial E}\right)|_{E = E_0}\right)\right] }\equiv e^{2i\phi}=\frac{E-E_{0}-i\Gamma/2}{E-E_{0}+i\Gamma/2} \,.
\ee
with the decay width given by
\be
\Gamma=\frac{1}{2\lambda^{2}}\left(\frac{\partial \theta}{\partial E}\right)^{-1}\bigg|_{E = E_0} \,.
\ee
The phase shift $\phi$ has then the standard form
\be
\tan\phi=\frac{\Gamma/2}{E-E_{0}} \,,\nonumber
\ee
with the lifetime of the resonance given by
\begin{empheq}[box=\fbox]{align}
& \nonumber \\
\qquad \Gamma^{-1} &=2\lambda^{2}\left(\frac{\partial \theta}{\partial E}\right)\bigg|_{E = E_0} \quad \nonumber \\
\qquad &=\left[\frac{\partial}{\partial E}\int_{a}^{b}d\tau\sqrt{2\left(E-U(\tau\right)})\bigg|_{0}\right]2\exp\left[2\int_{b}^{c}d\tau\sqrt{2\left(U(\tau\right)-E})\right]\,.\quad \nonumber \\
\label{eq:lifetime}
\end{empheq}
Note that $2\left(\frac{\partial \theta}{\partial E}\right)\bigg|_{E = E_0}$is the
classical period for oscillations between $a$ to $b$ and one could
have divided the transition probability i.e. the WKB factor in Eq.~\eqref{eq:DecayRateWKB} to get this formula heuristically, while here we have derived it purely from quantum mechanical considerations.

\subsection{Comparison to Coleman's formula}
\label{sec:ScalarFieldComparison}

We showed above that the exponential term in the life-time of the false vacuum is (for the particular case where the energy corresponds to the energy at the minimum of the potential) is the same as in Coleman's calculation. On the other hand the pre-factor  is quite different from that in  the well-known formula derived by Coleman~\cite{Coleman:1977py, Callan:1977pt}. However the derivation in the latter papers is not quite a straightforward application of WKB quantum mechanics but involve a series of additional assumptions whose status we discuss below. 
\begin{figure}[h!] 
\begin{center} 
\includegraphics[scale=1]{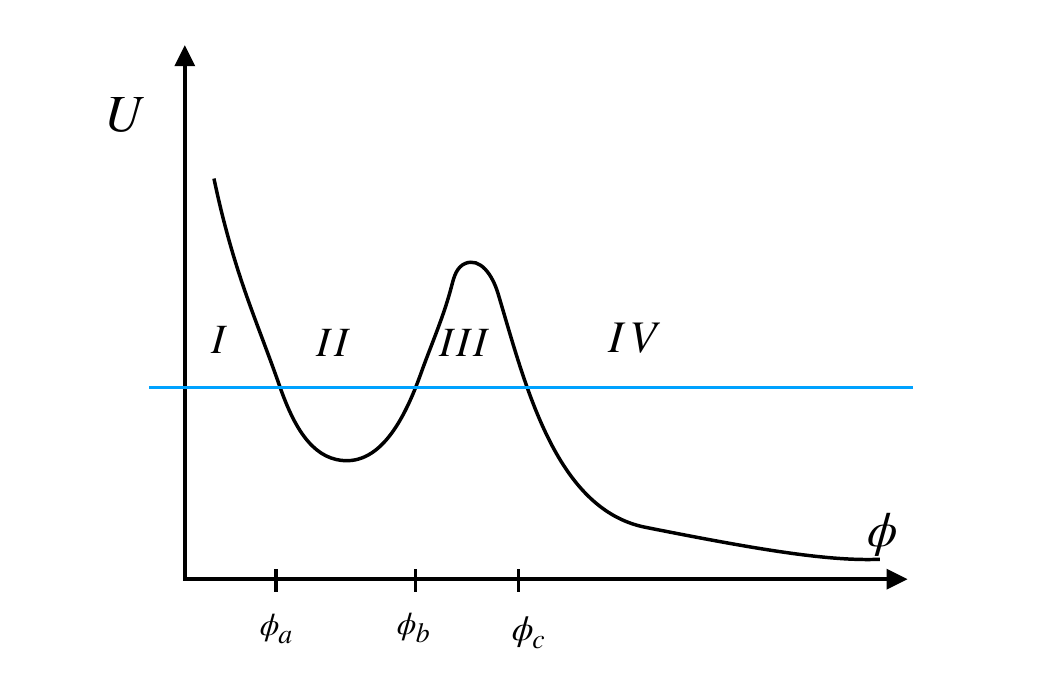}
\caption{The effective potential $U$ along the path $\phi_{\tau}$ and  for the point particle discussion in Sec.~\ref{sec:ScalarFieldComparison}.\label{fig:BouncePotential}}
\end{center} 
\end{figure}

\subsubsection*{Some issues with Coleman's argument}
Let us summarise some of the weak points with Coleman's original approach \footnote{See~\cite{Andreassen:2016cvx} for related critical comments.}.
\begin{itemize}
\item The Hamiltonian $H$ is Hermitian - its eigenvalues are necessarily real. Coleman makes essential use of the heuristic interpretation of $\Gamma$ as the imaginary part of an eigenvalue of $ H$ and the dilute gas approximation in order to get the decay rate. The calculation leading to Eq.~\eqref{eq:lifetime} used no such interpretation - $\Gamma$ is extracted from the $S$-matrix for the scattering of a field configuration from a potential which can accommodate quasi-bound states (resonances).
\item Coleman's calculation disagrees in the pre-factor from the standard WKB calculation (i.e. essentially Eq.~\eqref{eq:lifetime} with $U$ now being the quantum mechanical potential). In particular Coleman's formula for the decay rate $\Gamma$ involves dividing by the (infinite) translation symmetry of his instanton i.e. the pre-factor for the transition probability contains a factor $T\rightarrow\infty$ that is divided out to get the rate. In the calculation above there is no such factor and effectively the division is by the classical period for oscillations in the potential well.
\item The last factor of Eq.~\eqref{eq:lifetime} is the square of the standard WKB decay amplitude and naturally is the same as in Coleman's formula with also the identification of the exponent as the difference between the Euclidean classical action for a classical solution with the appropriate boundary conditions and the action for a particle whose position is localised in the well (at its minimum). 
\item In the field theory case Coleman gives an argument for the dominant contribution to this last factor to come from a Euclidean 4-sphere configuration. However in the actual evaluation of the decay amplitude the thin-wall approximation is used in which the under the barrier region effectively shrinks to a brane. On the other hand propagation in the classically allowed region to the right of the barrier is described in terms of the analytic continuation of this Euclidean instanton. The justification for the latter is unclear.
\item Of course one would like to have a quantum field theory argument for a physical picture of a first order phase transition with bubble nucleation and percolation. Coleman's argument is clearly motivated by this and indeed it would be nice to have a rigorous QFT justification for this. Unfortunately we do not see at this point how to achieve this.
\end{itemize}

\section{Vacuum decay in curved space}
\label{sec:DecayCurvedSpace}

\subsection{Review of Coleman-De Luccia}
\label{sec:CDLReview}

The application of Coleman's arguments to the case involving gravity are even more problematic. Apart from the issues highlighted above there is the problem that the notion of time (and hence that of transition probability per unit time) needs to be reinterpreted given that the WDW equation does not admit the usual notion of time. Let us first discuss the Euclidean mini-superspace case as given by CDL for the case $dS\rightarrow{\cal M}$ and generalised to $dS\rightarrow dS$
by Parke~\cite{Parke:1982pm}. Putting $t=i\tau$ and $a(t)=\rho(\tau)$ and gauge fixing to $N=1$ we have the metric $ds^{2}=d\tau^{2}+\rho^{2}(\tau)d\Omega_{3}^{2}$. The relevant Euclidean equation of motion then is the $\tau\tau$ component of the Einstein equation which reads ($\rho'=d\rho/d\tau,$etc.)
\begin{equation}
\rho'^{2}=1+\frac{1}{3}\left(\frac{1}{2}\phi'^{2}-V(\phi)\rho^{2}\right).\label{eq:rho}
\end{equation}
The Euclidean action is 
\begin{align}
S_{E} & =-2\pi^{2}\int_{0}^{\tau_{{\rm max}}}d\tau\left[3\rho+3\rho\rho'^{2}-V\rho^{3}-\rho^{3}\frac{1}{2}\phi'^{2}\right]\nonumber \\
 & =-12\pi^{2}\int_{0}^{\tau_{{\rm max}}}d\tau\left[\rho-\frac{1}{3}V\rho^{3}\right]\,.\label{eq:SE}
\end{align}
In the last step we used the equation of motion in Eq.~\eqref{eq:rho}. In this Euclidean argument this is supposed to be the instanton (bounce) action with $\phi(0)=\phi_{B}$ i.e. the value of the field at the so-called true minimum and $\phi(\infty)=\phi_{A}$, the value of the field in the false minimum. Of course there is complete symmetry between the two so the bounce action is the same for going from $A\rightarrow B$ or $B\rightarrow A$. The difference between up-tunneling (true to false vacuum) and down-tunneling (false to true) just comes from the fact that the background action which is subtracted to get the tunneling amplitude is different. So in the case of down-tunneling that we will consider here (i.e. $A\rightarrow B)$ the tunneling amplitude is
given by $e^{B/2}$ where,
\begin{equation}
\frac{B}{2}=S_{E}-S_{E}^{A}\label{eq:B1-1}
\end{equation}
 where the second term is the action for $\phi$ remaining at the
false minimum $\phi_{A}.$ i.e. 
\[
S_{E}^{A}=-12\pi^{2}\int_{0}^{\tau_{{\rm max}}}d\tau\left[\rho-\frac{1}{3}V_{A}\rho^{3}\right],
\]
where $V_{A}=V(\phi_{A})$. Hence we have 
\begin{align}
\frac{B}{2}& =-12\pi^{2}\int_{0}^{\tau_{{\rm max}}}d\tau\left[\rho-\frac{1}{3}V\rho^{3}\right]+12\pi^{2}\int_{0}^{\tau_{{\rm max}}}d\tau\left[\rho-\frac{1}{3}V_{A}\rho^{3}\right]\nonumber \\
 & =-12\pi^{2}\int_{0}^{\bar{\tau}-\delta\tau}d\tau\left[\rho-\frac{1}{3}V_{B}\rho^{3}\right]+2\pi^{2}\bar{\rho}^{3}T+12\pi^{2}\int_{0}^{\bar{\tau}-\delta\tau}d\tau\left[\rho-\frac{1}{3}V_{A}\rho^{3}\right]\label{eq:B2-1}
\end{align}
In the second line we have assumed that beyond the point $\bar{\tau}+\delta\tau$,
$V\simeq V_{A}$ so that the contribution from $\bar{\tau}+\delta\tau$
to $\text{\ensuremath{\tau_{{\rm max}}}}$ in the first term of the
first line cancels against the second term. Also $T$ in the middle
term is defined by 
\begin{equation}
\bar{\rho}^{3}T=2\int_{\bar{\tau}-\delta\tau}^{\bar{\tau}+\delta\tau}d\tau\rho^{3}(V(\phi(\tau)-V_{A}).\label{eq:T}
\end{equation}
 In the second line of Eq.~\eqref{eq:B2-1} we have taken the path
in $\tau$ such that for $0<\tau\le$ $\bar{\tau}-\delta\tau$, $\phi$
is held fixed at $\phi_{B}$ while in the interval $\bar{\tau}+\delta\tau\le\tau<\tau_{{\rm max}}$,
$\phi=\phi_{A}$. So in the first and third terms in Eq.~\eqref{eq:B2-1}
we can replace the integral over $d\tau=\frac{d\tau}{d\rho}d\rho$
using the Euclidean Eq.~\eqref{eq:rho} with $\phi$ fixed\footnote{Although not explicitly stated this seems to have been assumed also
in~\citep{Parke:1982pm}.}. This gives $\frac{d\tau}{d\rho}=\pm1/\sqrt{1-V_{B,A}\rho^{2}}$
in the first and third terms\footnote{In~\citep{Parke:1982pm} only the positive sign is kept here.}
so these integrations can be done giving us (in the thin wall limit
$\delta \tau\rightarrow0$),
\begin{equation}
\boxed{\quad \frac{B}{2}=-12\pi^{2}\left[\pm\frac{\left(1-\frac{1}{3}V_{A}\bar{\rho}^{2}\right)^{3/2}-1}{V_{A}}\mp\frac{\left(1-\frac{1}{3}V_{B}\bar{\rho}^{2}\right)^{3/2}-1}{V_{B}}\right]+2\pi^{2}\bar{\rho}^{3}T.\quad }\label{eq:B3}
\end{equation}
$\bar{\rho}$ is then determined by extremising $B$. Upon substituting
this value into the above one then gets the usual expressions which
we will quote later after re-deriving the above without invoking
Euclidean arguments with their corresponding interpretational issues.

\subsection{Vacuum transitions in mini-superspace}
\label{sec:CDLfromWDW}

An instructive exercise, that helps understanding the formalism outlined in Sec.~\ref{sec:WKBandWDW} and shows the differences between the Lorentzian and Euclidean appproaches, consists in studying vacuum transitions in a mini-superspace setup that includes a real scalar field. This calculation is a generalization of the ‘tunneling from nothing’ scenario~\cite{Vilenkin:1982de, Hartle:1983ai, Vilenkin:1984wp,Linde:1983mx}. For a recent discussion see for instance~\cite{Kristiano:2018oyv,deAlwis:2018sec,Halliwell:2018ejl}. The metric is 
\begin{equation}
ds^{2}=-N^{2}(t)dt^{2}+a^{2}(t)(dr^{2}+\sin^{2}rd\Omega_{2}^{2}) \,. \label{eq:minimetric}
\end{equation}
The action (setting $M_p=1/\sqrt{8\pi G}=1$) is given by the sum $S=S_{g}+S_{m}$, where 
\begin{eqnarray}
S_{g} & = & 2\pi^{2}\int_{0}^{1}dt\left(-N^{-1}3a\dot{a}^{2}+3kaN\right) \,,\label{eq:Sg}\\
S_{m} & = & 2\pi^{2}\int_{0}^{1}dt\left(N^{-1}\frac{1}{2}a^{3}\dot{\phi}^{2}-Na^{3}V(\phi)\right) \,.
\end{eqnarray}
Here $k=\pm1,0$ depending on whether the three-spatial slice is positively (negatively)
curved or flat. Of course  in the open $k=0,-1$ cases the factor $2\pi^{2}$ would have to be replaced by an appropriate compactified volume factor and the spatial metric in Eq.~\eqref{eq:minimetric} would need to be replaced by a flat or hyperbolic metric. Here we will focus on the $k=+1 $ case and for  convenience we will drop the $2\pi^{2}$ factor in the calculations below and restore it in the expressions for the classical action. We will make some remarks at end on the other two cases. The canonical momenta are
\begin{equation}
\pi_{N}=0\,, \qquad \pi_{a}=-N^{-1}6a\dot{a} \,,\qquad \pi_{\phi}=N^{-1}a^{3}\dot{\phi} \,,\label{eq:momenta}
\end{equation}
and the Hamiltonian constraint is 
\begin{equation}
{\cal H}=N\left(-\frac{\pi_{a}^{2}}{12a}+\frac{\pi_{\phi}^{2}}{2a^{3}}-3a+a^{3}V(\phi)\right)\approx 0 \,.\label{eq:calH}
\end{equation}

\begin{figure}[h!] 
\begin{center} 
\includegraphics[scale=0.3, trim=0cm 2cm 0cm 1.8cm,clip]{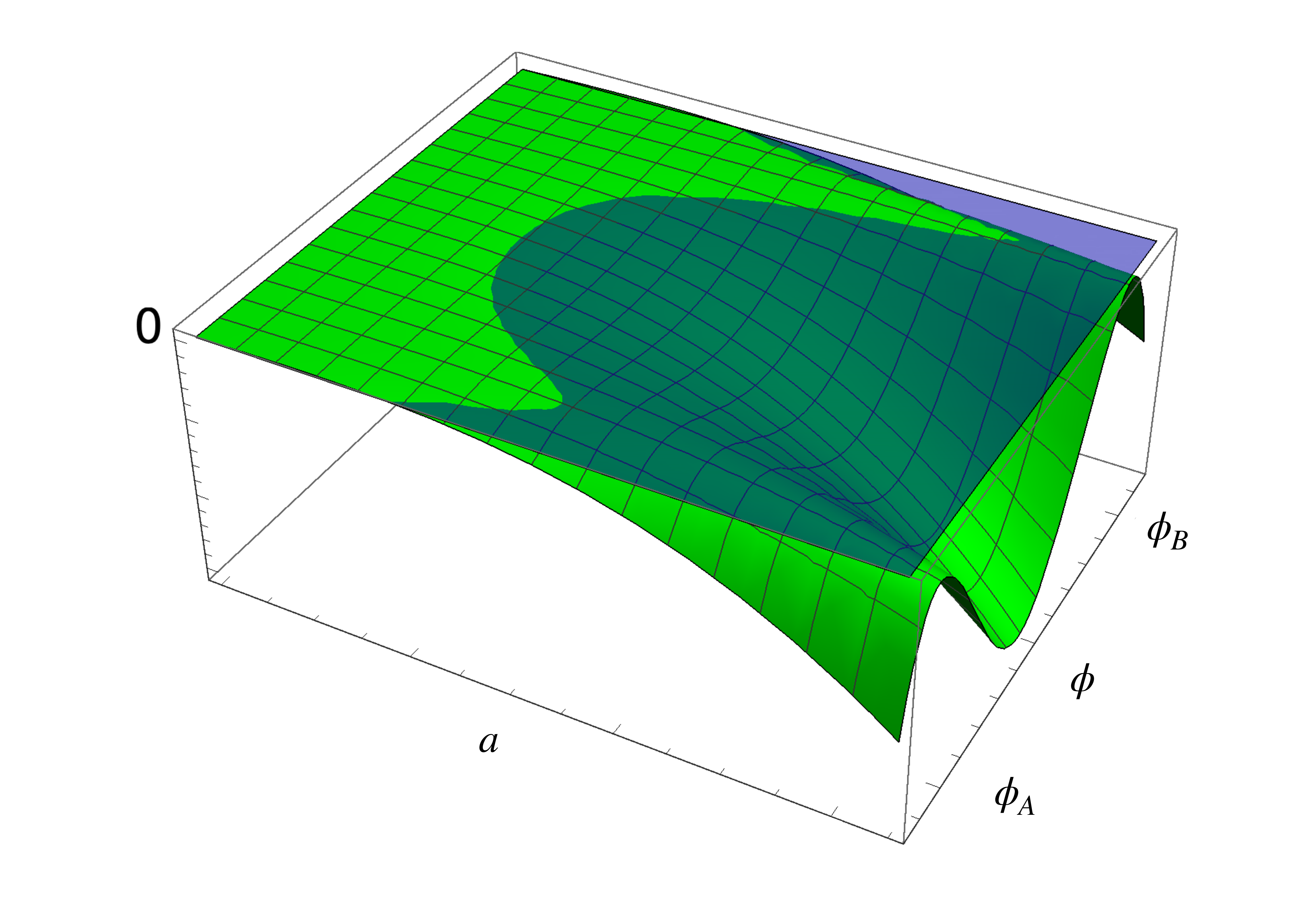}
\caption{\label{fig:Path}}
\end{center} 
\end{figure}

Comparing with Eq.~\eqref{eq:GeneralHamiltonian} we have
\begin{gather}
G^{aa} = -\frac{1}{6a} \,,\qquad G^{\phi\phi}=\frac{1}{a^{3}},\label{eq:G-1}\\
f(a,\phi) = -3a+a^{3}V(\phi).\label{eq:faphi}
\end{gather}
Consider a scalar potential with two dS minima in $\phi_A$ and $\phi_B$, with $V(\phi_A) \equiv V_A > V_B \equiv V(\phi_B)$. Then the general shape of the function $f(a, \phi)$ in Eq.~\eqref{eq:faphi} is plotted in Fig.~\ref{fig:Path}.

As we emphasised before, the superspace metric is not positive definite in the presence of gravity and this introduces significant differences to  WKB type tunneling arguments. As can be seen from this constraint equation in the absence of the scalar field (this is the `tunneling from nothing' model of~\cite{Vilenkin:1982de, Hartle:1983ai, Vilenkin:1984wp,Linde:1983mx}), one has a barrier at fixed $\phi$ for the scale factor $a$ when $a<\sqrt{3/V}$  whereas if the geometry is fixed (as in the original investigations of Coleman et al.) then there is a barrier for $\phi$ when $V$ is greater than its value at the point where $\pi_{\phi}$ becomes zero. However it is clear that these simple situations are not the only possibilities when both $a$ and $\phi$ are present. The following is to our knowledge the first time this more complex situation is discussed.

\subsection{Recovering Coleman-De Luccia}
\label{sec:RecoveringCDL}

It is instructive to try and recover the CDL expression in Eq.~\eqref{eq:B3} using the formalism of Sec.~\ref{sec:WKBandWDW}. Hopefully, this might help us in better understanding CDL. Let us choose the deformation parameter $s$ (in analogy with the Euclidean
time $\tau$) such that from some initial value (say at $s=0$) to
the point $\bar{s}-\delta s$ the field $\phi$ remains very close
to $\phi_{B}$ and for points $s_{\max}>s>\bar{s}+\epsilon$, $\phi$
becomes close to $\phi_{A}$. It should be emphsised here that $s$
has nothing to do with real time - it is simply a deformation parameter
that parametrises the path of integration. The range $\bar{s}-\epsilon<s<\bar{s}+\epsilon$
is the transitional region where in effect CDL used the thin wall
approximation. Thus the classical action (see Eq.~\eqref{eq:S0intf})
is 
\begin{small}
\begin{eqnarray}
S_{0}(a_{0},\phi_{B};a_{{\rm max}},\phi_{A}) & = & -12\pi^{2}\int_{0}^{s_{{\rm max}}}ds'C^{-1}(s')\left(-a+a^{3}\frac{V(\phi)}{3}\right)\nonumber \\
 & = & -12\pi^{2}\int_{0}^{\bar{s}-\delta s}ds'C^{-1}(s')\left(-a+a^{3}\frac{V_{B}}{3}\right)-12\pi^{2}\int_{\bar{s}-\delta s}^{\bar{s}+\delta s}ds'C^{-1}(s')\left(-a+a^{3}\frac{V(\phi)}{3}\right)\nonumber \\
 &  & -12\pi^{2}\int_{s+\delta s}^{s_{{\rm max}}}ds'C^{-1}(s')\left(-a+a^{3}\frac{V_{A}}{3}\right).\label{eq:SAB1}
\end{eqnarray}
\end{small}
Here in the first term we have used the fact that $\phi$ remains constant
and equal to $\phi_{B}$ while in the last term it remains equal to
$\phi_{A}.$ 

This path corresponds to the Euclidean path chosen by CDL and Parke. Thus in the Lorentzian case this action corresponds to ‘tunneling from nothing’, as in Hartle-Hawking/Vilenkin-Linde~\cite{Vilenkin:1982de, Hartle:1983ai, Linde:1983mx} wave function of the universe arguments, essentially keeping $\phi$ fixed, to the potentially emergent state $B$ (the true vacuum in CDL's language), then making a transition to the state $A$ (where both $\phi$ and $a$ can change), that then emerges as the classical background space time. This is then to be compared to the situation where the the state $A$ emerges from a ‘tunneling from nothing’ process. The latter gives an action 
\begin{equation}
S_{0}(a_{{\rm max}},\phi_{A};a_{0},\phi_{A})=-12\pi^{2}\int_{0}^{s}ds'C^{-1}(s')\left(-a+a^{3}\frac{V_{A}}{3}\right).\label{eq:SA}
\end{equation}
Now we have from Eq.~\eqref{eq:dtauds} 
\begin{equation}
\left(\frac{d\tau}{ds}\right)^{2}=-6a\left(\frac{da}{ds}\right)^{2}+2a^{3}\left(\frac{d\phi}{ds}\right)^{2}=-2C^{-2}(s)(-3a+a^{3}V(\phi))\label{eq:dtauds-1} \,.
\end{equation}
Thus as long as $a^{2}<3/V(\phi$), for a ``time-like'' trajectory in field space $C^{2}=-1$. In particular this would be the case for $d\phi/ds=0$. For the moment though we will leave this undetermined. 

The transition probability is given by (ignoring the pre-factors for
the moment),
\begin{equation}
P(A\rightarrow B)=\bigg|\frac{\Psi(a_{0},\phi_{B};a_{{\rm max}},\phi_{A})}{\Psi(a_{0},\phi_{A};a_{{\rm max}},\phi_{A})}\bigg|^{2}=\bigg|\frac{\alpha e^{iS_{0}(a_{0},\phi_{B};a_{{\rm max}},\phi_{A})}+\beta e^{-iS_{0}(a_{0},\phi_{B};a_{{\rm max}},\phi_{A})}}{\alpha e^{iS_{0}(a_{0},\phi_{A};a_{{\rm max}},\phi_{A})}+\beta e^{-iS_{0}(a_{0},\phi_{A};a_{{\rm max}},\phi_{A})}}\bigg|^{2}\equiv e^{-B}\label{eq:PAB} \,.
\end{equation}
The dominant term in this ratio will be exponentially larger than
the subdominant terms so the latter may be safely ignored.
\begin{eqnarray}
\frac{B}{2} & = & iS_{0}(a_{0},\phi_{B};a_{{\rm max}},\phi_{A})-iS_{0}(a_{0},\phi_{A};a_{{\rm max}},\phi_{A}) \label{eq:B1} \\
 & = & -12\pi^{2}i\int_{0}^{\bar{s}-\delta s}ds'C^{-1}(s')\left(-a+a^{3}\frac{V_{B}}{3}\right)-12\pi^{2}i\int_{\bar{s}-\delta s}^{\bar{s}+\delta s}ds'C^{-1}(s')\left(-a+a^{3}\frac{V(\phi)}{3}\right)\nonumber \\
 &  & -12\pi^{2}i\int_{\bar{s}+\delta s}^{s_{{\rm max}}}ds'C^{-1}(s')\left(-a+a^{3}\frac{V_{A}}{3}\right)+12\pi^{2}i\int_{0}^{s_{{\rm max}}}ds'C^{-1}(s')\left(-a+a^{3}\frac{V_{A}}{3}\right) \,, \nonumber
\end{eqnarray}
where we have chosen to keep $C$ - so the choice of phase is so far
undetermined. After some cancellations this may be rewritten as
\begin{align}
\pm \frac{B}{2} & =-12\pi{}^{2}i\int_{0}^{\bar{s}-\delta s}ds'C^{-1}(s')\left(-a+a^{3}\frac{V_{B}}{3}\right)+12\pi^{2}i\int_{0}^{\bar{s}-\delta s}ds'C^{-1}(s')\left(-a+a^{3}\frac{V_{A}}{3}\right)\nonumber \\
 & +2\pi^{2}\bar{a}^{3}T.\label{eq:B2}
\end{align}
where we have defined the tension $T$ in analogy with Eq.~\eqref{eq:T}, as the contribution to the action coming from the portion of the path such that $d\phi/ds \neq 0$
\begin{equation}
2\pi^{2}\bar{a}^{3}T=12\pi^{2}i\int_{\bar{s}-\delta s}^{s+\delta s}ds'C^{-1}(s')\left(a^{3}\frac{V(\phi)-V_{A}}{3}\right)\label{eq:T2}\,.
\end{equation}
Note that, despite the contribution in Eq.~\eqref{eq:T2} is similar to Eq.~\eqref{eq:T}, there is no physical wall in the process that we are considering, that preserves the full $O(4)$ symmetry of the minisuperspace model.
So far we have not made any approximation. The terms in the first
line of Eq.~\eqref{eq:B2} will now be evaluated (as in the corresponding
Euclidean case) keeping $\phi$ constant. So we may use $\frac{da}{ds'}=\pm\sqrt{1-\frac{V_{A,B}}{3}a^{2}}$
(see Eq.~\eqref{eq:dtauds-1} with $d\phi/ds=0$ which implies $C^{2}=-1$).
We will also assume that the last term of Eq.~\eqref{eq:B2} is also integrated over a time-like path in field space - so that we can choose $C^{2}=-1$ along this path as well, which requires of course that $da/ds$ is non-zero along this path. Hence
\begin{equation}
S_{0}^{A,B}=\pm i12\pi^{2}\int_{0}^{a}daa\sqrt{\left(1-a^{2}\frac{V_{A,B}}{3}\right)}=\mp i12\pi^{2}\frac{1}{V_{A}}\left\{ \left(1-a^{2}\frac{V_{A,B}}{3}\right)^{3/2}-1\right\} \label{eq:SA2}
\end{equation}
Thus we have (putting $a(\bar{s}\pm\delta s)=\bar{a}\pm\delta a$)

\begin{eqnarray}
\pm \frac{B}{2} & = & 12\pi^{2}\left\{ \frac{1}{V_{B}}\left[\left(1-\left(\bar{a}-\delta a\right)^{2}\frac{V_{B}}{3}\right)^{3/2}-1\right]-\frac{1}{V_{A}}\left[\left(1-\left(\bar{a}-\delta a\right)^{2}\frac{V_{A}}{3}\right)^{3/2}-1\right]\right\} \quad \nonumber \\
\quad  & + & 2\pi^{2}\bar{a}^{3}T\,.
\end{eqnarray}

\subsection{Concrete cases}
\label{sec:ThinWallApproximation}

Let us now consider several cases to identify the transition amplitudes using our formalism.

\begin{itemize}
\item{\bf Thin wall CDL}. 
Let us try to recover the CDL resukt in the thin wall approximation. We can evaluate the last equation in the thin wall approximation. In this
case the potential essentially reduces to a brane so that effectively
we may write the argument of the integral in Eq.~\eqref{eq:T2} in the limit $\delta s\rightarrow0$
as a delta function. Then the logarithm of the transition amplitude becomes
\begin{align}
\pm \frac{B}{2} & =12\pi^{2}\left\{ \pm\frac{1}{V_{B}}\left[\left(1-\bar{a}^{2}\frac{V_{B}}{3}\right)^{3/2}-1\right]\mp\frac{1}{V_{A}}\left[\left(1-\bar{a}^{2}\frac{V_{A}}{3}\right)^{3/2}-1\right]\right\} \nonumber \\
 & +2\pi^{2}\bar{a}^{3}T \,.\label{eq:Bthinwall}
\end{align}
This is the same expression as Eq.~\eqref{eq:B3}. It is also clear that
this is the sum of two Hartle-Hawking/Vilenkin-Linde terms and\footnote{Note that in the thin wall approximation the integrals in Eq.~\eqref{eq:T} and Eq.~\eqref{eq:T2} imply that in the region around $\bar{s}$ the potential takes the form $a^{3}(s)\left(V(\phi(s))-V_{A}\right)=2\bar{a}^{3}T\delta(s-\bar{s})$.} and a term coming from the portion of the path where $d\phi/ds \neq 0$, that in the CDL computation corresponds to the wall tension contribution. However, notice that we recovered the CDL expression for the tunnelling probability, Eq.~\eqref{eq:Bthinwall} without having to go to Euclidean space, but just using WKB quantum
mechanics and assuming that, for the path that extremizes the action, $C^{-1}$ in Eq.~\eqref{eq:T2} is imaginary. Extremizing Eq.~\eqref{eq:Bthinwall} gives the standard (generalised) CDL expression for transitions between two dS spaces (or AdS with
appropriate sign changes). The extremum is at 
\begin{equation}
\frac{1}{\bar{a}^{2}}=\frac{V_{B}}{3}+\frac{1}{4}\left(\frac{2}{T}\frac{\Delta V}{3}+\frac{T}{2}\right)^{2}=\frac{V_{A}}{3}+\frac{1}{4}\left(\frac{2}{T}\frac{\Delta V}{3}-\frac{T}{2}\right)^{2},\label{eq:abar}
\end{equation}
where $\Delta V=V_{A}-V_{B}$. Note that this expression shows that $\bar{a}$ is less than the horizon radius of both $A$ and $B$.
Putting this into Eq.~\eqref{eq:Bthinwall} gives the final expression (with $H^{2}\equiv V/3$)
\begin{equation}
\boxed{\quad B=\pm8\pi^{2}\left[\frac{\left\{ \left(H_{A}^{2}-H_{B}^{2}\right)^{2}+T^{2}\left(H_{A}^{2}+H_{B}^{2}\right)\right\} \bar{a}}{4TH_{A}^{2}H_{B}^{2}}-\frac{1}{2}\left(H_{B}^{-2}-H_{A}^{-2}\right)\right].\quad }\label{eq:Bthinwallextr}
\end{equation}
This is of course the well-known result. However its derivation and
interpretation is quite different from that of CDL (and subsequent
work which generalised the CDL result). Firstly we did not explicitly
use Coleman's tunneling formula - instead we directly solved the WDW
equation in the classical approximation, as a deformation of the solution
where the initial configuration is one in which the fields correspond
to a dS space (with vacuum energy  $V_{A}$ and compared it to
the undeformed configuration).

However, there are a few puzzles posed by this calculation as we discuss next.
\item{\bf Hartle-Hawking interpretation}. 

As the
last term in Eq.~\eqref{eq:Bthinwall} is positive, it increases the absolute value of $B$ and
hence decreases the tunneling probabilty. On the other hand, one can simply ask what is the relative probability of ‘tunneling from nothing’ to the state $B$ compared to ‘tunneling from nothing’ to the state $A$. In other words one might compute the following ratio
\begin{equation}
\frac{B}{2} = \frac{\left|\Psi(a_0, \phi_B; a_{\rm max}, \phi_B)\right|^2}{\left|\Psi(a_0, \phi_A; a_{\rm max}, \phi_A)\right|^2} \,,
\end{equation}
with $a_{\rm max} \geq \rm{max}(\sqrt{(3/V_A)}, \sqrt{3/V_B})$. The result would then be given by the top line of Eq.~\eqref{eq:Bthinwall} (i.e. the two Hartle-Hawking/Vilenkin-Linde terms) with no tension term. Note that, even though the integration of Eq.~\eqref{eq:Bthinwall} in this case extends till $a_{\rm max}$, only the integrals up to $\sqrt{3/V_A}$ and $\sqrt{3/V_B}$ contribute to the real part of $B$. Hence, for instance, the factor $\left(1 - \frac{\bar{a}^2 V_B}{3}\right)^{3/2}$ evaluated at $\bar{a}^2 > 3/V_B$ gives no contribution to $B$ and analogously for the term containing $V_A$ in Eq.~\eqref{eq:Bthinwall}. Hence the relative
probability is now given by $P=e^{-B}$ with 
\begin{equation}
\boxed{\quad B=24\pi^{2}\left\{ \mp\frac{1}{V_{B}}\pm\frac{1}{V_{A}}\right\} .\quad }\label{eq:HHV}
\end{equation}
This is simply the ratio of the Hartle-Hawking/Vilenkin-Linde (depending on the choice of sign) probabilities for tunneling from ‘nothing’ to the state $B$ compared to tunneling from ‘nothing’ to the state $A$. Note that Eq.~\eqref{eq:HHV} might be interpreted as the tunnelling rate for the transition dS A $\rightarrow$ ‘nothing’ $\rightarrow$ dS B (or the opposite, depending on the choice of the signs). Interestingly, it seems that this would give a greater probability for transition than the CDL calculation. In this connection it should be pointed out that in the Hartle-Hawking case the wave function in the classical region is indeed a superposition of wave functions for expanding and contracting universes. The process we are envisaging may then be thought of as the contracting branch with the field sitting at the $A$ minimum tunneling to ‘nothing’ and then reemerging as an expanding branch in the $B$ minimum or the reverse.

\item{\bf CDL vs Hartle-Hawking}. 

In the Lorentzian analog of the CDL calculation on the other hand,  it is not clear how the true vacuum (i.e. $B$) emerges into the classical region. In CDL, the tunnelling rate is computed exactly with the same integral as in Eq.~\eqref{eq:B1}. However, in CDL this is interpreted as the action for a Euclidean configuration given by a compound state that joins two portions of 4-spheres (corresponding to dS B and dS A) along a 3-sphere (the wall): the $SO(5)$ symmetry of the Euclidean four dimensional dS is broken to $SO(4)$ by the presence of the ‘Euclidean’ wall. Afterwards, as described in App.~\ref{sec:dSGeometry}, one of the angular variables that implements the $SO(4)$ symmetry, is analytically continued and becomes the usual Lorentzian time, breaking the symmetry to $SO(1,3)$. In this way, the initial under-the-barrier Euclidean configuration (the compound state) becomes the real three-dimensional equal-time slice of the nucleated spacetime. As the continued angular variable is one that preserves the Euclidean $SO(4)$ symmetry, the nucleated spacetime is still a compound state of dS A and dS B: both dS spacetimes enter the classical region and keep evolving according to the classical equations of motion.

In our computation, the integral of Eq.~\eqref{eq:B1} (that we chose in order to recover the same CDL expression, and try to give it a Lorentzian interpretation) is associated with a particular path in field space that, starting from the configuration $(a_0, \phi_B)$ (‘B nothing’) leads to the nucleation of a full dS A sphere, i.e. the configuration $(a_{\rm max}, \phi_A)$. This has to be contrasted, as described in Eq.~\eqref{eq:PAB} with the path that, starting from the configuration $(a_0, \phi_A)$ leads to the same full dS A sphere as above. Essentially we are comparing 3-geometries, which is all one can do in the context of quantum gravity as described by the WDW equation.

In our procedure we neither need a dilute gas approximation nor a single negative mode in the spectrum of fluctuations as in the CDL flat space argument. Furthermore, there is no notion of bubble nucleation. Both the numerator and the denominator in Eq.~\eqref{eq:PAB} correspond to spacetime configurations that preserve the $SO(4)$ symmetry. To be more explicit, the portion of the path that corresponds to the dS B is always under the barrier in the Lorentzian case: the computation relies on the fact that $\bar{a}$ in Eq.~\eqref{eq:Bthinwall} after extremizing is located in the under-the-barrier region, so that the integrands of Eq.~\eqref{eq:SA2} give an imaginary contribution and the resulting $B$ is real. In the Lorentzian approach, the dS spacetime B never sees the light of the classical region. Hence, in the present case built on the CDL argument how $B$ emerges as
a classical space-time is not at all clear. 

Our computation relies on the fact that the path in field space extremising the action can be split as described in Sec.~\ref{sec:RecoveringCDL}, and that it is such that $C^{-1}$ in Eq.~\eqref{eq:T2} is imaginary. If the latter condition holds, the portion of the path such that $d\phi/ds \neq 0$ contributes to the imaginary action, bringing a term which is the analogous of Eq.~\eqref{eq:T} in the Euclidean computation. In general, this is not necessarily the case and one should compute the value of the action for the path that solves the equation of motion. In order to compute the contribution to the imaginary part of the action of a given path, it is sometimes more convenient to use the distance on field space as the parameter $s$, in which case the action for the wave function in the numerator of Eq.~\eqref{eq:PAB} can be written as
\begin{eqnarray}
\label{eq:GeneralExplicitAction}
S_0[a_0, \phi_B; a_{\rm max}, \phi_A] = 2 \pi^2 \int_{a_0, \phi_B}^{a_{\rm max}, \phi_A} \left[-6a da^2 + a^3 d\phi^2\right]^{1/2} \, \left[6a\left(1 - \frac{a^2 V(\phi)}{3}\right)\right]^{1/2} \,.
\end{eqnarray}

Hence, only the portions of the path such that the product of the two square roots in Eq.~\eqref{eq:GeneralExplicitAction} is imaginary contribute to the imaginary part of the path, and hence to the tunnelling rate.
\item{\bf Standard classical paths}.  It is interesting to notice that, due to the presence of a non-positive definite superspace metric, it is possible to find a classical solution that connects two dS spacetimes A and B. To illustrate this point, consider first the usual Hartle-Hawking transition, as a special case of Eq.~\eqref{eq:GeneralExplicitAction}. If we include the scalar field, the model becomes richer. In fact, the first bracket in Eq.~\eqref{eq:GeneralExplicitAction} is proportional to the kinetic terms of Eq.~\eqref{eq:calH} and can become negative without resorting to an imaginary parameter
\begin{equation}
\frac{1}{2} \left(-6 a \dot{a}^2 + a^3 \dot\phi^2\right) = - \frac{\pi_a^2}{12 a} + \frac{\pi_\phi^2}{2 a^3} = 3 a \left(1 - \frac{a^2 V(\phi)}{3}\right) \,.
\end{equation}

Below, we show an example in which we observe a classical transition from dS A to dS B in the potential of the left panel of Fig.~\ref{fig:PotentialExample}. The initial conditions are given by $a(0) = \sqrt{3/V_A}$, $\phi(0) = \phi_A = -0.01$, $\dot\phi(0) = 0.2$, so that the field goes over the barrier and start oscillating in the dS B minimum at $\phi_B = 0.01$, with decreasing amplitude. In the example, the maximum of the barrier is located at $\phi_{\rm max} \simeq -0.00375$ and the value of the potential is $V_{\rm max} \simeq 0.1068$. The initial speed necessary for the field to classically overcome the barrier and make the transition possible is roughly given by the height of the barrier $\Delta V = V_{\rm max} - V_A \simeq 0.0068$, so that $\frac{1}{2} \dot\phi^{2} \gtrsim \Delta V$. This is analogous to the recently proposed ‘fly-over’ scenario~\cite{Blanco-Pillado:2019xny} \footnote{We can equally get Hawking-Moss configurations after adjusting some of the parameters.}. Asymptotically, the dS B solution is recovered. In the example we used $V_A = 0.1$ and $V_B = 0.05$. In Fig.~\ref{fig:BracketTerms} we report the values of the brackets in Eq.~\eqref{eq:GeneralExplicitAction}, showing that they change sign at the same time even though the parameter $s$ of the evolution is always kept real.
\begin{figure}[h!] 
\begin{center} 
\includegraphics[scale=0.55]{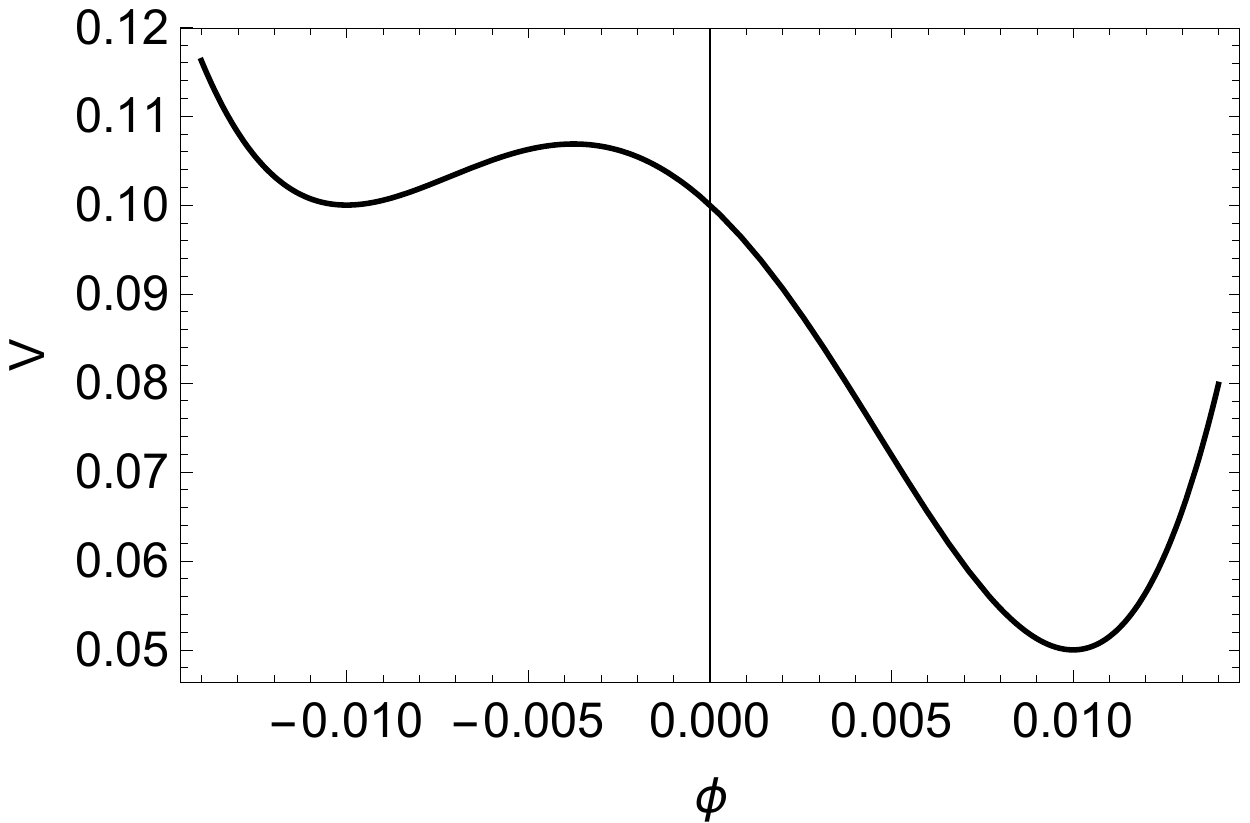} \qquad
\includegraphics[scale=0.55]{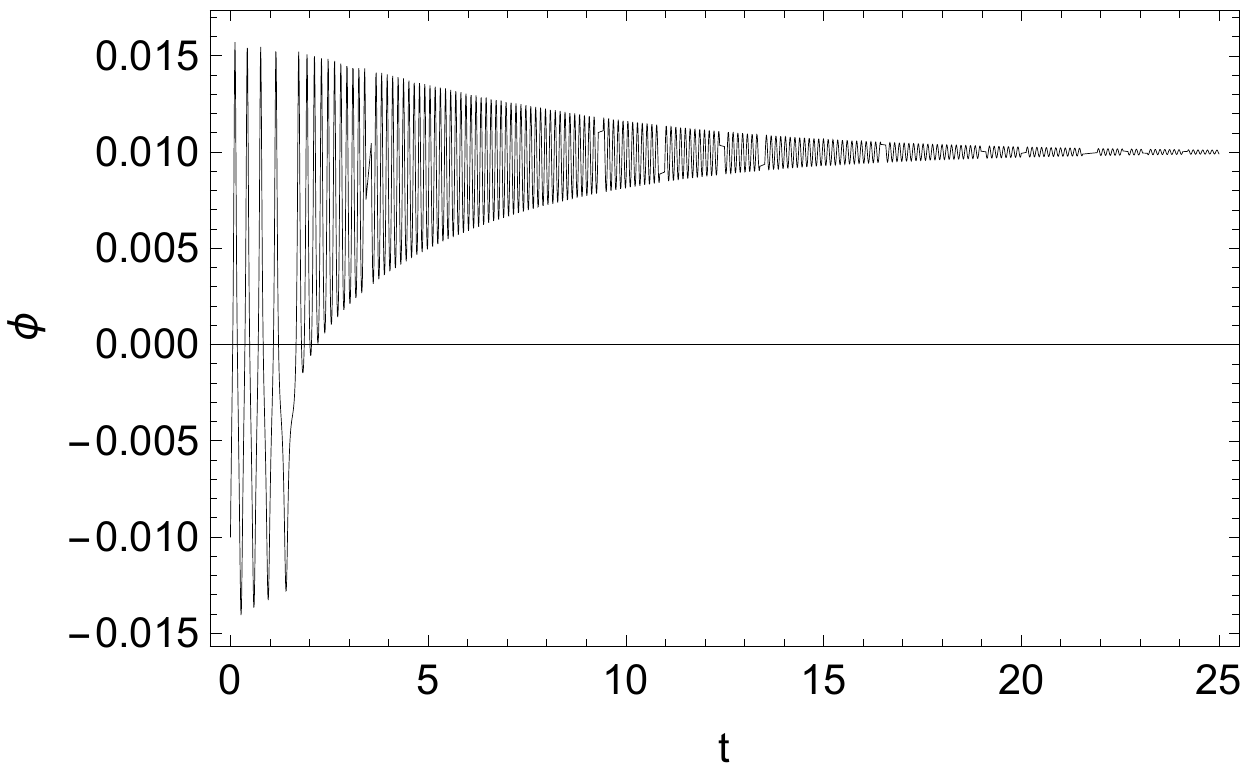}
\caption{\footnotesize{The potential in the left panel is a quartic polynomial in which four parameters are fixed by requiring two minima as described in the main text, while the remaining parameter fixes the height of the barrier. The right panel shows the trajectory of the field that starts in the A minimum and oscillates around the B minimum with decreasing amplitude. Asymptotically it goes to rest in the B minimum at $\phi = 0.01$.} \label{fig:PotentialExample}}
\end{center} 
\end{figure}
\begin{figure}[h!] 
\begin{center} 
\includegraphics[scale=0.55]{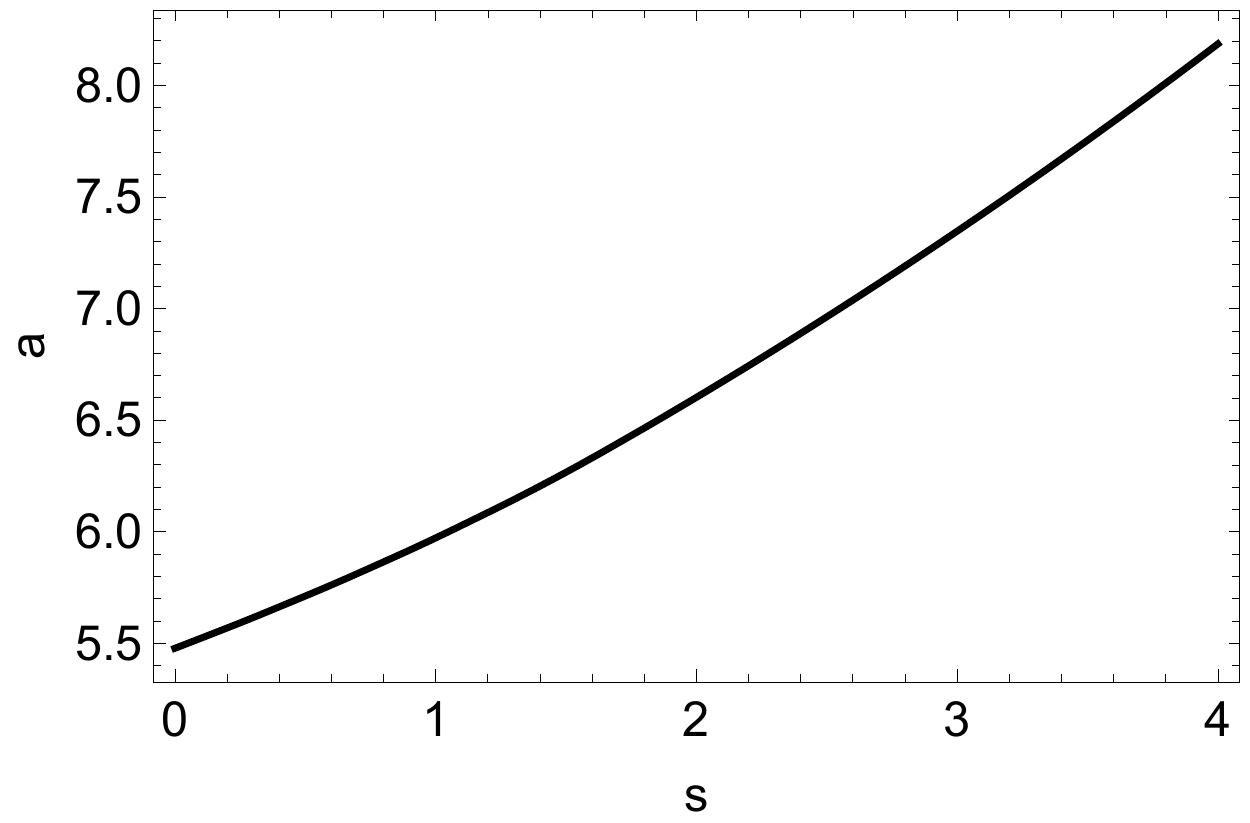} \qquad
\includegraphics[scale=0.55]{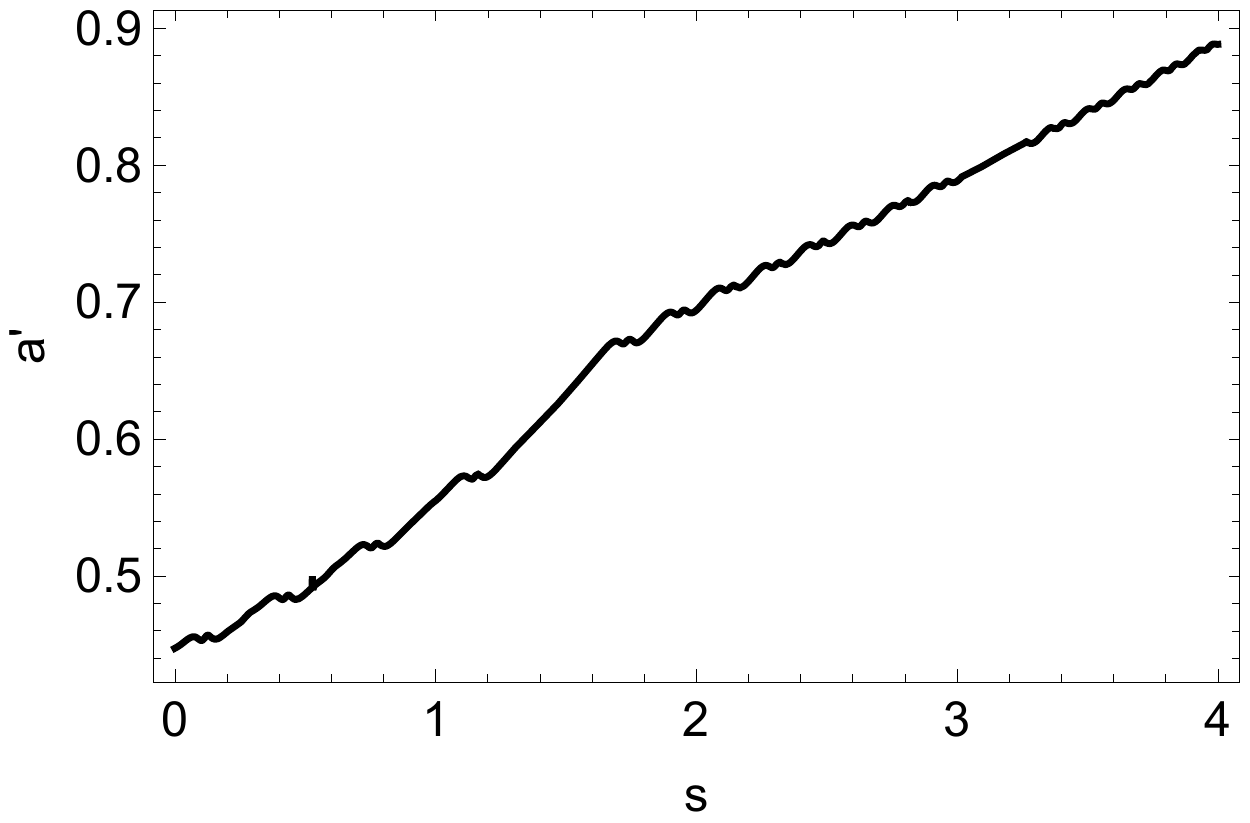}
\caption{\footnotesize{In the left panel we plot the scale factor. In the right panel we plot the derivative of the scale factor with respect to time $t$. After a contracting phase and a few large oscillations, the system settles into a regime of exponential growth once it is in the lower vacuum. } \label{fig:ScaleFactors}}
\end{center} 
\end{figure}
\begin{figure}[h!] 
\begin{center} 
\includegraphics[scale=0.45]{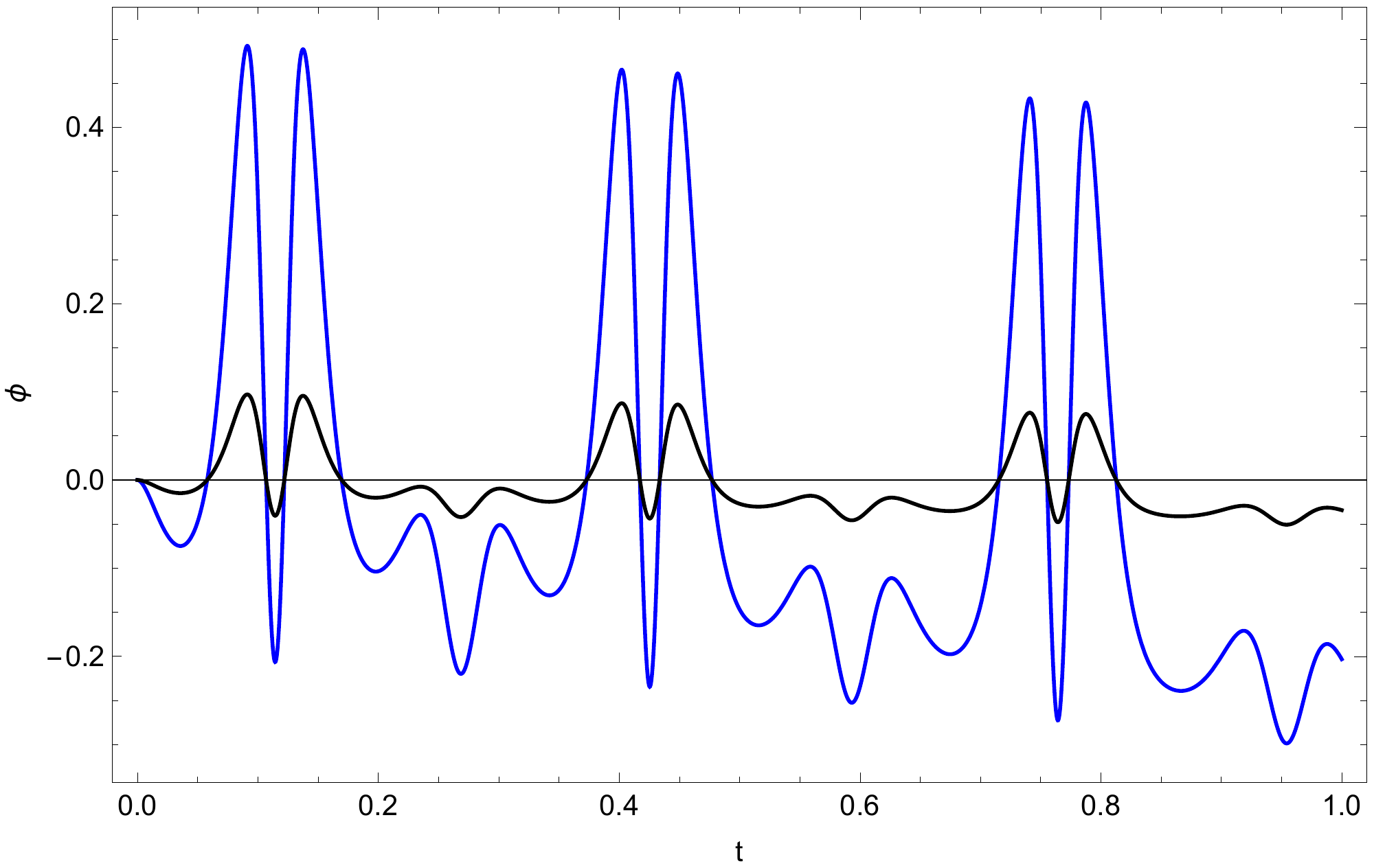} 
\caption{\footnotesize{The black line represents the term $\frac{1}{2} \left(a^3 \dot\phi^2 - 6 a \dot{a}^2\right)$, while the black line represents $1- \frac{a^2 V(\phi)}{3}$.}\label{fig:BracketTerms}}
\end{center} 
\end{figure}

\item{\bf Non-standard classical paths}. 
The Hamiltonian and momentum constraints can be expressed as the standard FLRW equation:

\be
\left(\frac{\dot a}{a}\right)^2=\frac{8\pi G}{3}\left(\frac{1}{2}\dot\phi^2+V(\phi)\right)-\frac{1}{a^2}\label{eq:hamiltonian}
\ee

and the scalar field equation

\be
\ddot\phi+3\left(\frac{\dot a}{a}\right)\, \dot\phi + V'(\phi)=0\label{eq:scalarequation} 
\ee

We want to explore if there are paths in the $\phi,a$ space that can connect the two sides of the barrier for a scalar potential with more than one minimum. A simple path can be starting with $a=$ constant. In this case the FLRW equation (Hamiltonian constraint) becomes:
\be
\frac{1}{2}\dot\phi^2+V(\phi)=\frac{3}{8\pi Ga^2}\equiv E
\ee
which is the same as the  energy conservation equation for a particle of energy $E$. As long as $E>V$ the field passes through classically over the barrier with no need of tunneling. Note that this happens only for a closed universe. Note also that the scalar field equation is satisfied automatically as usual for energy conservation. This is Einstein's static universe solution. 

For non-static cases, the standard classical solution would correspond to an initial velocity $\dot\phi$ to be big enough so that the kinetic energy can overcome the difference in potential energies to cross the barrier.  In cosmological setting this would correspond to the standard scenarios such as Hawking-Moss and the fly-over of the previous bullet point. However considering trajectories in the  $a(t), \phi(t)$ space can be more general than this. Due to the negative signature of the superspace metric there is no need to climb the potential energy barrier but we can move in the 2d $(a, \phi)$ space. 

For this, note that in Eq.~\eqref{eq:hamiltonian} the curvature term adds a negative term to the scalar potential and therefore if the scale factor decreases with time the initial speed needed to cross the barrier is smaller than that required to compensate for the potential difference. This can also be seen from Eq.~\eqref{eq:scalarequation} since for a contracting universe $H<0$  the ‘friction’  term $3H\dot\phi$ has the opposite sign, accelerating the field $\phi$ rather than slowing it down. This means that the classical path would have a contracting phase while the scalar field climbs through the barrier and then starts expanding after the transition. A sequence of contracting and expanding classical paths connecting different vacua would seem to be generic in a multi-minima scalar potential and could provide the basis of a novel scenario for early universe cosmology beyond old and new inflation. For a recent discussion of bouncing cosmologies see for instance~\cite{Gungor:2020fce}.

\subsubsection*{A toy model} 

To identify the classical path in a more concrete way let us consider the simplest square scalar potential.
\begin{figure}[h!] 
\begin{center} 
\includegraphics[scale=0.3,trim=3cm 12cm 0cm 5cm,clip]{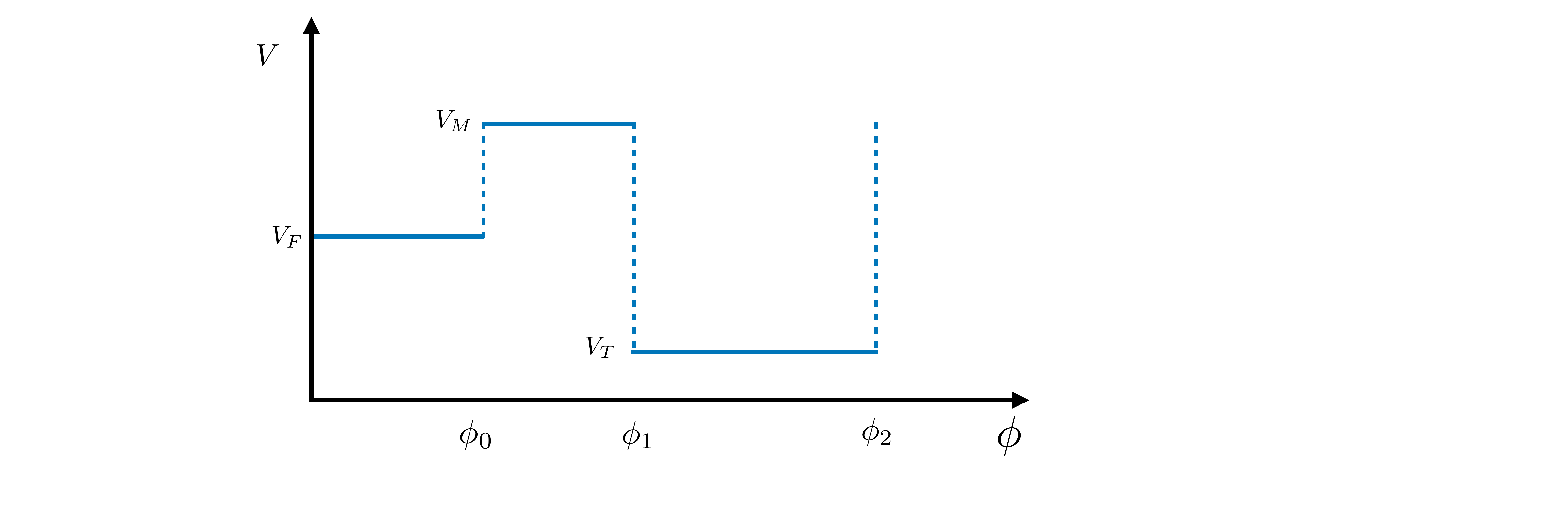}
\caption{Square potential.\label{fig:squarepotential}}
\end{center} 
\end{figure}
For a square potential as in the figure the second equation simplifies as $V'=0$. This can then be integrated to give:

\be
\dot\phi=\frac{K}{a^3}
\ee
with $K$ a constant that will be different for each of the regions $0<\phi<\phi_0$, $\phi_0<\phi<\phi_1$ and $\phi_1<\phi<\phi_2$.

Plugging this in the first equation (and setting $8\pi G=1$) we get:

\be
\dot a= \pm\sqrt{\frac{K^2+2Va^6-6a^4}{6a^4}}
\ee

Taking the ratio of these two equations we get:

\be
\phi= \pm \sqrt{6}K\int \frac{da}{a\sqrt{K^2+2Va^6-6a^4}}\label{eq:phiintegral}
\ee
where $V=V_F,V_M,V_T$ depending of the region as shown in the figure.

In principle this equation defines the path in the $\phi,a$ space. There will be  one expression like Eq.~\eqref{eq:phiintegral}  for each of the three regions with the integration constants and the values of the constants $K$ could be fixed by matching the solutions at the points $\phi_0$ and $\phi_1$.

\subsubsection*{A numerical example}

Let us consider the scalar potential of Fig.~\ref{fig:PotentialExample} with different initial conditions. Taking $\phi(0) = \phi_A$, $a(0) = 6.5$ and $\dot\phi(0) = 0.1$, we find using the Friedmann equation that $\dot{a}(0) \simeq \pm 0.69$. Let us then consider a contracting universe as initial condition: $\dot{a}(0) \simeq -0.69$. In this case, we get the scale factor and field evolutions in Fig.~\ref{fig:NonStandardFieldEvolution} and Fig.~\ref{fig:EvolutionFieldSpace}. Note that this is a different scenario with respect to standard  paths through the barrier such as Hawking-Moss and  ‘fly-over’ decay, as the initial kinetic energy ($\frac{1}{2} \dot\phi^{2} \simeq 5 \times 10^{-3}$ in the present example) is smaller than the barrier height ($\Delta V \simeq 6.8 \times 10^{-3}$). This illustrates explicitly our claim that due to the non-positive nature of the superspace metric there are more configurations connecting  the two different vacua.

\begin{figure}[h!] 
\begin{center} 
\includegraphics[scale=0.55]{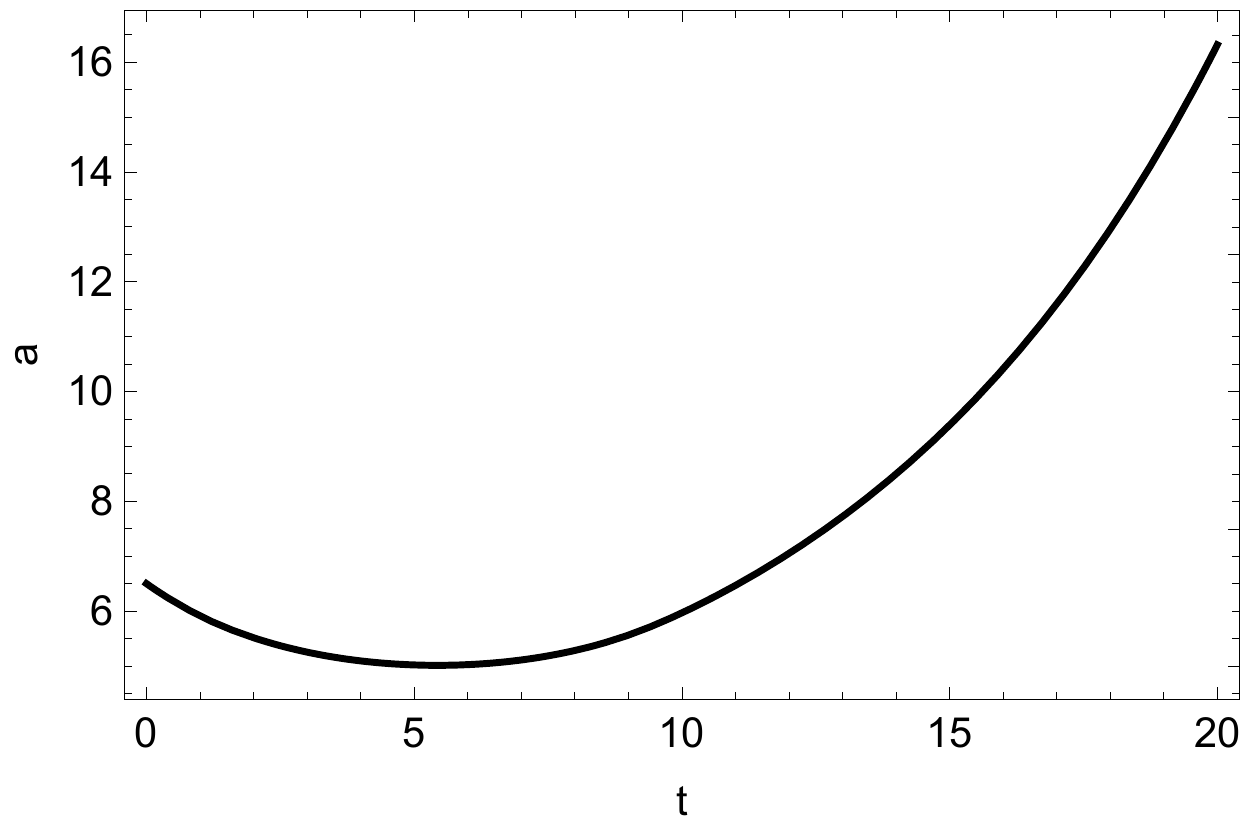} \qquad
\includegraphics[scale=0.59]{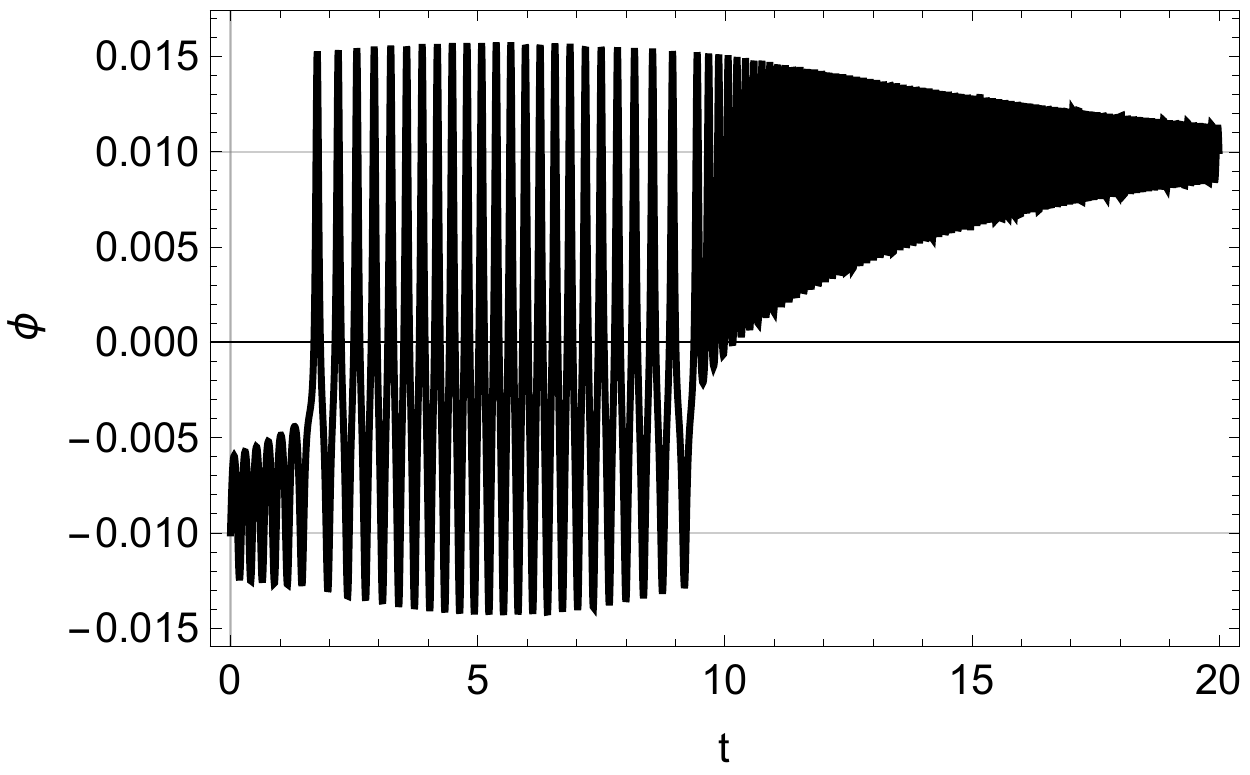}
\caption{\footnotesize{In the left panel we plot the scale factor  which is initially shrinking but then start growing.  In the right panel we plot the value of the scalar field with respect to  $t$. Initially  the field starts oscillating around the higher vacuum, then it bounces between the two vacua and finally it settles in  the lower vacuum.} \label{fig:NonStandardFieldEvolution}}
\end{center} 
\end{figure}

\begin{figure}[h!] 
\begin{center} 
\includegraphics[scale=0.8]{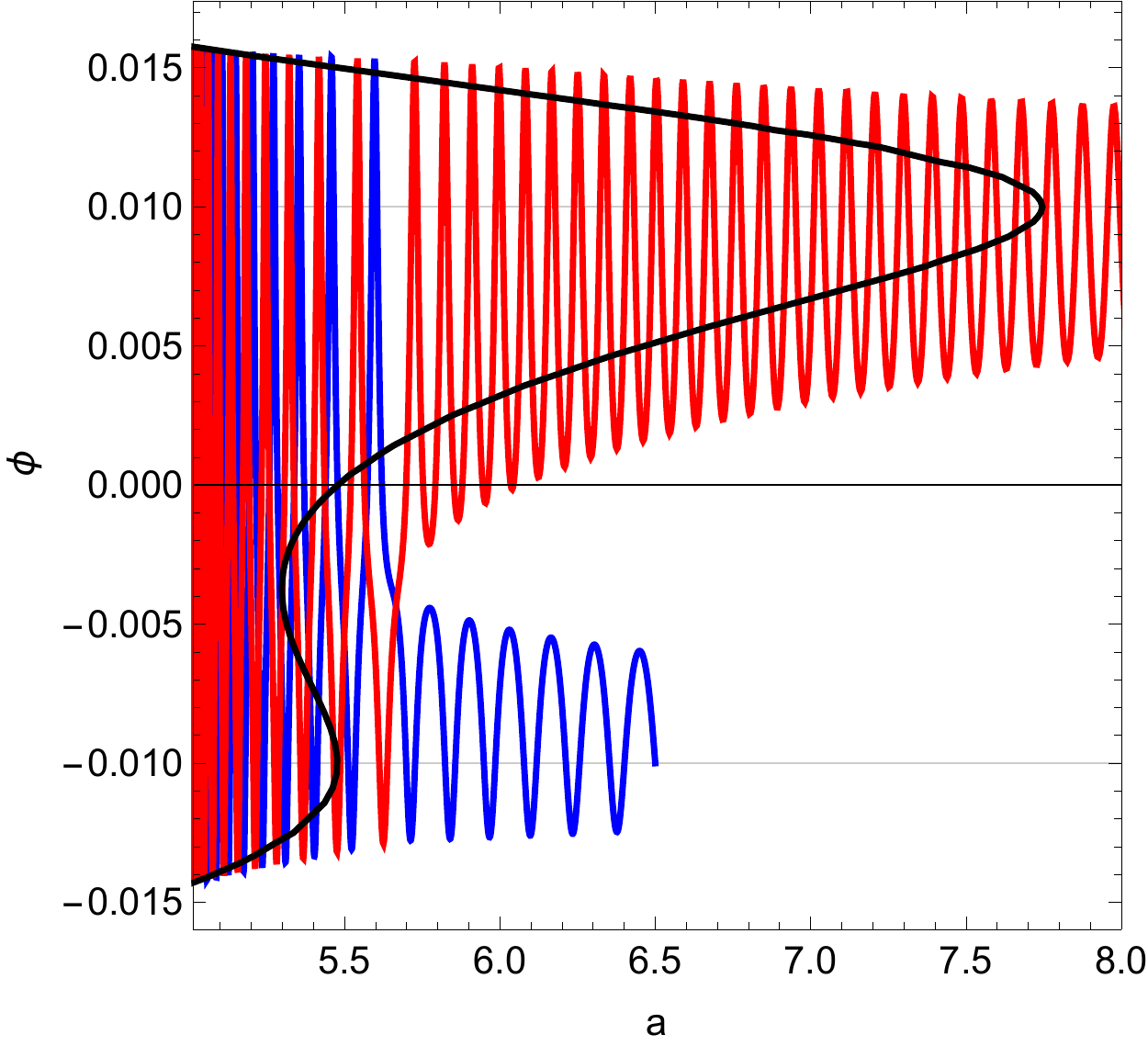}
\caption{\footnotesize{We show the evolution of the system in field space: after a contracting phase (blue), the scale factor starts expanding (red) and the system gets trapped in the vacuum B. The black line corresponds to the locus $a^2 = 3/V(\phi)$, hence it is the locus where the function $f(a,\phi) = 0$, see Fig.~\ref{fig:Path}.} \label{fig:EvolutionFieldSpace}}
\end{center} 
\end{figure}
\end{itemize}

\section{Open or Closed Universe?}
\label{sec:OpenClosed}

As we have seen in the previous sections, following  a Lorentzian approach the spacetime geometry is established from the beginning. For simplicity and computational control we started with a closed universe and the end result is then also a closed universe with $k=1$ LFRW  metric\footnote{Note that this is also the starting point in the Hartle-Hawking and Vilenkin approaches towards defining the wave function of the universe. The fact that closed universes are finite whereas flat and open universes have infinite volume makes closed universes better suited to define the probabilities associated to wave functions.}:
\begin{equation}
ds^2= -dt^2+a^2(t) d\Omega_3^2
\end{equation} 
which differs clearly from the open universe conclusion of CDL. This metric is by construction Lorentzian and has the $O(4)$ symmetry corresponding to a closed universe with no need of any analytic continuation.
 For $a\propto \cosh(\lambda t)$ this corresponds to the $SO(4,1)$ invariant closed  dS universe in global coordinates. In this case the natural foliation corresponds to horizontal surfaces of constant time. 
 
\begin{figure}[h!] 
\begin{center} 
\includegraphics[scale=0.8, trim=0cm 0.5cm 0cm 0.5cm,clip]
{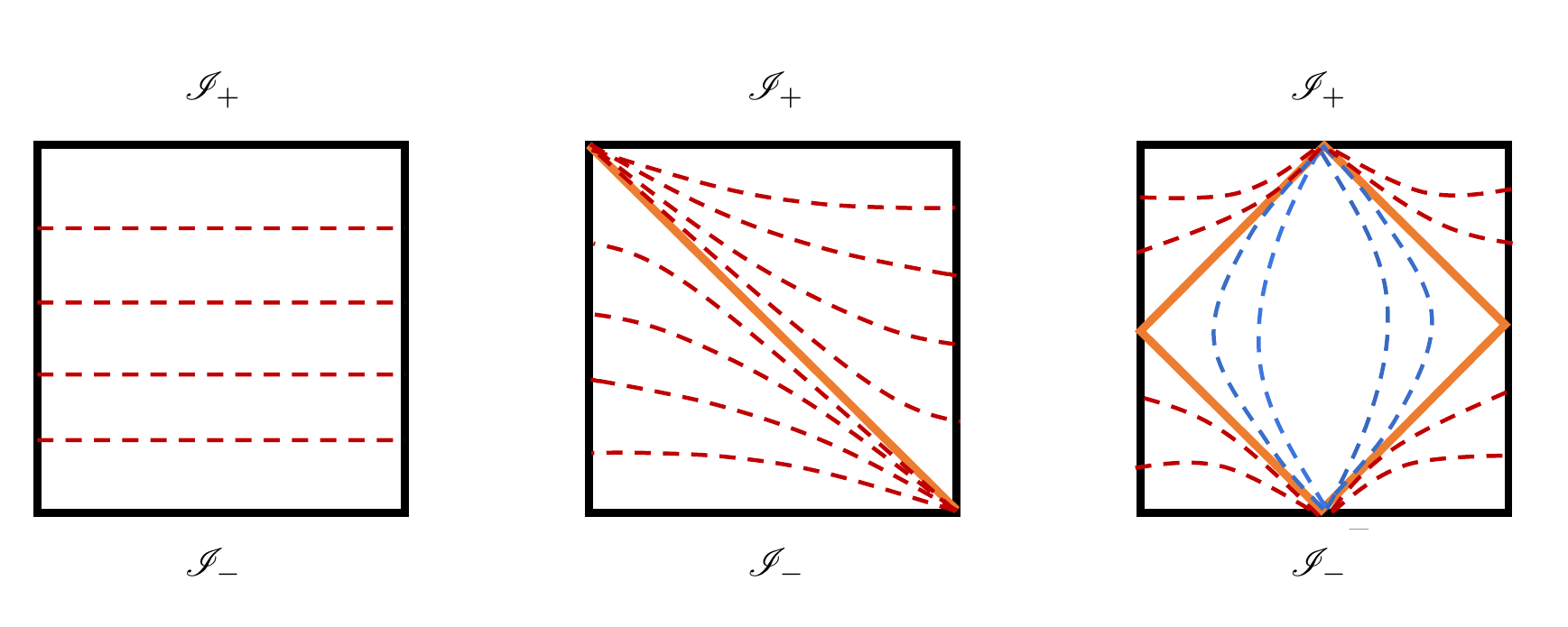}
\caption{Penrose diagrams for dS space with slicing corresponding from left to right to closed, flat and open slicings, respectively. Notice that the horizontal closed universe slicing is global.\label{fig:slicings}}
\end{center} 
\end{figure}

Here it is important to remark that dS space allows open, flat and closed slicings, see Fig.~\ref{fig:slicings}. Therefore, geometrically all slicings are equally allowed. Which foliation is preferred depends on the coupling to matter. For instance slices of constant inflaton field would naturally determine the proper time slicing and fix the curvature of the expanding universe within the nucleated bubble. In CDL this fixes the open universe but only after analytic continuation for which the original $O(4)$ symmetry becomes $O(3,1)$ of the open slicing. In the Lorentzian mini-superspace approach that we have followed here, the original $O(4)$ symmetry remains and implies the closed slicing. Next we will see that this remains true beyond mini-superspace as in the FMP Hamiltonian approach to quantum transitions.

\subsection{Beyond mini-superspace}

We have seen that our Lorentzian treatment led naturally to a closed slicing of dS space contrary to the CDL arguments. However, this may be an artefact of using the mini-superspace approximation in which the metric is only a function of time and there is no concrete description of the emergence of a bubble. In the Euclidean approach, even though the original calculations are also in mini-superspace,  the presence of the bubble and its spacetime trajectory after tunneling  is obtained from the proposed analytic continuation. In the Lorentzian approach there  is at present no explicit formalism to describe the quantum transitions between different vacua of a  scalar field potential. However, in the thin wall approximation in which the relevant quantities are the vacuum energy of the two vacua, the Hamiltonian formalism developed by Fischler, Morgan and Polchinski~\cite{Fischler:1990pk} can be used. In this case the spherically symmetric metric depends on both time and the radial coordinate. This allows us to describe the dynamics of the wall and its trajectory. Here  we recall the basics of this approach, which like the Lorentzian mini-superspace model, naturally implies a closed universe slicing of the spacetime as seen by an observer inside the bubble.

The spherically symmetric metric takes the form:
\be
ds^2=-N_t^2(t,r)dt^2+L^2(t,r)(dr+N_r(t,r)dt)^2+R^2(t,r)d\Omega_2^2
\ee
with $N_t,N_r$ the lapse and shift functions respectively and $d\Omega_2^2$ the line element for the 2-sphere. The system consists of two dS spaces with cosmological constants $\Lambda_I, \Lambda_O$ separated by a wall of tension $\sigma$  at $r=\hat r$. The bulk and boundary actions are the standard gravitational ones and the  matter action is given by the two cosmological constants, so the total action is:
\be
S=\frac{1}{16\pi G}\int_{\mathcal M} d^4x\sqrt{-g}\mathcal{R}+\frac{1}{8\pi G}\int_{\partial \mathcal M} d^3y\sqrt{-h} K+S_M+S_W,
\ee
where $K$ is the extrinsic curvature of the wall and
\bea
S_M &= & -4\pi  \int dt dr L N_t R^2\left(\Lambda_O\theta(r-\hat r)+ \Lambda_I \theta(\hat r-r) \right), \nonumber \\
S_W &= & -4\pi T\int dt dr \delta(r-\hat r)\left[N_t^2-L^2(N_r+\dot{\hat{r}})^2\right].
\eea
 In the above we defined $T\equiv 4\pi G\sigma$. Following the standard Dirac prescription for this Hamiltonian system, the Hamiltonian and momentum constraints can be found and the matching conditions at the wall lead to an equation for the wall trajectory of the form:
 \be
 \dot{\hat{R}}^2+V=-1; \qquad V=-\frac{\hat{R}^2}{R_0^2},
 \label{rdot}
 \ee
 where $\hat R=R(\hat r)$ and $R_0$ is the turning point:
 \be
 R_0^2=\frac{4T^2}{\left[(H_O^2-H_I^2)^2+2T^2(H_O^2+H_I^2)+T^4 \right]}.
 \ee
 With $H^2_{I,O}=8\pi G \Lambda_{I,O}/3$. The classical trajectory of the wall is then given by:
 \be
 R(t)=R_0\cosh\frac{t}{R_0}
 \ee
 The quantum probabilities are determined from the solutions of the WDW equation $\mathcal{H}\Psi=0$ with ${\mathcal P}$ the relative probability of the configuration of the two dS spaces and the wall compared to that for just one dS: 
 \be
 {\mathcal P}({\rm dS}\rightarrow{\rm dS}/\rm{dS}\oplus W)=\frac{|\Psi({\rm dS}/\rm{dS}\oplus W)|^2}{|\Psi(\rm{dS})|^2}
\ee
The detailed calculation using the WKB method including a discussion of the matching of the under-the-barrier wave function to that in the classical region is given in~\cite{deAlwis:2019dkc} and the result reproduces the standard exponential factor $e^{-B}$ with $B$ given by Eq.~\eqref{eq:Bthinwallextr}. This provides yet another Lorentzian way to derive the same decay rate. But contrary to the mini-superspace approach, the presence of the wall and its classical trajectory after the transition is made quite explicit

\begin{equation}
\label{eq:GeneralWF}
\Psi = a e^{I} + b e^{- I} \,,
\end{equation}
where, given a configuration with action $S$, we have denoted the combination $iS = I$ and the action $S$ is evaluated on a classical solution. 
The total action away from the turning point (but still under the barrier) is

\begin{small}
\begin{equation}
\begin{gathered}
\label{eq:B1i}
I_{\rm tot} = \, \frac{\pi}{4H_{\rm I}^{2}}\left[1-\epsilon(\hat R^{\prime}_{-}) \frac{2}{\pi} \left(\cos^{-1}\left(\frac{\hat R}{R_{\rm o}}\sqrt{1 - H_{\rm I}^{2} R_{\rm O}^{2}}\right)\right)\left(\frac{R_{\rm o}^{2} - \hat R^{2}}{R_{\rm o}^{2} - \hat R^{2}(1 - H_{\rm I}^{2} R_{\rm o}^{2})}\right)^{3/2}\right] - \\  
- \frac{\pi}{4 H_{\rm O}^{2}} \left[1 - \epsilon(\hat R^{\prime}_{+}) \left(2 + \frac{2}{\pi} \cos^{-1}\left(\frac{\hat R}{R_{\rm o}}\sqrt{1-H_{\rm O}^{2} R_{\rm o}^{2}}\right)\right)\left(\frac{R_{\rm o}^{2}-\hat R^{2}}{R_{\rm o}^{2} - \hat R^{2}(1 - H_{\rm O}^{2} R_{\rm o}^{2})}\right)^{3/2} \right] + \\ 
+ \frac{\hat R^{3}}{2 R_{\rm o}} \sqrt{R_{\rm o}^{2} - \hat R^{2}}  \left[\frac{(H_{\rm O}^{2} - H_{\rm I}^{2} + T^{2})}{\sqrt{1 - c_{-}^{2}\hat R^{2}}} - \frac{(H_{\rm O}^{2} - H_{\rm I}^{2}-T^{2})}{\sqrt{1 - c_{+}^{2}\hat R^{2}}}\right] - \\
- \left[\frac{H_{\rm O}^{2} - H_{\rm I}^{2} + T^{2}}{4 T H_{\rm I}^{2}} - \frac{H_{\rm O}^{2} - H_{\rm I}^{2} - T^{2}}{4 T H_{\rm O}^{2}}\right] R_{\rm o} \sin^{-1} \left(\frac{\hat R}{R_{\rm o}}\right) \,.
\end{gathered}
\end{equation}
\end{small}

The background action is obtained by setting $\hat{r}_{\pm}=0$ in the above expressions (corresponding to having the complete dS space with Hubble parameter $H_{\rm O}$), giving us
\begin{equation}
\overline{I} = -\frac{\pi}{2 G H_{\rm O}^{2}}\left[(1 - H_{\rm O}^{2} a_{\rm O}^{2})^{3/2} - 1 \right] \,,\label{eq:HHV}
\end{equation}
which gives the Hartle-Hawking (under the barrier) wave function when
substituted into Eq.~\eqref{eq:GeneralWF} with $b=0$ and gives the Vilenkin version when $b=2ia$.

Now as pointed out in~\cite{Blau:1986cw}  when the background geometry is a black hole there are two classically allowed (I and III) and one classically forbidden regions (II) as in the usual tunneling problem in quantum mechanics, as discussed for instance in Sec.~\ref{sec:FlatSpace}. Classically, the wall expands (or contracts) up to a classical turning point and then re-collapses (or re-expands), but quantum mechanically it can tunnel under the barrier and resurface after the second turning point.  In the dS to dS case however there is no region I~\cite{Bachlechner:2016mtp,deAlwis:2019dkc} and the situation as discussed in more detail in~\cite{deAlwis:2019dkc} is similar to tunneling from ‘nothing’. In either case the WDW equation has two independent solutions as in Eq.~\eqref{eq:GeneralWF} in each of these regions that need to be matched at the classical turning points. In the dS to dS case there is  just the one turning point, the coefficients $a$ and $b$ will determine the two coefficients in region III  through WKB matching conditions. In `tunneling from nothing' discussions one usually imposes an additional boundary condition, either the outgoing wave condition of  Vilenkin or the real wave function (coming from the so-called `no-boundary' condition') of Hartle and Hawking. However, if no such  condition is imposed, in general one of the two solutions in Eq.~\eqref{eq:GeneralWF} will dominate in each region, depending on the sign of the real part of   the action $I$. 

\noindent The ratio
\begin{equation}
\label{eq:GeneralDefProbability}
\mathcal{P}(\mathcal{B} \rightarrow \mathcal{N}) = \left|\frac{\Psi_{\mathcal{N}}}{\Psi_{\mathcal{B}}}\right|^2 \,,
\end{equation}
gives the probability of finding the system in the `nucleated' state $\mathcal{N}$ versus being in the `background' state $\mathcal{B}$, Notice that the two states $\mathcal{B}$ and $\mathcal{N}$ do not always have the meaning of `initial' and `final' states (see Sec.~\ref{sec:Introduction}): the transition can be clearly interpreted as happening in time if there is an initial classical motion of the bubble wall.  In the cases in which there is no initial classical motion of the wall the interpretation is less clear. For this reason we will refer to the state $\mathcal{B}$ as the `background' spacetime, instead of initial spacetime. Given that this is a  relative probability, it does not have to be smaller than one, and we avoid the problem of the normalisation of wave functionals.

The semi-classical wave function in the region III (i.e. the classical region where the brane spontaneously emerges from ‘nothing’) is obtained by analytically continuing the expression in Eq.~\eqref{eq:B1i}. The denominator in Eq.~\eqref{eq:GeneralDefProbability} corresponds to the emergence from ‘nothing’ of dS space with the `background' radius $H_O^{-1}$ and the relative probability is given by 

\begin{align}
\label{eq:Pout}
& \mathcal{P}_{{\rm {\rm out}}}(\mathcal{B} \rightarrow \mathcal{N}) = \frac{|\Psi^{{\rm out}}_{{\cal N}}|^{2}}{|\Psi^{{\rm out}}_{{\cal B}}|^{2}} =  \\
& = \frac{|a+i\frac{b}{2}|^{2} e^{2 \text{Re}\left(I_{\rm out}(\hat{R})\right)} + |a-i\frac{b}{2}|^{2}e^{-2 \text{Re}\left(I_{{\rm out}}(\hat{R})\right)}+2\text{Re}\left[(a+i\frac{b}{2})(a^{*}-i\frac{b^{*}}{2})\right]e^{2i\text{Im} \left(I_{\rm out}(\hat{R})\right)}}{|a+i\frac{b}{2}|^{2}e^{2 \text{Re}\left(I_{{\rm out}}(0)\right)} + |a-i\frac{b}{2}|^{2} e^{-2 \text{Re}\left(I_{{\rm out}}(0)\right)} + 2 \text{Re}\left((a+i\frac{b}{2})(a^{*}-i\frac{b^{*}}{2})\right) e^{2i \text{Im} \left(I_{{\rm out}}(0)\right)}} \,.\nonumber
\end{align}.

As argued in~\cite{deAlwis:2019dkc}  the dominant contribution comes from the ratio of the first terms in the numerator and denominator and gives precisely the expression that was obtained by the generalisations of CDL~\cite{{Weinberg:2012pjx}} and by Brown and Teitelboim~\cite{Brown:1988kg}, namely the expression in Eq.~\eqref{eq:Bthinwallextr}.
 
 However it is important to emphasise that even though the final  exponential term for the relative probability   is the same as that obtained by the Euclidean instanton/dilute-instanton-gas method the expression before minimising with respect to the wall radius i.e. Eq.~\eqref{eq:Bthinwall} is very different from the (analytic continuation of) expression in Eq.~\eqref{eq:B1i}. This suggests that the Lorenzian continuation of the CDL or BT Euclidean argument is strictly speaking just the mini-superspace calculation of Sec.~\ref{sec:CDLfromWDW} whilst the next in order of complication - namely the spherically ($S_2$) symmetric Lorentzian calculation~\cite{deAlwis:2019dkc} based on~\cite{{Fischler:1990pk}} gives a completely different amplitude - even though the  exponential term is in agreement. Furthermore as pointed out in Sec.~\ref{sec:FlatSpace} even in the flat space case the direct WKB calculation gives a different pre-factor.
 
Finally, it is interesting to remark that in the Hamiltonian approach of~\cite{Fischler:1989se,Fischler:1990pk}, in order to  compare their results with the Euclidean approach of FGG~\cite{Farhi:1989yr}, they describe a canonical Euclideanisation of their approach by working on a static path and determine the relevant functions as $R,L$ in terms of the bubble location $\hat R$. Then they use $\hat R$ as the parameter that plays the role of Euclidean time in the Euclidean formalism. By doing this they successfully explain the Euclidean results of FGG in terms of a singular instanton that in the Hamiltonian approach corresponds to well behaved geometries. This Euclideanisation essentially corresponds to the standard $t\rightarrow it$ Wick rotation and does not correspond to the analytic continuations performed by CDL.

\subsection{Classical bubble trajectory  after nucleation}
Now we want to determine the Penrose diagram for the trajectory of the bubble after nucleation.

The equations of motion for the wall are given by the junction conditions which are obtained by embedding the wall coordinates in dS spacetime. Assuming rotational invariance  the metric of the wall is given by
\be
ds^2_\Sigma=-d\tau^2+R^2(\tau)\, d\Omega^2 \,. \label{eq:wall_metric}
\ee
We will first choose static coordinates for  dS, 
\be
ds^2=-(1-H^2r^2)\, dt^2+\frac{dr^2}{1-H^2r^2}+r^2\, d\Omega_2^2 \,,
\ee
where $H^{-1}$ is the radius of dS. In these coordinates the wall radius is given by $r(\tau)=R(\tau)$. The junction conditions are 
\begin{equation}
(K^+)^{ i}_{\ j}-(K^-)^{ i}_{\ j}=-4\pi G \sigma\delta^{i}_{\ j} \,,
\end{equation}
where $K_{ij}^\pm$ is the extrinsic curvature at each side  and $\sigma$ is the tension of the wall. In order to compute the extrinsic curvature it is helpful to use Gaussian normal coordinates in which $K_{ab}$  takes the simple form
\begin{equation}
K_{ab} = - \Gamma^n_{ab} = - \frac{1}{2} \partial_n g_{ab} = -\frac{1}{2} n^\mu \partial_\mu g_{ab} \,,
\end{equation}
where $n^\mu$ denotes the unit vector orthogonal to the wall and $g_{ab}$ is the induced metric at the wall. The whole computation then boils down  to calculate the normal vector $n^\mu$ in an appropriate coordinate systems, from which we can compute the extrinsic curvature on the two sides of the wall and then enforce the junction conditions. To do so let us first denote the four-velocity of a point on the wall as $U^\mu$. In the static patch coordinate system, due to the spherical symmetry of the wall we have
\begin{equation}
U^\mu_S = \left(\dot{t}_{dS},\dot{R},0,0\right) \,,
\end{equation}
where $\dot{}$ denotes the derivative with respect to proper time. Note that the  normalisation of the four-velocity $g_{\mu \nu} U^\mu_S U^\nu_S = - 1$ implies the following relation between $\dot{t}_{dS}$ and $\dot{R}$:
\begin{equation}
\label{eq:TdRdS}
(1-H^2 R^2) \dot{t}_{dS}^2 = 1 +(1-H^2 R^2)^{-1} \dot{R}^2 \,.
\end{equation}
To compute the  normal vector to the wall  first notice that this is orthogonal to the four-velocity $U^\mu$, i.e. $g_{\mu \nu} n^\nu U^\mu_S = 0$. Using this  condition and that  $g_{\mu \nu} n^\mu n^\nu = 1$ implies

\begin{equation}
n^\mu = \left((1-H^2 R^2)^{-1} \dot{R}, \pm \sqrt{1-H^2 R^2 + \dot{R}^2},0,0\right) \,.
\end{equation}
One can now compute the junction condition. The $\theta$ component gives,
\begin{equation}
\sqrt{1-H_+^2 R^2+\dot R}-\sqrt{1-H_-^2 R^2+\dot R^2}=4\pi\sigma R \,,
\end{equation}
where the subscript $\pm$ indicates the side of the wall where we are evaluating.
After some manipulation this equation leads to Eq.~\eqref{rdot}, and so to a solution for the radius of the wall $R$. One also has the  $\tau$ component of the extrinsic curvature,\
$K_{\tau\tau}=U^{\mu}U^{\nu}\nabla_{\nu}n_{\mu}=-n_\mu U^{\nu}\nabla_\nu U^\mu $, which can be interpreted as the normal acceleration of the wall.  This implies that the trajectory followed by the  wall is not a geodesic unless $K_{\tau\tau}$ vanishes. This can be  evaluated in the static patch coordinates,
\begin{equation}
K_{\tau\tau}=-\frac{\ddot R-H^2 R}{\sqrt{1-H^2 R^2+\dot R^2}}=-\frac{\sqrt{1-H ^2R_0^2}}{R_0} \,, \label{eq:normalcurvature}
\end{equation}
where in the last step we have used Eq.~\eqref{rdot}. This last equation implies that the normal acceleration is a non-vanishing constant (since $R_0H<1$).
\subsection*{Global slicing\footnote{This part follows the last part of appendix C of~\cite{Blau:1986cw}.}}
We will now embed the wall in global coordinates, which are described in Eq.~\eqref{foliation:global}. The metric is given by
\begin{equation}
 ds^2=\frac{1}{H^2\cos^2T}\, \left(-dT^2+d\rho^2+\sin^2\rho d\Omega^2 \,, \label{eq:globalconformal}
\right)
\end{equation}
where $T$ is conformal time that varies from $-\pi/2$ at $\mathscr{I}_-$ to  $\pi/2$ at $\mathscr{I}_+$.  Embedding the wall metric in Eq.~\eqref{eq:wall_metric} into the global coordinates implies that $T(\tau)$ and $R(\tau)=H^{-1}\sec T(\tau)\sin\rho(\tau)$.  Also, plugging this back into Eq.~\eqref{eq:globalconformal} we have,
\begin{equation}
-\dot T^2+\dot\rho^2=-H^2\cos^2 T \,,
\end{equation}
where dots are derivatives with respect to proper time. Using this relation we find
\begin{equation}
\dot T^2=\frac{H^2\cos ^2T}{1-\rho'^2},~~\dot {\rho}^2=\frac{H^2\cos ^2T}{1-\rho'^2}\rho '^2 \,, \label{eq:tdot}
\end{equation}
with $\rho'\equiv d\rho/d T$. To describe the trajectories of the wall in  global coordinates we would like to find an expression for  $\rho(T)$.  Let us start by noticing that
\begin{eqnarray}
\dot R&=&\frac{1}{H} \frac{d T}{d\tau}\left(\tan T\sec T \sin\rho+\sec T\cos\rho \rho' \right)\nonumber\\
&=& \frac{\tan T\sin\rho+\cos\rho\rho'}{\sqrt{1-\rho'^2}}=\sqrt{\frac{\sin^2{\rho}}{H^2R_0^2\cos^2T}-1} \,,
\end{eqnarray}
where we have  used Eq.~\eqref{eq:tdot} and in the last step Eq.~\eqref{rdot}. The last equality is a first order non-linear differential equation which determines the wall trajectory in conformal coordinates. The solution turns out to be remarkably simple: $\cos(\rho)=\sqrt{1-H^2R_0^2}\cos T \label{eq:sol1}$ as the reader may easily verify. 

In practice it turns out that to obtain this solution it is more convenient to use Eq.~\eqref{eq:normalcurvature}. This is straightforward since it  is also possible to write  $\ddot R$,  in terms of $\rho$, $T$ and derivatives of $\rho$ with respect to $T$.  After substituting into Eq.~\eqref{eq:normalcurvature} we get \footnote{Alternatively we can use that  is possible to write $K_{\tau\tau}=-\frac{\dot\beta}{\dot R}$, with $$\beta\equiv\sqrt{1-H^2R^2+\dot R^2}=\frac{\cos\rho+\sin\rho\tan T\rho'}{\sqrt{1-\rho'^2}}\,.$$},
\begin{equation}
K_{\tau\tau}=-\frac{\cos (T) \rho''-\sin (T)\rho'^3+\sin (T) \rho'}{\left(1-\rho'^2\right)^{3/2}}H \,,
\end{equation}
given that $K_{\tau\tau}$ is constant this is a   second order ODE for $\rho$ as a function of $T$. 
Furthermore this expression does not depend explicitly on $\rho$ and  can be easily integrated if we rewrite it as,
\begin{equation}
\frac{\sqrt{1-H ^2R_0^2}}{H R_0}=\cos^2 (T)\frac{d}{dT}\left(\frac{ \sec T \rho'}{\sqrt{1-\rho'^2}}\right) \,,
\end{equation}
which leads to

\begin{equation}
\rho'=\pm\frac{\sqrt{1-H^2R_0^2}\sin(T)}{\sqrt{H^2R_0^2+(1-H^2R_0^2)\sin ^2(T)}}\,,\label{eq:rho'}
\end{equation}
where to fix one integration constant we have imposed $\rho'=0$ at $T=0$, which comes from Eq.~\eqref{rdot}.
We will keep the positive signs as it means that the wall speed increases. Notice  that  $\rho'<1$ and that at $\mathscr{I}_+$, $\rho'=\sqrt{1-H^2R_0^2}<1$. This expression can be integrated to obtain
\begin{equation}
\boxed{\quad \cos(\rho)=\sqrt{1-H^2R_0^2}\, \cos T \quad} \label{eq:sol1}
\end{equation}
where we have used that at $T=0$, $R_0=\cos(\rho(0))$. Eq.~\eqref{eq:sol1} determines the trajectory of the wall in global coordinates. Now let us analyse this expression, first
note that this the trajectory never crosses the light cone $\rho=T$ since $T<\arccos(\sqrt{1-HR_0^2}\cos T)=\rho$.    Also note that   all trajectories end at $\rho=\pi/2$, since   at $T=\pi/2$, $\cos\rho=0$. Moreover the world sheet of the trajectory is a time-like  hyperboloid having $SO(3,1)$ invariance. To see this we can substitute Eq.~\eqref{eq:sol1} into the equations for the embedding of global dS from Eq.~\eqref{foliation:global}.
Hence the equation for the brane world volume in embedding coordinates is,

\begin{eqnarray}
X_0&=& H^{-1}\tan T,\nonumber \\
 X_1&=&H^{-1} \frac {\cos\rho}{\cos T}=H^{-1}\sqrt{1-H^2R_0^2},\nonumber \\
X_2^2+X_3^2+X_4^2&=&R^2=\frac{\sin^2{\rho}}{H^2\cos^2 T} =   \frac{1}{H^2\cos^2T}-\frac{1-H^2R_0^2}{H^2}.
\end{eqnarray}

Hence we have the equation for the world sheet of the brane,

 \begin{equation}
-X_0^2+X_2^2+X_3^2+X_4^2=R_0^2.
\end{equation}\label{eq:braneworld}

This is the equation of a hyperboloid with  $SO(3,1)$ symmetry\footnote{In fact one could have guessed the solution simply by demanding that $X_1$ is a constant since that is the simplest choice for the embedding given that $R^2$ cannot be set to a constant, and then fixing the constant from the fact that at $T=0=X_0$, $R=R_0$.}, in other words it is 3 dimensional dS space with radius $R_0$   and  corresponds to the Lorentzian rotation of Coleman's~\cite{Coleman:1980aw} Euclidean bounce solution.

In summary we have explicitly found a closed expression for the classical trajectory of the wall after nucleation given by Eq.~\eqref{eq:sol1}. It corresponds to an $SO(3,1)$ symmetric  hyperboloid. The speed is determined by Eq.~\eqref{eq:rho'} which even though it increases it never reaches the speed of light ($|\rho'|<1$). However it can easily be seen that if gravity is decoupled ($G\rightarrow 0$) the turning point goes as $R_0\rightarrow 3\sigma/(\Lambda_O-\Lambda_I)$ but $H\rightarrow 0$ and then $\rho'\rightarrow 1$ reproducing the flat space results~\cite{Aguirre:2008wy}. In general, the limiting speed differs from the speed of light by a small amount of order $M^2/M_P^2$ with $M$ the reference scale of the scalar field potential. Finally, from the Penrose diagram it can be seen that the trajectory is such that a signal from the centre of the bubble cannot reach the wall but in principle radiation from the wall can reach the observer at the centre.

\begin{figure}[h!] 
\begin{center} 
\includegraphics[scale=0.4,trim=6cm 10cm 10cm 5cm,clip	]{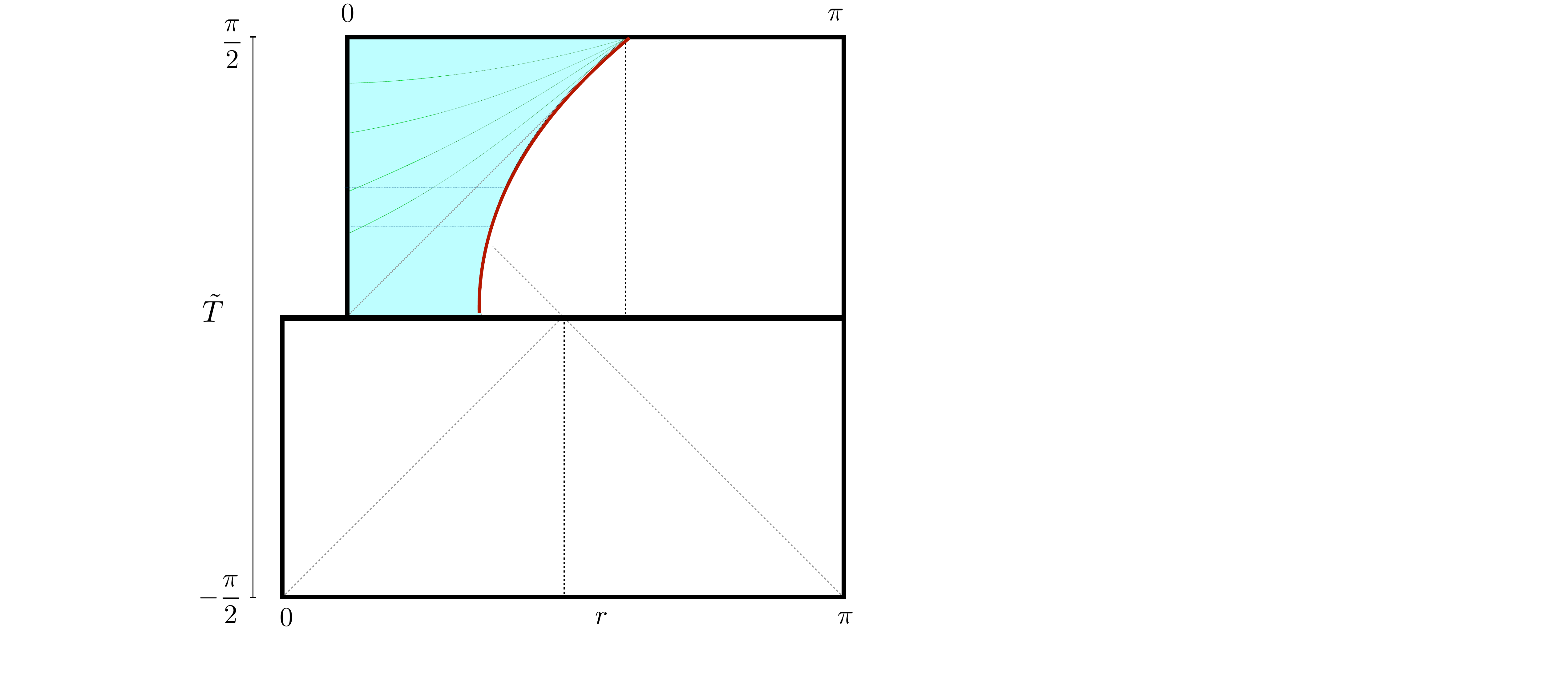}
\caption{Penrose diagram for the FMP dS to dS transition.
The lower part is the universe before the transition. The upper part  is the universe after the transtion composed of two  regions with different vacua separated by a wall, which is  the red line. The equation of the wall is given by Eq.~\eqref{eq:sol1}. The pale blue region is the part of the universe with the true vacuum, where the green dotted lines are open universe constant time slices and the blue dotted lines are closed universe constant time slices.\label{fig:Potential}}
\end{center} 
\end{figure}

\label{sec:LorentzianGeometry}

\subsection{Cosmological implications}

Let us now compare the cosmological differences between CDL and a closed universe after the tunneling transition. This revives the old question regarding the spatial curvature of the universe. After inflation this issue is considered less urgent since whatever the original curvature is, a short period of inflation is enough to render the universe essentially flat. However at least as a question of principle, especially concerning  the question of whether the universe is infinite or finite  and also potential observational effects, it is still relevant to address these differences. See for instance references~\cite{Efstathiou:2003hk,Aghanim:2018eyx,DiValentino:2019qzk,Handley:2019tkm, Efstathiou:2020wem, DiValentino:2020hov} for a recent debate regarding current observations.

\subsection*{Initial conditions and Inflation}
Vacuum decay offers a unique physical mechanism to provide the initial conditions for the evolution of the universe. The initial conditions for the classical cosmological evolution are the configurations of $\phi(t)$ and $a(t)$ right after tunnelling that we will define as  $t=0$.
\begin{itemize}
\item{\bf Open Universe\footnote{See~\cite{Gott:1982zf,Ratra:1994vw,Bucher:1994gb,Linde:1999wv,Freivogel:2005vv,Bousso:2013uia,Freivogel:2014hca,Bousso:2014jca} for related discussions.}.}
The Lorentzian equations of motion after CDL tunneling  are those for an open $k=-1$ LFRW model with the standard cosmological evolution for a scalar field with canonical kinetic terms and potential $V(\phi)$,
\begin{eqnarray}
 \left(\frac{\dot a}{a}\right)^2 & =&\frac{8\pi G}{3}\left(\frac{1}{2}\dot\phi^2+V(\phi)\right)+\frac{1}{a^2} \,, \\
0&=&\ddot{\phi}+3H\dot\phi  +V_{,\phi} \,. \nonumber
\label{eqn:fieldinflation}
\end{eqnarray}
Initial conditions consistent with the smoothness of the CDL instanton are 
\begin{equation}
\dot\phi(0)=\phi(0)=0 \,, \qquad  a(t)=t +{\mathcal O(t^3)} \,.\nonumber
\end{equation}
 Here $a(0)=0$ is a coordinate singularity. It is clear from Eqs. \eqref{eqn:fieldinflation} that $\dot a$ could not be chosen to vanish at $t=0$  for positive potentials. We can see that initially the dynamics is dominated by the curvature term $1/a^2$. The friction ($3H$) diverges at $t\to 0$, and then the initial $\dot\phi$ does not increase by much. 
 
From Eq.~\eqref{eqn:fieldinflation} we obtain the
 $\dot H$ equation for $k=-1$:
 \begin{equation}
 \dot H= \frac{\ddot a}{ a} -\left(\frac{\dot a}{a}\right)^2=-4\pi G \dot\phi^2 -\frac{1}{a^2} < 0 \,,
 \label{acceleration}
 \end{equation}
 which guarantees in general that there is no local minimum value for $a(t)$ ($\dot a=0 \implies \ddot a<0$) and that there should be at least one point for which $a=0$ (the coordinate singularity).
 
 After a critical time of order $t^*\sim (\alpha V)^{-1/2}$ with $\alpha^{-1}=3/8\pi G=3M_p^2$, the potential  starts to dominate, and the curvature term becomes less important as the universe expands.  This could mark the onset of inflation as long as the slow roll conditions are satisfied.

There are some conditions on the potential in order to achieve this scenario. In particular, it is required to have that $V''/V>1$ to be able to have a solution that satisfies the instanton boundary conditions~\cite{Jensen:1983ac} (see fig.\ref{fig:potential}). This being the opposite of the slow-roll condition, there has to be a curvature change in the scalar potential to eventually give rise to inflation.

We may be more quantitative and follow the scale factor $a(t)$ and $\phi(t)$ for times smaller than $t^*$. In this case we can expand around the initial points
$V(\phi(0)+\delta\phi) \sim \Lambda+ \beta\delta\phi $ with $\beta=V'(\phi(0)) $ and find
\begin{equation}
a(t)=\frac{\sinh (\lambda t)}{\lambda} \,, \qquad \lambda^2=\alpha\Lambda=\frac{\Lambda}{3M_p^2} \,,
\end{equation}
and the scalar field:
\begin{equation}
\phi(t)=\frac{\beta}{3\lambda^2}\left[\frac{\cosh(\lambda t)-1}{\cosh(\lambda t)+1}+ \log\left(\frac{\cosh(\lambda t)+1}{2}\right)\right].
\end{equation}
From here we can explicitly verify that for the domain of validity of this regime ($t\leq t^*$) the scale factor starts at zero and then increases linearly with time whereas the scalar field increases as  $\phi(t)\lesssim \beta t^2$. For $t>t^*\equiv 1/\sqrt{\lambda}$,  $\Lambda$ dominates as long as $\beta$ is small enough ($\beta<\Lambda/M_p$) such that $\Lambda$ dominates over both $\beta\phi$ and $\dot\phi^2$. In that case the universe starts a standard inflationary period, otherwise the field rolls fast and depending on the potential there may or may not be a standard period of inflation.

\begin{figure}[h!] 
\begin{center} 
\includegraphics[scale=1,trim=0cm 0.2cm 1cm 0cm,clip]{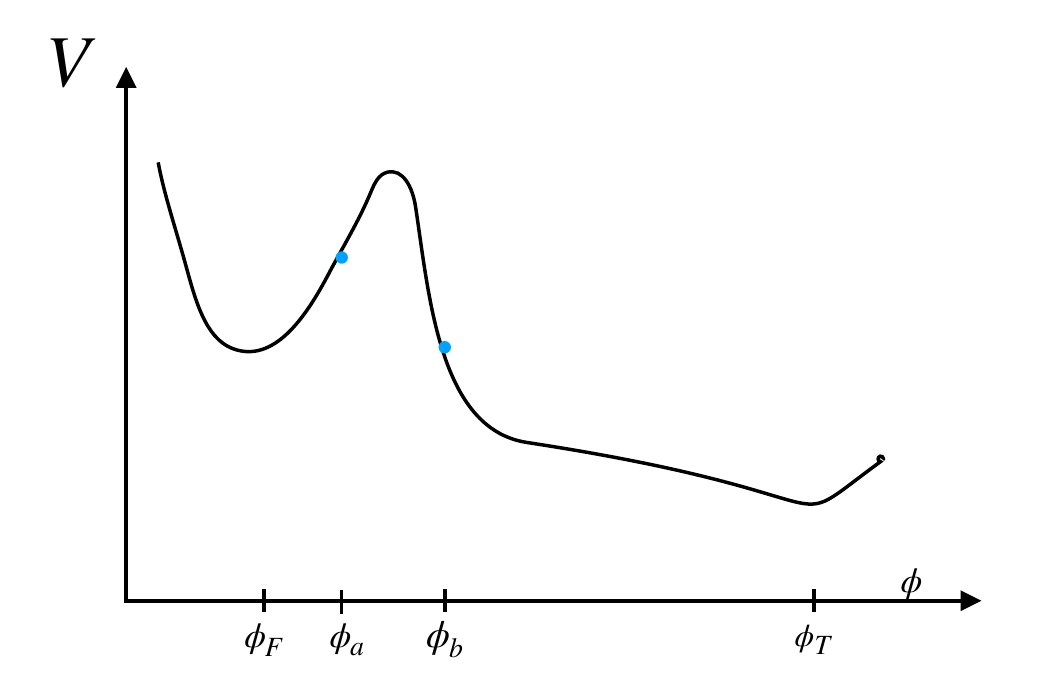}
\caption{The scalar field potential has a false vacuum at $\phi_f$ and an inflationary region on the right of $\phi_T$. In the standard CDL picture the field tunnels from $\phi_a$ to $\phi_b$.\label{fig:potential}}
\end{center} 
\end{figure}

 \item {\bf Closed Universe}. 
In the closed universe after tunneling the equations are \footnote{For relatively recent discussions on the cosmology of closed universes see for instance~\cite{White:1995qm, Ellis:2001ym, Linde:2003hc, Uzan:2003nk, Lasenby:2003ur, Masso:2006gv, Bonga:2016iuf, Ratra:2017ezv}.}:
\begin{eqnarray}
 \left(\frac{\dot a}{a}\right)^2 & =&\frac{8\pi G}{3}\left(\frac{1}{2}\dot\phi^2+V(\phi)\right)-\frac{1}{a^2} \,,\\
0&=&\ddot{\phi}+3H\dot\phi  +V_{,\phi} \,. \nonumber
\label{eqn:fieldcosmology}
\end{eqnarray}
 The negative sign in the first equation due to the positive curvature ($k=1$) changes the picture substantially. First note that the initial conditions can be fixed by imposing that at the turning point  $\pi_a,\pi_\phi=0$. This implies that the right hand side of the Friedmann equation above vanishes and the the natural initial conditions can be (making a convenient choice for $\phi(0)$)\footnote{There are  more general initial conditions for  $G_{MN}\pi^N\pi^M\neq 0$ where $G^{MN}$ is the non-positive metric in superspace. As we have described before (see Sec.~\ref{sec:ThinWallApproximation}) when this is the case there also classical solutions that go from the false to the true vacua.  In this case,  the field oscillates around the true minima for a finite time until it reaches the points where $\dot\phi=\dot a=0$ that could be used as initial conditions for the rest of the evolution of the universe.}
 
\begin{equation}
\dot\phi(0)=\phi(0)=0\,, \qquad  \dot a(0)=0\,. 
\label{initial}
\end{equation}

Contrary to the open case, there is no curvature dominated period since the curvature term has at best  to balance the energy density term in order to have $(\dot a/a)^2\geq 0$. In particular at $t=0$ we have $a(0)=3/(8\pi G V(0))\neq 0$. Different from the open universe case, the scale factor does not vanish.

Also, the equation for $\dot H$ now reads:
\begin{equation}
 \dot H=\frac{\ddot a}{ a} -\left(\frac{\dot a}{a}\right)^2=-4\pi G \dot\phi^2 +\frac{1}{a^2} \,,
 \label{acceleration2}
 \end{equation}
which, unlike the open and flat cases, can be positive or negative depending on the relative size of the kinetic terms and the curvature. Note that a positive curvature adds a positive contribution to the equation for $\dot H$. This means in particular that even though the universe is closed it does not necessarily recollapse~\cite{Barrow:1988xi}.

Repeating the quantitative analysis as for the open case, assuming the energy density is dominated by $\Lambda$, we find:
\begin{equation}
a(t)=\frac{\cosh (\lambda t)}{\lambda} \,, \qquad \lambda^2=\alpha\Lambda=\frac{\Lambda}{3M_p^2} \,.
\end{equation}
Contrary to the open case the scale factor {\it does not} vanish at any point and after nucleation it starts with a minimum value of order $a(0)\geq a_{min}\sim M_p/\sqrt{\Lambda} $ which is large enough to be in the  regime for which the classical evolution equations are valid. The scalar field evolves as:
\begin{equation}
\phi(t)=-\frac{\beta}{3\lambda^2}\left[\log\cosh(\lambda t)-{\rm sech}^2(\lambda t)+1\right] \,.
\end{equation}
Now the Hubble parameter is $H=\dot a/a=\lambda \tanh(\lambda t)$, and
\begin{equation}
\frac{\ddot a}{a}=\frac{8\pi G}{3}\left(\Lambda-\dot\phi^2\right) \,, \qquad  -\frac{\dot H}{H^2}=-{\rm csch}^2(\lambda t) \,.
\end{equation}
Unlike the open case there is no critical time  before which the curvature term dominates over the $\Lambda$ contribution in the Friedmann equation. But in order to have the potential energy to be dominated by $\Lambda$ we need $\beta\phi<\Lambda$ which happens for times $t<t^*\left(\Lambda/\beta M_p\right)^2$ and for having the kinetic energy suppressed with respect to the potential energy $\dot\phi^2\ll \Lambda$ which happens for times $t<t^*\left(\Lambda/\beta M_p\right)^2$ after that the universe stops accelerating. Both of these conditions are satisfied if $\beta\ll\Lambda/M_p$ which is the equivalent of slow-roll condition.

So we have two different possible outcomes after nucleation depending on the value of $\beta$: a short period of relative fast roll and a few e-foldings before the standard slow-roll inflation or an inflationary period right after nucleation if $\beta\ll\Lambda/M_p $. The maximum number of e-foldings from this period would be:
\begin{equation}
N_{max}= \int_0^{t_c} H dt=\log \cosh\left(\frac{t_c}{t^*}\right)\sim \frac{t_c}{t^*}\sim\frac{2}{\epsilon} \,,\qquad t_c=t^*\left(\frac{\Lambda}{\beta M_p}\right)^2 \,.
\end{equation}
Where $\epsilon=M_P^2V_{,\phi}^2/2V^2\simeq M_P^2\beta^2/2\Lambda^2< 1$ is the usual slow roll parameter. This is the maximum number of e-foldings since  at $t=0$ the nucleation happens at the minimum value of $a(t)$ by imposing $\dot a(0)=0$, implying $a(0)=a_{min}=1/\lambda$~\cite{Ellis:2001ym}. Unlike the flat and open cases in which the initial value of $a$ at the start of inflation can be as small as possible, i.e. as small as the Planck scale $l_P$, in the closed case the existence of  a lower bound for $a$ with $a_{min}$ much bigger than the Planck length implies a a much stronger upper bound in the number of e-folds in order to fit with the present value  size of the observable universe $a_0$ which could be estimated if $\delta=|\Omega -1|=1/a_0H_0 $ is measured: $N_{max}\leq \ln{a_0/a_{min}}$. Note that this is independent of the standard argument for $N\sim 60$ setting bounds on $\Omega$ today which if measured with enough precision  may differentiate between open or closed universes.

 \subsection*{Density Perturbations}
 
The magnitude of density perturbations measured by  $\delta\rho/\rho\sim H^2/\dot\phi $ grows from zero at $t=0$ to order $\delta\rho/\rho\sim \lambda^3/\beta $ close to $t=t_c$ which can be smaller or larger than standard slow-roll inflation depending on the values of $\beta$ and  $\Lambda$\footnote{Note that in the open universe case,  this quantity diverges at $t=0$ since $H$ diverges and $\dot\phi(0)=0$ there, and then increases with time.}. 

If $\beta$ does not satisfy the slow-roll condition then $N=\int_0^{t^*} H dt=\log \cosh 1\sim 1$. Therefore it is clear that in this case the scalar potential will need to flatten up through an inflection point in order to have an adequate period of inflation afterwards.

In both cases (slow or fast roll after bubble nuecleation) the fact is that $k=1$ not only provides initial conditions for inflation but also contributes to the density perturbations since in general the presence of curvature provides  a new scale. It affects the power spectrum in the sense that  the long wavelength modes which exit the horizon during the early stages of inflation may carry imprints of the spatial curvature, whereas the short wavelength modes which exit the horizon later are not affected by the spatial curvature. 

This can also be seen if we compute the power spectrum for the inflationary perturbations. As it is well known the scalar field during inflation produces an adiabatic power spectrum of scalar and tensor perturbations. Whereas for flat universes the power spectrum is nearly scale invariant, this is not the case for  open and closed universes. As the long wavelength modes leave the horizon carrying the imprint of the curvature they deviate from scale invariance at  large scales. For small curvature $\Omega_k=k/(a_0H_0) <1$ the power spectrum can be written as \footnote{A more appropriate treatment of the power spectrum after tunneling needs to take into account the fluctuations of the wall. In the case of open inflation these translate into an excited initial state which also imply deviations from scale invariance~\cite{Bucher:1994gb}. For the  closed universe solution obtained in Sec.~\ref{sec:CDLfromWDW} we leave the analysis of the inflationary  perturbations for future work.}~\cite{Cheung:2007st,Creminelli:2013cga},
\begin{equation}
P(q)=A_s q^{(n_s-1)-3}\left(1-\frac{19}{8}\frac{k}{q^2}\right)+\mathcal{O}(\Omega_k^2)) \,,
\label{eq:inflpowerspectrum}
\end{equation}
where $A_s$ is the amplitude of the scalar fluctuations   and $q$ is the comoving wavenumber. We have also neglected self interactions of the curvature perturbation. From Eq.~\eqref{eq:inflpowerspectrum} we can read that at large scales, or smaller $q$ , the power spectrum is suppressed for a closed universe ($k=+1$), but is enhanced for the same scales for an open universe.

 These deviations from scale invariance can have an effect on the CMB. At large scales the main contribution to the angular power spectrum $C_l$ is given by the Sachs-Wolfe effect\footnote{ For a derivation of the effect of the primordial power spectrum over Sachs-Wolfe effect see formula (2.6.19 ) of~\cite{Weinberg:2008zzc}}
\begin{equation}
l(l+1)C_l=\frac{4\pi}{25}A_s\left(1-\frac{19}{8}\frac{kr_L^2}{3(l-1)(l+2)}\right) \,.\label{eq:Cls}
\end{equation}
Where $r_L$ is the radial coordinate of the surface of last scattering. Then we see that at linear order in the curvature there is a suppression/enhancement of the low $l$ modes of the temperature anisotropy of the CMB depending on the sign of $k$.

Note that this  computation  assumes that the only contribution from the curvature comes from inflation and, although this is  not quite accurate, it works as a qualitative  approximation. Another  point is that the effect  described is model independent but, as we mentioned  before, there are other signatures that depend on each model that may have important consequences. For example for the case of open inflation after CDL,  there is a fast roll phase before slow roll, by which  the authors of~\cite{Freivogel:2005vv,Bousso:2013uia} have argued that because of an  anthropic bound on the duration of inflation the fast roll phase translates into a potentially observable suppression of the low $l$ modes of the CMB. This  implies that a negatively curved universe can have an effect which is  indistinguishable from  a positively curved universe, so in  order to break this degeneracy it might be necessary to study higher order correlation functions of cosmological observables. 

At small scales it can be seen from Eq.~\eqref{eq:inflpowerspectrum} that the power spectra coincide. This means that the CMB  power spectrum at large angles or small multipoles is suppressed with respect to the standard flat $\Lambda$ CDM model. Since the effect  through inflation is only  present at large scales, the power spectrum coincides with the flat case for large multipoles ($\ell \gtrsim 30$). This  is the regime that has been tested most successfully, although recently several articles have found some evidence for the closed universe inflationary model from the latest CMB observations~\cite{DiValentino:2019qzk,Handley:2019tkm,Efstathiou:2020wem}.
\end{itemize}

\subsection*{Observational Implications and the String Landscape}

The implications of a closed universe after bubble nucleation may have important observational implications and would radically affect the dynamics of the string landscape. Let us list some of them.

\begin{itemize}
\item {\bf General Prediction}. Due to the  richness of the string landscape, it has been a serious challenge to identify concrete and general predictions that could be tested with the potential to rule out the landscape paradigm. The standard belief that bubble nucleation after vacuum decay gives rise only to an open universe, has been identified as the most concrete general prediction that could be subject to experimental test at some point. However, if the outcome is a closed universe, the prediction would be exactly the opposite. At the moment we cannot rule out the possibility that an open universe could also be allowed. Furthermore note that, in principle, the idea behind the landscape is that universes are continuously produced from a series of quantum tunneling among the many different vacua. If a parent universe is closed and a daughter universe is open, this chain of universe creation would not be possible. However if both parent and daughter universes are closed this is natural. Furthermore as emphasized before in this work, the string theory landscape is a result of brane nucleation rather than a CDL like process, and should be treated as in FMP.

\item{\bf Bubble collisions}. In the landscape picture, the continuous creation of inflationary bubbles may give rise to the possibility of bubble collisions that may have left some imprint in the CMB, gravitational waves, etc. See for instance~\cite{Kleban:2011pg, Aguirre:2009ug} for an overview. One important aspect of the treatment of bubble collisions is the symmetry after the collision. Assuming open universes, each bubble spatial section has an $SO(3,1)$ symmetry which breaks to $SO(2,1)$ after the collision. Numerical relativity techniques have been developed during the past few years to address this problem (see for instance~\cite{Johnson:2015gma} and references therein). If the universe is closed the situation differs substantially, the finiteness of the volume of spatial sections affects the probability of collisions and the natural symmetry breaking would be from the $SO(4)$ symmetry of the corresponding three-spheres to $SO(3)$. This should modify the description of  bubble collisions. A detailed discussion of this interesting effect lies beyond the scope of this article.

\item{\bf Beyond bubbles and CDL}. The natural outcome of a phase transition, such as vacuum decay, is through the nucleation of bubbles of the new vacuum. Bubbles are required if the universe has infinite volume as in flat and open spacetimes. If the spacetime is closed there is the possibility that the full space and not only a region within it  can change to the new vacuum with non-vanishing probability since the volume is finite.  Note also that, as we have seen in the previous section, the fact that in mini-superspace the kinetic energy for the scale factor is negative whereas that for the scalar field is positive (Eq.~\eqref{eq:calH}), then the Hamiltonian constraint ${\cal H}=0$ allows to pass classically through the barrier of the scalar potential. This does not appear without gravity nor in the Euclidean approach nor in the Hartle-Hawking case without scalar fields. This would then be the leading contribution to the ratio of probabilities for the creation of each of the two dS spaces. Even though these conclusions may be artefacts of the mini-superspace approximation they would affect the structure of the landscape and deserve further study.

\item{\bf Number of e-folds}. For the open universe case,  reference~\cite{Freivogel:2005vv}  extracted a lower bound on the number of e-foldes during inflation $N>59.5$ similar to the observed bound $N>62$ (modulo logarithmic corrections due to the different epochs of matter/radiation domination after inflation). For $k=1$ as we discussed before, there is an {\it upper} bound on the number of e-folds (again with $\Delta N\sim 2$ e-folds with respect with the observed one)~\cite{Uzan:2003nk}. The reason being, as mentioned above, that for the closed universe, there is a minimum value of the scale factor $a_{min}\sim M_p/\sqrt{\Lambda}$ that already starts large (contrary to the flat and open cases in which $a\sim 0$ before inflation and have in principle no limit on the maximum number of e-folds before reproducing the standard cosmology after inflation). This tends to favour concave models of inflation and disfavour models, such as chaotic inflation, that may have an essentially unlimited number of e-folds.

\item{\bf Power suppression}. As we have seen, both open and closed universes introduce a new cosmological scale, the curvature, and affect the density perturbations observed from the CMB. The net effect is  a suppression of the power  spectrum at large angles or small multipoles. This effect may have two different origins, one due to the curvature and the other if there is a period of fast rolling before inflation.

\end{itemize}

\section{Conclusions}

We have presented here an extension of the standard quantum mechanics WKB approximation to field theory and gravity. Expanding on previous approaches we developed a general geometric formalism to extend the WKB to wave functions in Wheeler's superspace and found explicit expressions for vacuum transition probabilities interpreted as ratios of the square of wave functions of the two different configurations for both field theory and gravity. 

In field theory we presented explicitly two different cases with two vacua. The first is the standard scalar potential with two minima at finite field value and the second one with the second minimum corresponding to a runaway. In both cases we reproduce the standard Coleman results at leading order. In the second case we  found the (potential) bound states and corresponding resonances to provide the explicit expression for the decay rate Eq.~\eqref{eq:lifetime} which agrees with the Euclidean approach at leading order but differs in the pre-factor. We further emphasised our approach is the natural generalisation of the standard quantum mechanical WKB calculation and does not have subtle issues such as the  handling of negative modes and trusting the dilute instanton approximation.

 In the gravity case, our results can be summarised in two directions. In one way they confirm the results obtained by Euclidean methods. In particular we provide a Lorentzian perspective for the estimation and interpretation of the decay rates and the wall trajectory after tunneling, illustrating the validity of the Euclidean techniques at least as far as getting the leading exponential behavior goes. On the other hand we point out that there are substantial differences from the Euclidean approach. In particular for the mini-superspace case we found that:
 
 \begin{itemize}
 \item Selecting a very particular path we can reproduce exactly the decay rate as computed by CDL, however the interpretation of this path is not at all clear.
 
 \item The natural transition rate gives simply the ratio of the two corresponding Hartle-Hawking wave functions, without a tension term as in Eq.~\eqref{eq:HHV} which gives dominant decay rate as compared to CDL. It seems then that this mini-superspace approach is actually a generalisation of Hartle-Hawking/Vilenkin/Linde transitions than to CDL.
 
 \item The fact that the metric in superspace is not positive definite, as manifested for instance in the Hamiltonian constraint Eq.~\eqref{eq:calH}, allows for classical paths to connect the two dS vacua without the need to pass through or across the barrier. This will not have an exponentially suppressed decay rate but it depends on the initial conditions.
 
 \item The natural dS foliation corresponds to the global slicing which leads to a closed universe as in Hartle-Hawking/Vilenkin-Linde unlike the open universe claimed by CDL. Here, recall that the driving argument was based on the analytic continuation triggered by the argument of the scalar field. Surfaces of constant $\phi$ would give rise to the hyperbolic foliation of spacetime. However, as noted already in~\cite{Coleman:1980aw} in the extreme thin-wall approximation we only have two values of the cosmological constant and there is no preferred dS space foliation. This means that already in the Euclidean framework described for instance in~\cite{Brown:1988kg} horizontal slices of dS could have been chosen for the regions inside and outside the bubble with a closed universe.
 
 \item In mini-superspace there is no way to discuss bubble nucleation and therefore the transition should be interpreted between two entire dS spaces. This computation really makes sense only for $k=1$  since the spatial volume in global slicing is finite. A full description beyond mini-superspace is needed in order to properly include the bubble. 
In the Hamiltonian approach to quantum tunneling developed in~\cite{Fischler:1989se,Fischler:1990pk} it is clear that the geometry inside and outside the bubble is also of a closed universe~\cite{deAlwis:2019dkc}. In this approach  the metric depends not only on time but also on the radial coordinate $r$. This reinforces our conclusion. This formalism only compares the transition between two different cosmological constants without considering a scalar field with the corresponding potential. This corresponds to the extreme thin-wall approximation or brane nucleation in string theory. A full Hamiltonian approach including the scalar field dynamics is not yet available.

\item Our approach and that of~\cite{Fischler:1989se,Fischler:1990pk}  started with a spherically symmetric geometry that was natural  to describe the bubble after the transition but in principle we could have chosen a flat or negatively curved spacetime to start with. The fact that these spaces have infinite volume renders  the volume integrals problematic. For instance in Eq.~\eqref{eq:Sg} the $2\pi^2$ factors come from integrating the volume of the three-sphere  which in the flat and open cases would diverge. This would require a proper volume regularisation before extracting physical information. Similar volume integrals for closed slicings appear in the approach of~\cite{Fischler:1989se,Fischler:1990pk}. Furthermore, in~\cite{Fischler:1989se,Fischler:1990pk} a prescription is provided to Euclideanise their results, the corresponding analytic continuation is simply $t\rightarrow it$ which is not the analytic continuation proposed by CDL. 

\item We believe that  the possibility that the geometry of the bubble after nucleation corresponds to a closed FLRW universe, contrary to CDL,  should be  seriously considered. Indeed this is the natural implication of brane nucleation in string theory as we have argued in this paper. This may have important physical implications if our universe is described in terms of a bubble after vacuum transitions as discussed at the end of the previous section. Besides the deep implications of having a finite against an infinite universe, with finite number of stars and galaxies, it may eventually be tested if there is a definite way of determining the curvature of the universe. For the string theory landscape, it will at least eliminate the standard claim that detecting a closed universe would rule out the multiverse. These are important cosmological questions that deserve further scrutiny.
 
 \end{itemize}

\section*{Acknowledgements}
We thank Stefano Ansoldi, John Bohn, Cliff Burgess,  Jim Hartle,  Veronica Pasquarella,  Jorge Santos, Andreas Schachner and Alexander Westphal for useful discussions. We also thank Andreas Schachner for  comments on the manuscript. We thank Thomas Hertog for comments on a previous version. The work of SC is supported through MCIU (Spain) through contract PGC2018-096646-A-I00 and by the IFT UAM- CSIC Centro de Excelencia Severo Ochoa SEV-2016-0597 grant. FM is funded by a UKRI/EPSRC Stephen Hawking fellowship, grant reference EP/T017279/1 and partially supported by STFC consolidated grant ST/P000681/1. The work of FQ has been partially supported by STFC consolidated grants ST/P000681/1, ST/T000694/1.
\appendix

\section{De Sitter geometry}
\label{sec:dSGeometry}

\subsection{Different foliations}
dS spacetime can be defined on a hypersurface,
\begin{equation}
-X_0^2+X_1^2+X_2^2+X_3^3+X_4^4=\Lambda^2 \,,
\end{equation}
in five dimensional Minkowski
\begin{equation}
ds^2=-dX_0^2+dX_1^2+dX_2^2+dX_3^3+dX_4^4 \,.
\label{eqn:Lorentzian5metric}
\end{equation}
\subsection*{Global Slicing}
The entire dS space can be  covered by,
\begin{eqnarray}
X_0&=&\Lambda\sinh(t/\Lambda) \,,\nonumber\\\label{foliation:global}
X_1&=&\Lambda \cosh(t/\Lambda)\cos(\rho)\\
X_2&=&\Lambda \cosh(t/\Lambda)\sin(\rho)\sin\theta\cos\phi \,, \nonumber\\
X_3&=&\Lambda  \cosh(t/\Lambda)\sin(\rho)\sin\theta\sin\phi \,, \nonumber\\
X_4&=&\Lambda \cosh(t/\Lambda)\sin(\rho)\cos\theta \,, \nonumber
\end{eqnarray}
and the metric in Eq.~\eqref{eqn:Lorentzian5metric} becomes,
\begin{equation}
ds^2=-dt^2+\cosh^2( t/\Lambda)\left(d\rho^2+\sin^2(\rho) d\Omega_2^2\right) \,,
\end{equation}
where constant time slices have $O(4)$ invariance.
Unlike the other foliations this parametrisation has the full $SO(4,1)$ invariance. The global conformal metric in Eq.~\eqref{eq:globalconformal} is related to this
by the relation $\cosh(t/\Lambda)=1/\cos(T)$ with $H\equiv 1/\Lambda$.
\subsection*{Causal patch}
The geometry as observed by an observer moving along a constant time hypersurface on one of the hemispheres, is given by the following foliation,
\begin{eqnarray}
X_0&=&\sqrt{\Lambda^2-r^2}\cosh(\tau/\Lambda) \,, \label{foliation:causalpatch}\\
X_1&=&\sqrt{\Lambda^2-r^2}\sinh(\tau/\Lambda) \,, \nonumber\\
X_2&=&r\sin\theta\cos\phi \,, \nonumber\\
X_3&=&r\sin\theta\sin\phi \,, \nonumber\\
X_4&=&r\cos\theta \,.\nonumber
\end{eqnarray}
The metric becomes,
\begin{equation}
ds^2=\left(1-\frac{r^2}{\Lambda^2}\right)d\tau^2+\left(1-\frac{r^2}{\Lambda^2}\right)^{-1}d\tau^2+r^2d\Omega_2^2 \,,
\end{equation}
where $r$ varies between $0$ and $\Lambda$ and $\tau$ goes from $0$ to $\infty$. Notice that at $r=\Lambda$ the metric is not singular but there is a horizon.
\subsection*{Open slicings}
\begin{figure}[h!] 
\begin{center} 
\includegraphics[scale=0.35,trim=7cm 4.5cm 0cm 5cm,clip]{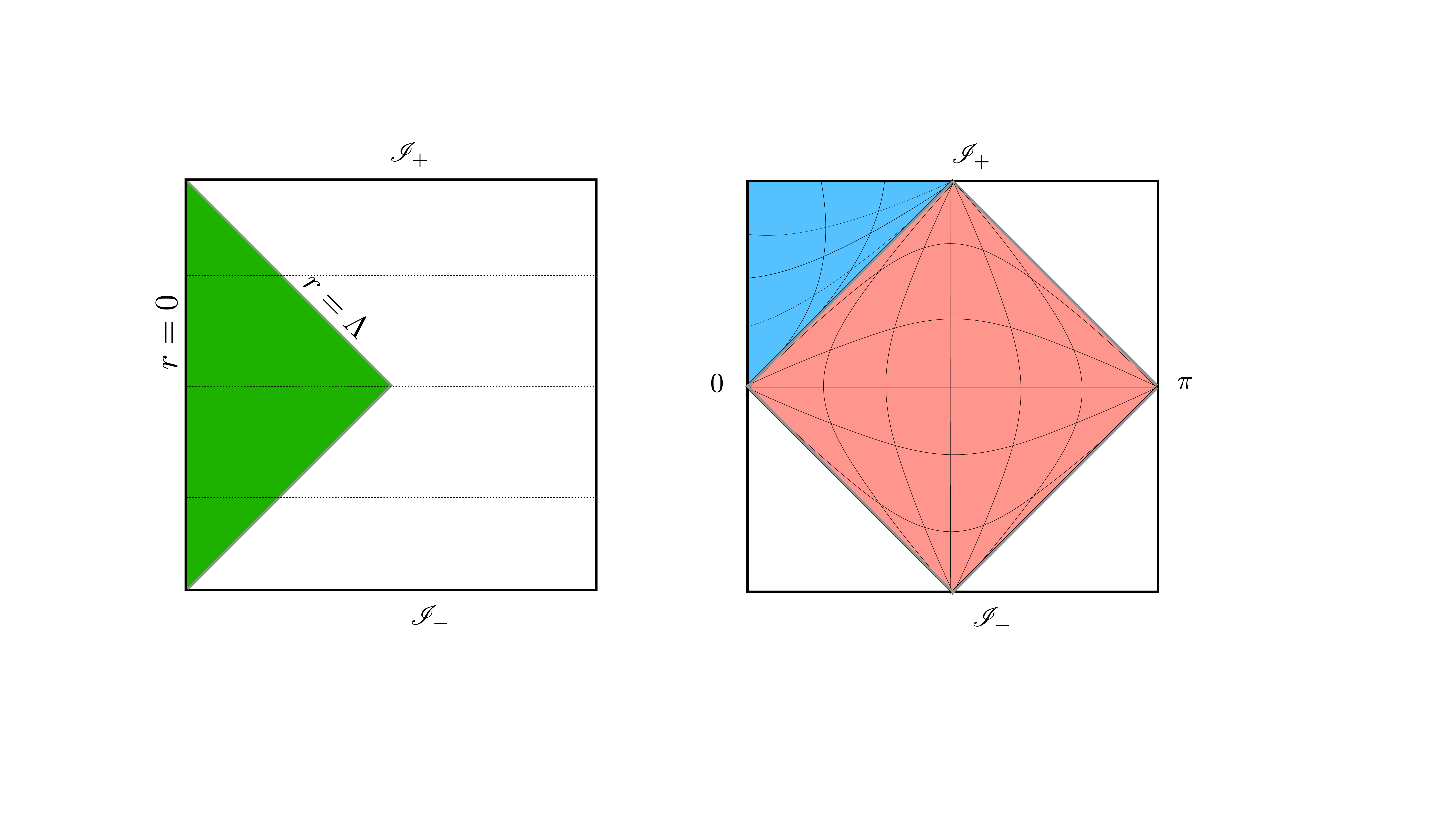}
\caption{Left panel: The green part is the region covered by the static patch of an observer at the south pole. The horizontal dotted lines are the constant time slices in the global coordinates. Right panel:  Penrose diagram for the open slicing metrics in Eq.~ \eqref{foliation:timelikeslices} and Eq.~\eqref{foliation:openslicing}. The timelike lines are hypersurfaces of constant $\xi$ and spacelike lines are hypersurfaces of constant $\chi$ \label{fig:deSitter_hyperbolic_slicing}}
\end{center} 
\end{figure}

 Picking the coordinates,
\begin{eqnarray}
X_0&=&\Lambda \sin(\xi/\Lambda)\sinh(\chi)\,, \label{foliation:timelikeslices}\\
X_1&=&\Lambda\cos(\xi/\Lambda) \,, \nonumber\\
X_2&=&\Lambda \sin(\xi/\Lambda)\cosh(\chi)\sin\theta\cos\phi \,, \nonumber\\
X_3&=&\Lambda \sin(\xi/\Lambda)\cosh(\chi)\sin\theta\sin\phi \,, \nonumber\\
X_4&=&\Lambda \sin(\xi/\Lambda)\cosh(\chi)\cos\theta \,, \nonumber
\end{eqnarray}
the metric in Eq.~\eqref{eqn:Lorentzian5metric} becomes,
\begin{equation}
ds^2=d\xi^2+\Lambda^2\sin^2(\xi/\Lambda)(-d\chi^2+\cosh^2\chi^2 d\Omega_2^2)\,.
\end{equation}
This foliation does not describe the full dS but only the causal patch of an observer at the centre of the hyperboloid. The region covered is  depicted in red in Fig.~\ref{fig:deSitter_hyperbolic_slicing}. Note that  the analytical  continuation $\chi\to i \chi$ leads to
\begin{equation}
ds_E^2=d\xi^2+\Lambda^2\sin^2(\xi/\Lambda)(d\chi^2+\cos^2\chi^2 d\Omega_2^2) \,,
\label{metric:euclidean}
\end{equation}
which is the euclidean metric of a 4 sphere. This  continuation was equivalent to make $X_0\to i X_0$.
Another useful foliation is given by,
\begin{eqnarray}
X_0&=& \Lambda \cosh(\xi/\Lambda)\nonumber \,, \\\label{foliation:openslicing}
X_1&=&\Lambda \sinh(\xi/\Lambda)\sinh(\chi) \,, \\
X_2&=&\Lambda \sinh(\xi/\Lambda)\cosh(\chi)\sin\theta\cos\phi \,, \nonumber\\
X_3&=&\Lambda  \sinh(\xi/\Lambda)\cosh(\chi)\sin\theta\sin\phi \,, \nonumber\\
X_4&=&\Lambda  \sinh(\xi/\Lambda)\cosh(\chi)\cos\theta \,, \nonumber
\end{eqnarray}
which foliates dS  with spacelike hyperboloids. This foliation covers the  region that lies between the outside region of  the causal patch and the asymptotic future. It intersects the region covered by Eq.~\eqref{foliation:timelikeslices}, at $\xi=0$ which are the null surfaces for the observer on the hemisphere . The metric in Eq.~\eqref{eqn:Lorentzian5metric} becomes,
\begin{equation}
ds^2=-d\xi^2+\Lambda^2\sinh^2(\xi/\Lambda)\left(d\chi^2+\sinh^2\chi d\Omega_2^2\right) \,,
\label{eqn:opendeSitter}
\end{equation}
which is also  pictured as the region in blue in Fig.~\ref{fig:deSitter_hyperbolic_slicing}.
To get  the Euclidean continuation we need to do the transformations, $\chi\to i \chi$  and $\xi\to i \xi$, which lead to Eq.~\eqref{metric:euclidean}. finally note that  this transformation are equivalent to $X_0\to iX_0$ .

\subsection{Analytic Continuation in CDL }
\label{sec:CDLReview}

Given two vacua $\phi_F$ and $\phi_T$ such that $V(\phi_F)>V(\phi_T)\geq 0$, according to~\cite{Coleman:1980aw}  to calculate the scalar field that interpolates between the two minima first we need to consider the Euclidean solution that extremises the action in order to derive the remaining geometry inside and outside the bubble. This was done in several steps:

\begin{itemize}
\item In field theory, the Euclidean bounce solution is  $O(4)$ invariant in such a way that the scalar field depends on the Euclidean distance 
$\xi^2=|x|^2+\tau^2$. Analytic continuation changes this to a $O(3,1)$ and $\xi^2\rightarrow |x|^2-t^2$. Once gravity is included the corresponding line element is assumed to share that symmetry. 
\begin{figure}[h!] 
\begin{center} 
\includegraphics[scale=0.4,trim=0cm 0.5cm 1cm 0cm,clip]{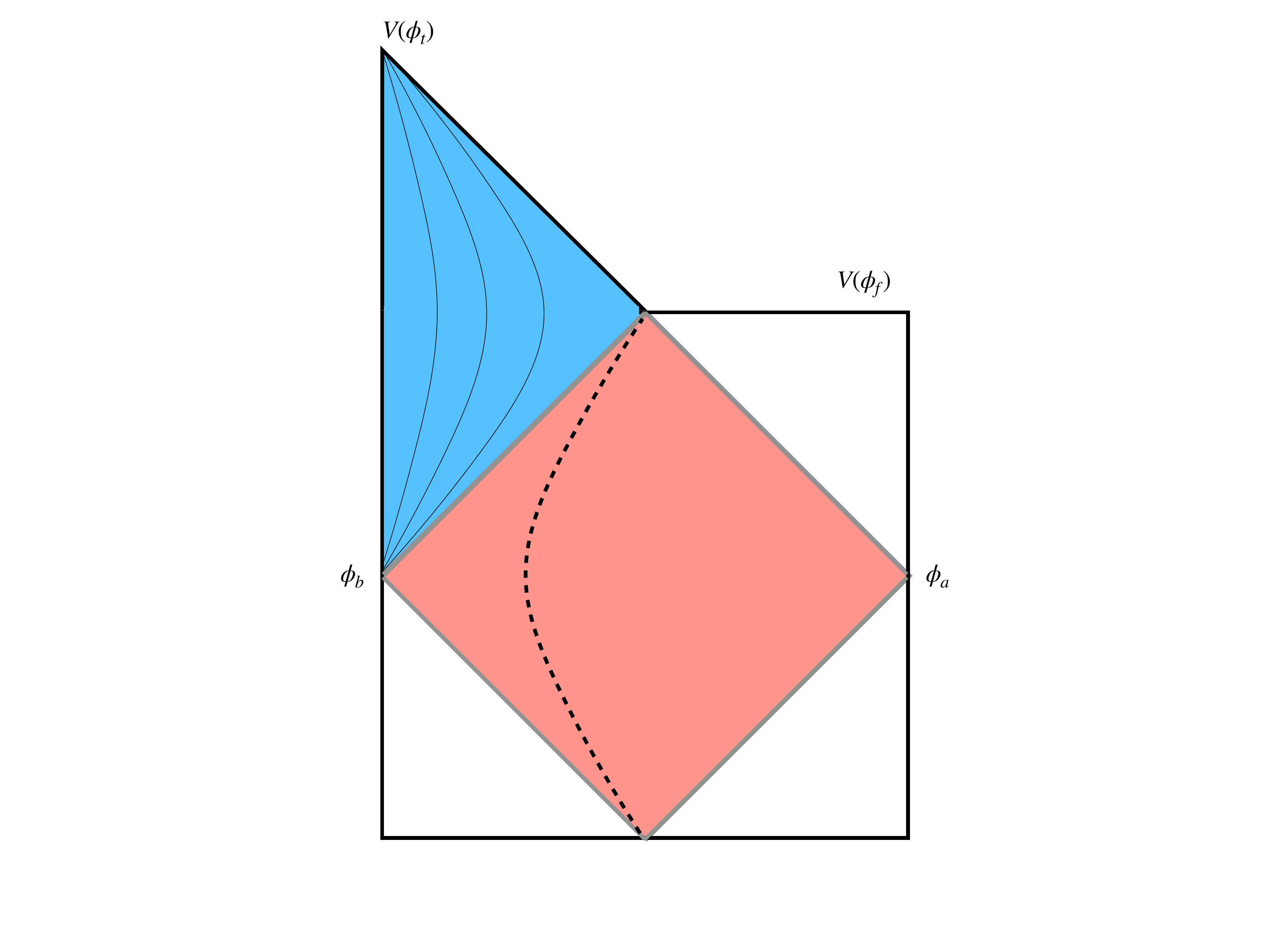}
\caption{Penrose diagram for the dS to Minkowski transition mediated by a  CdL instanton. The red causal diagram is obtained after analytically continue the $O(4)$ instanton to $O(3,1)$ whose metric is given in Eq.~\eqref{eq:CDLCausalDiamond}. The dotted line is the wall which is also a  constant  $\sigma$ slice.The blue region is  open FRW with  $V(\phi_t)=0$ obtained after analyticaly continue Eq.~\eqref{eq:CDLCausalDiamond} past the lightcone. In this region we also draw the constant radius hypersurfaces.\label{fig:CdLPenroseDiagram}}
\end{center} 
\end{figure}

\item Starting with the $O(4)$ symmetric Euclidean dS  metric,
\begin{equation}
ds^2=d\xi^2+\rho^2(\xi)(d\psi^2+\sin^2\psi d\Omega_2^2)
\end{equation}
and  writing $\psi\to \pi/2+i\sigma$ we get the $O(3,1)$ invariant metric,
\begin{equation}
ds^2=d\xi^2+\rho^2(\xi)(-d\sigma^2+\cosh^2\sigma d\Omega_2^2)
\end{equation}  
where $\sigma$ now runs from $-\infty$ to $\infty$. This spacetime  is dS foliated by timelike hyperboloids.
If we fix the angles and write $d\xi^2=\rho^2(\xi)(dy^2)$ we get,
\begin{equation}
ds^2=\rho^2(y)\, (-d\sigma^2+dy^2) \label{eq:CDLCausalDiamond}
\end{equation}
whose geometry is represented by the causal diamond in Fig.~\ref{fig:CdLPenroseDiagram}. The wall lies within this region as indicated by the dotted timelike hypersurface in Fig.~\ref{fig:CdLPenroseDiagram}. An observer in the hemisphere will see the wall moving at a speed approaching the speed of light.

\item This space is not geodesically complete, because for timelike geodesics it is possible to go past the lightlike hypersurfaces. To complete the description we can make a further analytical continuation $\sigma\to i \pi/2+\chi$ and a rotation $\xi\to it$ which leads to,
\begin{equation}
ds^2=-dt^2+\rho^2(-i \xi)(d\chi^2+\sinh^2\chi d\Omega_2^2).
\end{equation}
This region describes an  FRW open space, and covers the blue patch in Fig.~\ref{fig:deSitter_hyperbolic_slicing}. Note that  the analytical continuation  $\sigma\to -i \pi/2+\chi$  describes the upper right region of the diagram.    Constant time hypersurfaces end on  $\mathscr{I}_+$ at $\xi=\pi/2$

\item To  interpret the diagram in terms of dynamics of the scalar field, let us assume that the tunneling points where $\phi_b$ and $\phi_a$, as indicated in Fig.~\ref{fig:potential}. Both points are hemispheres of dS where $\rho$ vanishes. The wall  separating the two regions is inside the causal diamond as indicated in Fig.~\ref{fig:CdLPenroseDiagram} by the dotted line. The  left region outside the causal diamond describes the dynamics after the tunneling while the right region describes the false vacuum dynamics.  Spacelike hypersurfaces in these region  represent  constant field surfaces.  Then after the tunneling the field rolls down to the true vacuum $V(\phi_t)$ at ${\mathcal I}_+$.
After analytic continuation the surfaces of constant field values correspond to constant values of $|x|^2-t^2$ which are hyperbolae. This defines the natural foliation of the spacetime.
\end{itemize}

In Fig.~\ref{fig:CdLPenroseDiagram} we have assumed that the true vacuum is Minkowski.  In the case of dS  this only extends up to the horizontal line at $T=\pi/2$.

 \bibliographystyle{elsarticle-num}

\end{document}